\documentclass[11pt, oneside]{article}   	
\usepackage{jheppub}
\usepackage{amsmath}
\usepackage{amssymb}
\usepackage{amsfonts}
\usepackage{amsthm}
\usepackage{amsbsy}
\usepackage{dsfont}
\usepackage{array}
\usepackage{mathtools}
\usepackage[usenames,dvipsnames]{xcolor}
\usepackage{subcaption}
\usepackage{dsdshorthand}
\usepackage{graphicx}
\usepackage[vcentermath]{youngtab}
\usepackage{multirow}
 \usepackage{relsize}
 \usepackage{marginnote}

\usepackage{tikz}
\usetikzlibrary{trees}
\usetikzlibrary{decorations.pathmorphing}
\usetikzlibrary{decorations.markings}
\usetikzlibrary{shapes.misc} 

\usepackage[vcentermath]{youngtab}
\newcommand{\myng}[1]{\,{\tiny\yng #1}\,}

\newcommand{\symtensor}{\<{\myng{(2)},\bullet}\>}
\newcommand{\antisymtensor}{\<{\myng{(1,1)},\bullet}\>}
\newcommand{\bifundamental}{\<{\myng{(1)},\myng{(1)}}\>}

\newcommand\ONc{O(N/2)^2\rtimes \Z_2}
\newcolumntype{L}{>{$}l<{$}} 
 
 \newcolumntype{R}{>{$}r<{$}}
\newcolumntype{M}{R@{$|{}\>$}L}

\newcommand{\thalf}{\tfrac{1}{2}}

\newcommand{\eps}{\epsilon}
\newcommand{\GN}{{\rm GNY}}

\newcommand{\up}{\uparrow}
\newcommand{\dn}{\downarrow}
\newcommand{\nmax}{n_\text{max}}
\usepackage{multirow}

\makeatletter
\def\@fpheader{\ }
\makeatother

\title{The Gross-Neveu-Yukawa archipelago}
\author[a]{Rajeev S. Erramilli,}
\author[b]{Luca V. Iliesiu,}
\author[c,d,e]{Petr Kravchuk,}
\author[f]{Aike Liu,}
\author[a]{David Poland,}
\author[f]{David Simmons-Duffin}
\affiliation[a]{Department of Physics, Yale University, 217 Prospect St, New Haven, CT 06520, USA}
\affiliation[b]{Stanford Institute for Theoretical Physics, Stanford University, 382 Via Pueblo Mall, Stanford, CA 94305, USA}
\affiliation[c]{School of Natural Sciences, Institute for Advanced Study, 1 Einstein Dr, Princeton, NJ 08540, USA}
\affiliation[d]{Simons Center for Geometry and Physics, Stony Brook University, 100 Nicolls Road, Stony Brook, NY 11794-3840, USA}
\affiliation[e]{Department of Mathematics, King's College London, Strand, London, WC2R 2LS, UK}
\affiliation[f]{Walter Burke Institute for Theoretical Physics, Caltech, 1200 East California Blvd, Pasadena, CA 91125, USA}

\emailAdd{rajeev.erramilli@yale.edu}
\emailAdd{liliesiu@stanford.edu}
\emailAdd{petr.kravchuk@kcl.ac.uk}
\emailAdd{aliu7@caltech.edu}
\emailAdd{david.poland@yale.edu}
\emailAdd{dsd@caltech.edu}

\date{}
\abstract{We perform a bootstrap analysis of a mixed system of four-point functions of bosonic and fermionic operators in parity-preserving 3d CFTs with $O(N)$ global symmetry. Our results provide rigorous bounds on the scaling dimensions of the $O(N)$-symmetric Gross-Neveu-Yukawa (GNY) fixed points, constraining these theories to live in isolated islands in the space of CFT data. We focus on the cases $N=1,2,4,8$, which have applications to phase transitions in condensed matter systems, and compare our bounds to previous analytical and numerical results.}
\preprint{CALT-TH 2022-027}

\begin{document}

\maketitle
\pagenumbering{roman}
\setcounter{page}{2}
\pagenumbering{arabic}
\setcounter{page}{1}

\section{Introduction}

The conformal bootstrap \cite{Polyakov:1974gs,Ferrara:1973yt,Mack:1975jr} has emerged as a powerful tool to rigorously\footnote{Throughout this paper, when stating that our bootstrap results are rigorous we mean that they do not rely on any unstated assumptions about QFTs. However, the bounds are not completely rigorous in the mathematical sense since they rely on some technical assumptions about our search algorithms (for example, over OPE space). \label{footnote:define-rigor}} constrain CFT data of strongly-coupled fixed points. Based solely on unitarity, symmetry, and assumptions about gaps in the spectrum of scaling dimensions, the bootstrap has produced stringent bounds on  critical exponents of several universality classes describing real-world statistical and quantum phase transitions. These include the 3d critical Ising model \cite{ElShowk:2012ht,El-Showk:2014dwa,Kos:2014bka,Simmons-Duffin:2015qma,Kos:2016ysd,Simmons-Duffin:2016wlq} which describes liquid-vapor transitions and uniaxial magnets, and the $O(N)$ models \cite{Kos:2013tga,Kos:2015mba,Kos:2016ysd} --- in particular the $O(2)$ model \cite{Chester:2019ifh} which describes the superfluid  transition in ${}^4$He, and the $O(3)$ model \cite{Chester:2020iyt} which describes classical Heisenberg ferromagnets.

The Ising and $O(N)$ models are perhaps the simplest 3d universality classes that can be reached via a renormalization group (RG) flow from a scalar theory. In this work, we focus on perhaps the simplest 3d universality class involving {\it fermions}: the $O(N)$-symmetric Gross-Neveu-Yukawa (GNY) model. The GNY model contains $N$ Majorana fermions $\psi_i$ transforming in the vector representation of $O(N)$, interacting with an $O(N)$-singlet pseudoscalar $\phi$  \cite{Gross:1974jv}. 
The Lagrangian is\footnote{Here we are following the conventions in appendix A of \cite{Iliesiu:2015qra} and contracting the indices of the two components of the Majorana fermions by $\psi_i \psi_i = \Omega^{\alpha \beta} \psi_{i,\alpha} \psi_{i, \beta}$.   Note that $\psi_i\psi_i$ is parity-odd --- hence the Yukawa term $\f \psi_i \psi_i$ preserves parity, since $\f$ is a pesudoscalar.}
\be
 \label{eq:lagrangian}
	\cL_\GN = -\frac12 (\partial \phi)^2 - i \frac{1}{2}  \psi_i \slashed{\partial} \psi_i -\frac{1}{2}m^2\phi^2 -\frac{\lambda}{4}\phi^4 - i \frac{g}{2}  \phi \psi_i\psi_i .
\ee 
This theory has a critical value of \(m^2\) below which \(\phi\) spontaneously gets a nonzero vacuum expectation value (VEV), giving a mass to $\psi_i$ and, consequently, breaking parity. Above the critical value of \(m^2\), the VEV of $\f$ vanishes, parity is preserved, and the fermions are massless. At the critical value of $m^2$, this theory is expected to flow to a CFT with a single relevant parity-even $O(N)$-singlet scalar operator $\epsilon \sim \phi^2$.

Beyond serving as one of the simplest models of scalar-fermion interactions in quantum field theory, the GNY universality classes have been proposed to describe a variety of quantum phase transitions in condensed matter systems with emergent Lorentz symmetry. For example, this model and some of its variations have been proposed to describe phase transitions in graphene (using $N=8$) \cite{herbut2006interactions,Herbut:2009qb, herbut2009relativistic, Mihaila:2017ble},  time-reversal symmetry breaking in d-wave superconductors (also for $N=8$) \cite{vojta2000quantum, vojta2003quantum}, and time-reversal-symmetry breaking transition of edge-modes in topological superconductors  (for several low values of $N$) \cite{Ziegler:2021yua}. For the special case $N=1$, the GNY critical point is expected to exhibit emergent supersymmetry at the transition \cite{Grover:2013rc}.    

The GNY models have been studied previously with the conformal bootstrap in \cite{Iliesiu:2015qra,Iliesiu:2017nrv}. Those works performed a bootstrap analysis of a single four-point correlator of fermionic operators $\< \psi_i \psi_j \psi_k \psi_l\>$.\footnote{To be more precise, in \cite{Iliesiu:2015qra} no global symmetry was assumed, while \cite{Iliesiu:2017nrv} considered external fermions that transformed in the vector representation of an $O(N)$ global symmetry as in the GNY theories.} The resulting bounds exhibited a sequence of kinks on the boundary of the space of allowed CFT data, which showed good agreement with perturbative estimates of the scaling dimensions of the GNY critical points (such as the $\epsilon$ \cite{gracey1990three, rosenstein1993critical, zerf2016superconducting, Gracey:2016mio, fei2016yukawa, Mihaila:2017ble, Zerf:2017zqi, Ihrig:2018hho} and large-$N$ expansions \cite{Gracey:1992cp,Derkachov:1993uw,Gracey:1993kc, Petkou:1996np, Moshe:2003xn, Iliesiu:2015qra, fei2016yukawa, Manashov:2017rrx, Gracey:2018fwq}). However, bootstrapping the four-fermion correlator was not enough to constrain the GNY theories to lie within isolated islands. 

In this paper we expand this study to include a mixed system of scalar and fermionic operators, characterized by their representations under parity and the global $O(N)$ symmetry. We focus on the critical points with $N=1,\,2,\,4, $ and $8$, which capture the experimentally-relevant transitions described above and are outside the perturbative control of the large-$N$ expansion. The four-point functions we analyze include all combinations allowed by the symmetries of $\psi_i$ (the lowest dimension fermion in the vector representation of $O(N)$), $\sigma \sim \phi$ (the lowest dimension, parity-odd, $O(N)$ singlet scalar), and $\epsilon \sim \phi^2$ (the lowest dimension, parity-even, $O(N)$ singlet scalar). We refer to $\s,\psi_i$, and $\e$ as ``external operators" (as opposed to the ``internal operators" that appear in their OPEs). By simultaneously imposing crossing symmetry and unitarity for all four-point functions of external operators, we show that each GNY critical point lies within an isolated island that severely constrains its low-lying scaling dimensions. 

\begin{table}[t!]
\begin{center}
	\resizebox{\columnwidth}{!}{
	\begin{tabular}{l | L L L | L L L}
		\hline\hline
		 & \hspace{0.5cm} \Delta_\psi & \hspace{0.5cm} \Delta_\sigma &\hspace{0.5cm}  \Delta_\epsilon&\hspace{0.5cm}  \eta_\psi &\hspace{0.5cm}  \eta_\phi &\hspace{0.5cm}  \nu^{-1} \\
		\hline
		\(\mathbf{N=2}\) & &  &  \\
		\(n_\text{max}=18, \Delta_{\sigma'}>2.5\)  
		 & 1.0672\mathbf{(25)} & 0.657\mathbf{(13)} & 1.74\mathbf{(4)} & 0.134\mathbf{(5)} & 0.313\mathbf{(25)} & 1.26\mathbf{(4)} \\
		\(n_\text{max}=18, \Delta_{\sigma'}>3\)  
		 & 1.06861\mathbf{(12)} & 0.6500\mathbf{(12)} & 1.725\mathbf{(7)} & 0.13722\mathbf{(24)} & 0.3000\mathbf{(23)} & 1.275\mathbf{(7)}\\
		$\epsilon$-exp w/DREG$_3$ \cite{Ihrig:2018hho} &  1.07(2)  & 0.6467(21)  &  1.724(15) & 0.1400(39) & 0.2934(42) & 1.276(15)\\
		Monte Carlo \cite{Tabatabaei:2021tqv} & 1.068(3)  & 0.655(5) &  1.81(3) & 0.136(5) &  0.31(1) &1.19(3) \\
		\hline 
		\(\mathbf{N=4}\) & & & \\ 
		\(n_\text{max}=18, \Delta_{\sigma'}>3\)  
		  & 1.04356\mathbf{(16)}& 0.7578\mathbf{(15)} & 1.899\mathbf{(10)} & 0.08712\mathbf{(32)} & 0.5155\mathbf{(30)} & 1.101\mathbf{(10)}  \\
		$\epsilon$-exp w/DREG$_3$ \cite{Ihrig:2018hho} &  1.051(6)  & 0.744(6) & 1.886(33) & 0.102(12) & 0.487(12) & 1.114(33) \\
		Monte Carlo* \cite{Huffman:2019efk} & - & 0.755(15) & 1.876(13) & - & 0.51(3)  & 1.124(13) \\
		\hline
		\(\mathbf{N=8}\) & & & \\
		\(n_\text{max}=18, \Delta_{\sigma'}>3\)  
		 & 1.02119\mathbf{(5)} & 0.8665\mathbf{(13)} & 2.002\mathbf{(12)} & 0.04238\mathbf{(11)} & 0.7329\mathbf{(27)} & 0.998\mathbf{(12)} \\
		$\epsilon$-exp w/DREG$_3$ \cite{Ihrig:2018hho} & 1.022(6)  & 0.852(8) & 2.007(27) & 0.043(12) & 0.704(15) & 0.993(27) \\
		Monte Carlo* \cite{Liu:2019xnb} & 1.025(10)  & 0.79(1) & 2.0(1) & 0.05(2) & 0.59(2)  & 1.0(1)  \\
		\hline\hline
	\end{tabular}}
\caption{A summary of the bootstrap estimates obtained in this paper for the three external operator that we study. Error bars in \textbf{bold} are rigorous.${}^\text{\ref{footnote:define-rigor}}$ We compare these result to those obtained from the $\epsilon$-expansion  and previous Monte Carlo studies. Methods denoted by * indicate that they study the closely related chiral Ising fixed point as opposed to the model studied in this work. The \(\epsilon\)-expansion work \cite{Ihrig:2018hho} relies on the DREG\(_3\) prescription to analytically continue spinors away from $d=4$.
\label{tab:bootstrap-GNY-summary}
}
\end{center}
\end{table}

\begin{figure}[t!]
	\centering
	\includegraphics[width=.95\textwidth]{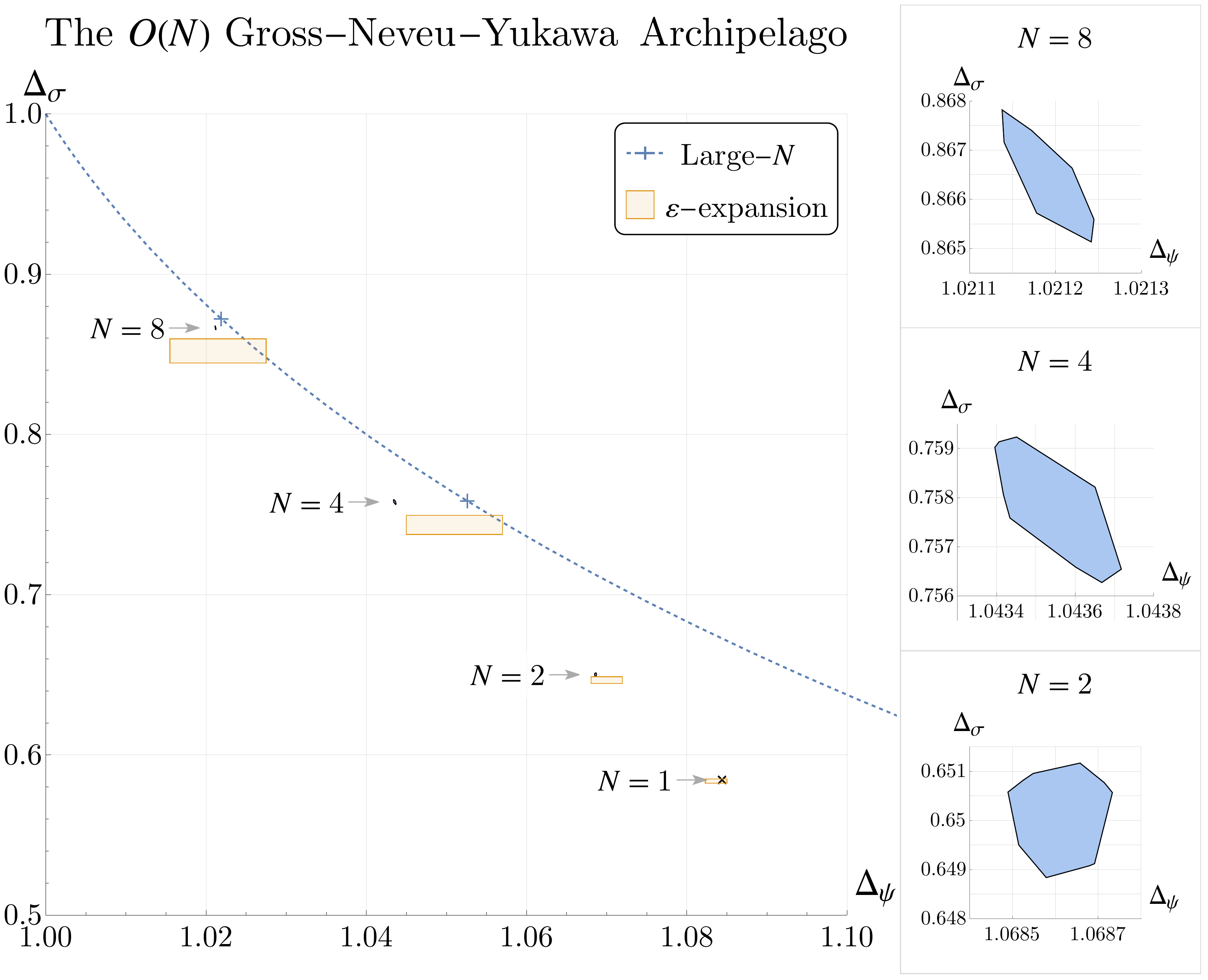}
	\caption[GNY archipelago in \((\Delta_\psi,\Delta_\sigma)\)]{A compilation of our \(N=2,4,8\) islands at \(n_\text{max}=18\), projected onto the \((\Delta_\psi,\Delta_\sigma)\) plane, compared against the perturbative estimates in the large-\(N\) expansion (represented by the dotted blue curve), Borel-resummations of the \((4-\epsilon)\)-expansion \cite{Ihrig:2018hho} (represented by the orange boxes), and the location of the \(N=1\) island from the \(\cN=1\) supersymmetric Ising bootstrap \cite{Rong:2018okz, Atanasov:2018kqw, Atanasov:2022bpi} (represented by the x). \label{fig:archipelago-sig-psi-intro}}
\end{figure}

\begin{figure}[t!]
	\centering
	\includegraphics[width=.95\textwidth]{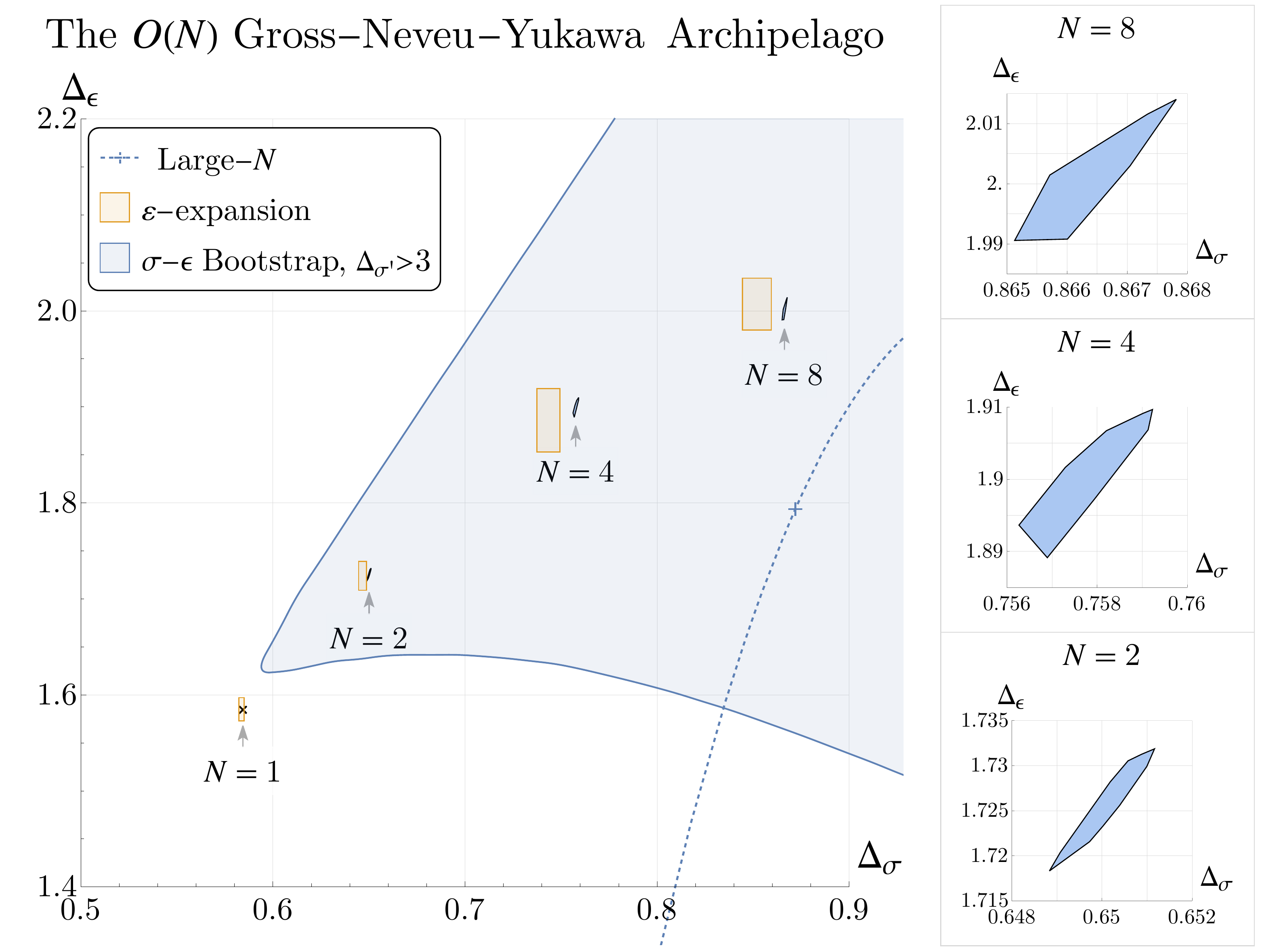}
	\caption[GNY archipelago in \((\Delta_\sigma,\Delta_\epsilon)\)]{A compilation of our \(N=2,4,8\) islands at \(n_\text{max}=18\), projected onto the \((\Delta_\sigma,\Delta_\epsilon)\) plane, compared against the perturbative estimates in the large-\(N\) expansion (represented by the dotted blue curve), Borel-resummations of the \((4-\epsilon)\)-expansion \cite{Ihrig:2018hho} (represented by the orange boxes), and the location of the \(N=1\) island from the \(\cN=1\) supersymmetric Ising bootstrap \cite{Rong:2018okz, Atanasov:2018kqw, Atanasov:2022bpi} (represented by the x). We also superpose the general \(\sigma\)-\(\eps\) bootstrap bounds with the assumption \(\Delta_{\sigma'}>3\) from \cite{Atanasov:2018kqw}.\label{fig:archipelago-sig-eps-intro}}
\end{figure}

Our bounds on the scaling dimensions of $\s,\psi_i,$ and $\e$ for the theories with $N=2,\,4$, and $8$ are illustrated in figures~\ref{fig:archipelago-sig-psi-intro} and~\ref{fig:archipelago-sig-eps-intro}, and we give a numerical summary in table \ref{tab:bootstrap-GNY-summary}. Results for the $N=1$ theory using similar methods are shown later in figure \ref{fig:susy-emergence-plot}. For all values of $N$, our results are close to perturbative estimates from resummations of the $\epsilon$-expansion. 
For $N=8$, the large-$N$ estimates are also close to the island that we find. Monte Carlo estimates for some of the scaling dimensions are also available for $N=2,\,4,\,8$~\cite{Tabatabaei:2021tqv,Huffman:2017swn,Liu:2019xnb}; these results, while close to our islands, have error bars that, in most cases, are disallowed by the rigorous bounds obtained from the bootstrap (see section~\ref{sec:results}). In all cases, our work presents a significant jump in the precision of scaling dimension determinations.
 
An important subtlety is that there are in fact two different GNY models that are often confused in the literature, having the same number of fermions but different global symmetry groups. In addition to the $O(N)$ GNY models discussed above, there are also ``chiral'' GNY models that possess an $O(N/2)^2 \rtimes \Z_2$ global symmetry. These models, sometimes referred to as being in the ``chiral Ising'' universality class, are nearly degenerate with the $O(N)$ GNY models for the most common low-lying operators, being only distinguishable at high perturbative order. The Monte Carlo estimates mentioned above at $N=4,8$ are believed to fall in this class. However, due to the expected near-degeneracy, we posit that our bootstrap results for the leading operators also provide good (albeit non-rigorous) estimates of the scaling dimensions in the $O(N/2)^2 \rtimes \Z_2$  GNY models. In this work we review some of the existing perturbative estimates for scaling dimensions in both models and provide a number of new ones that will be useful in our bootstrap study. We will also do our best to differentiate which of the two critical models is known to describe various phase transitions discussed in the condensed matter literature.
 
Another interesting case is the $N=1$ GNY model which, as previously mentioned, is believed to have emergent supersymmetry at criticality. The island that we find with mild assumptions about gaps in various sectors of this model, shown later in figure \ref{fig:susy-emergence-plot}, is fully consistent with this picture. Without making any assumptions about supersymmetry in the bootstrap setup, we find:
 (i) The island lies right on a line along which the low-lying scaling dimensions are related due to supersymmetry, $\Delta_{\epsilon} = \Delta_\psi + \frac{1}2 = \Delta_\sigma+1$. (ii) The lowest dimension operator with spin-$3/2$ is very close to the unitarity bound across the entire island, suggesting the existence of (at least) an approximate supercurrent for any theory that lies within it. (iii) The $\cN=1$ super-Ising CFT, whose scaling dimensions have been tightly constrained in previous bootstrap studies by a priori assuming supersymmetry, can also be seen to live right at a tip of the island. These computations provide a nice consistency check of our implementation and suggest that there are no non-supersymmetric fixed points at $N=1$ (consistent with our gap assumptions).

Our work is a culmination of a series of developments in the numerical bootstrap that extend practical limits to allow studies of a wider set of 3d CFTs. To set up the crossing equations, we used the formalisms developed in \cite{Kravchuk:2016qvl,Erramilli:2019njx} and the conformal block computation algorithms developed and implemented in \texttt{blocks\_3d} \cite{Kravchuk:2017dzd, Karateev:2018oml,Erramilli:2019njx,Erramilli:2020rlr}. We also implemented several Haskell libraries (described in appendix~\ref{app:software}) to efficiently and robustly set up systems of mixed correlators and compute the resulting bounds. We explored the space of external scaling dimensions and external OPE coefficients using the Delaunay search and ``cutting surface'' OPE search algorithms developed in   \cite{Chester:2019ifh}. Finally, we solved large-scale SDPs using the solver \texttt{SDPB} \cite{Simmons-Duffin:2015qma,Landry:2019qug}. We hope that these technologies can be used to place comparable constraints on more complicated 3d CFTs with fermionic degrees of freedom, including extensions of the GNY models with different global symmetries or Chern-Simons matter theories whose monopole operators carry half-integer spin.

The structure of this paper is as follows. In section~\ref{sec:theoretical-background-GNY}, we give theoretical background, including details about the perturbative expansions of the GNY models. We also discuss the gap assumptions that we impose as well as the differences between the $O(N)$ GNY models and the $O(N/2)^2 \rtimes \Z_2$ GNY models. In section~\ref{sec:numerical-setup}, we discuss the numerical setup of our bootstrap problem; the reader interested solely in results for the GNY model can continue straight to the next section. There, in section \ref{sec:results} we discuss the main results of the paper and show a series of bounds on scaling dimensions of the low-lying operators in the theory. We discuss possible future directions in section \ref{sec:discussion}.

\section{Theoretical background and spectrum assumptions}

\label{sec:theoretical-background-GNY}

\subsection{Large-$N$ and $\epsilon$-expansion }
\label{sec:large-N-and-epsilon-expansion}

\begin{table}[t!]
\begin{center}
\renewcommand{\arraystretch}{1.1}
\resizebox{\columnwidth}{!}{
\begin{tabular}{l|c c|l|l }
\hline\hline Operator~& Parity & $O(N)$ & $\Delta$ at large $N$.&$\Delta$ in $\e$-exp.\\
\hline $\psi_i$ & $+$ &V & $1+ \frac{4}{3\pi^2 N} + \frac{896}{27\pi^4 N^2}+ \frac{\#}{N^3} +\dots $  & $\frac{3}2 - \frac{N+5}{2(N+6)} \epsilon + \dots$ \\
 $\psi_i' \sim \phi^2 \psi_i$ & $+$ &V & $3+\frac{100}{3\pi^2 N}+ \dots$  & - \\
  $\chi_i \sim \phi^3 \psi_i$ & $-$ &V & $4+ \frac{292}{3\pi^2 N}+\dots$  & - \\
 $\sigma \sim \phi$ & $-$  &S& $1- \frac{32}{3\pi^2 N} +\frac{32 \left(304-27 \pi ^2\right)}{27 \pi ^4 N^2}+ \dots$ & $1-\frac{3}{N+6} \epsilon+ \dots$ \\
  $\epsilon \sim \phi^2$ & $+$ &S& $2+\frac{32}{3\pi^2 N} -\frac{64 \left(632+27 \pi ^2\right)}{27 \pi ^4 N^2}  + \dots$& $2+ \frac{\sqrt{N^2 +132 N + 36} - N-30}{6(N+6)}\epsilon + \dots$\\ 
  $\sigma' \sim \phi^3 $ & $-$ &S& $3+\frac{64}{\pi^2 N} -\frac{128 \left(770-9 \pi ^2\right)}{9 \pi ^4 N^2}  + \dots$ & $3+\frac{\sqrt{N^2 +132N +36} - N - 30}{6(N+6)}\epsilon+\dots$  \\
  $\epsilon' \sim \phi^4 $ & $+$ &S& $4+\frac{448}{3\pi^2 N} -\frac{256 \left(3520-81 \pi ^2\right)}{27 \pi ^4 N^2} + \dots$ & -  \\
  $\phi^k $ & $(-1)^k$ &S& $k +\frac{16 (3 k-5) k}{3 \pi ^2 N} - \frac{\#}{N^2}+ \dots$
& -  \\
  $\sigma_T \sim \psi_{(i}\psi_{j)} $ & $-$ &T& $2+\frac{32}{3\pi^2 N} +\frac{4096}{27 \pi ^4 N^2} +\dots$ & -\\
  $J^\mu\phi^2 \partial_\mu \partial^2 \phi $ & $-$ &A& $8+\dots$& -\\
\hline\hline 
\end{tabular}}
\end{center}
\caption{Large-$N$  \cite{Gracey:1992cp,Derkachov:1993uw,Gracey:1993kc, Petkou:1996np, Moshe:2003xn, Iliesiu:2015qra, fei2016yukawa, Manashov:2017rrx, Gracey:2018fwq} and $\epsilon$-expansion estimates \cite{gracey1990three, rosenstein1993critical, zerf2016superconducting, Gracey:2016mio, fei2016yukawa, Mihaila:2017ble, Zerf:2017zqi, Ihrig:2018hho}  for the scaling dimensions in the GNY model. The results for $\Delta_{\psi'}$ and $\Delta_{\chi}$ are new and derived in appendix~\ref{sec:fermion-op-in-large-N-exp}. The numerators denoted by \(\#\) indicate known expressions that have been omitted here for concision. $\Delta_{\psi}$ has a positive correction at $O(1/N^3)$ that can be found in~\cite{Gracey:1993kc} and $\Delta_{\phi^k}$ has a negative correction (for $k>1$) at $O(1/N^2)$ that can be found in~\cite{Manashov:2017rrx}.} 
\label{tab:large-N-and-e-exp-GNY}
\end{table}

Since we need a baseline expectation for the scaling dimensions of the external operators $\s,\psi_i,\e$, and we  must also impose gaps for some low-lying internal operators, we collect the leading estimates for scaling dimensions of the $O(N)$ GNY model obtained from the large-$N$ expansion and the $\epsilon$-expansion in table \ref{tab:large-N-and-e-exp-GNY}. In addition to the known perturbative results, we give new calculations of the leading correction at large-$N$ for $\Delta_{\psi_i'}$ and $\Delta_{\chi_i}$ in appendix \ref{sec:fermion-op-in-large-N-exp}. Some of these scaling dimensions have been computed to higher order, primarily in the context of the closely related $O(N/2)^2 \rtimes \Z_2$ GNY models (discussed further below). However, the results for the leading scaling dimensions $\Delta_{\psi}, \Delta_{\sigma}, \Delta_{\epsilon}$ are degenerate between the $O(N)$ theory and the $O(N/2)^2 \rtimes \Z_2$ theory up to 3-loop order. For large-$N$ estimates see \cite{Gracey:1992cp,Derkachov:1993uw,Gracey:1993kc, Petkou:1996np, Moshe:2003xn, Iliesiu:2015qra, fei2016yukawa, Manashov:2017rrx, Gracey:2018fwq}, for $\epsilon$-expansion estimates see \cite{gracey1990three, rosenstein1993critical, zerf2016superconducting, Gracey:2016mio, fei2016yukawa, Mihaila:2017ble, Zerf:2017zqi, Ihrig:2018hho}, or see \cite{fei2016yukawa} for a two-sided Pade expansion in the $2+\epsilon$ and $4-\epsilon$ expansion. We will primarily compare our results to the 4-loop $\epsilon$-expansion resummations performed in~\cite{Ihrig:2018hho}, done using a computation scheme that is believed to be applicable to the $O(N)$ GNY models.

\subsection{The $N=1$ theory and emergent supersymmetry}

 We also get additional information for our gap assumptions from the $N=1$ case which is of interest by itself. 
This fixed point is expected to exhibit emergent supersymmetry in the IR and, while this has not been rigorously checked, there have  been numerous perturbative tests of this proposal. The basic argument relies on the $N=1$ critical point having a single relevant singlet scalar in the IR. That is, we expect that only a single coupling --- the mass $m$ --- needs to be tuned in~\eqref{eq:lagrangian} to $m=m_*(g,\lambda)$ in order to reach the $O(1)$ GNY IR fixed point, and the same fixed point is reached regardless of the values of $g$ and $\lambda$.  On the other hand, for $N=1$ and $\lambda = g^2/2$, the interaction in~\eqref{eq:lagrangian} is explicitly $\cN=1$ supersymmetric. After tuning the scalar mass to the critical point the RG flow will preserve $\cN=1$ supersymmetry.\footnote{This is slightly more subtle than the Lagrangian~\eqref{eq:lagrangian} being supersymmetric. Depending on the sign of the scalar mass term, we have either a phase with spontaneously broken supersymmetry and a massless goldstino, or a phase with spontaneously broken time-reversal symmetry and a mass gap. Supersymmetry is spontaneously broken in the $\<\phi\>=0$ phase because the fermion remains massless while the scalar gets a mass. The RG flow is not supersymmetric in this case, and so it is important to tune the effective scalar mass to zero to obtain a time-reversal invariant and supersymmetric IR theory.} Since, as all other critical RG flows of~\eqref{eq:lagrangian}, it terminates at the $O(1)$ GNY CFT, this implies that this CFT has $\cN=1$ supersymmetry. This argument has been more concretely probed by observing the expected supermultiplet relations $\Delta_{\epsilon} = \Delta_\psi+1/2 = \Delta_\sigma + 1$ between the $\epsilon$-expansion results for  $\Delta_\sigma$, $\Delta_\psi$, and $\Delta_\epsilon$ obtained at four loop order \cite{Zerf:2017zqi}. These relations  were also  probed using a 2-sided Pad\'e approximation in the $2+\epsilon$ and $4-\epsilon$ expansions \cite{fei2016yukawa}.

 The emergence of supersymmetry was also probed non-perturbatively with the conformal bootstrap in \cite{Iliesiu:2017nrv}, where, without explicitly imposing supersymmety,  a kink was seen close to the line relating $\Delta_\psi = \Delta_\sigma + 1/2$ when imposing the appropriate bound for $\Delta_{\sigma'}$. Nevertheless, since the CFT associated to the $O(1)$ GNY critical point has no theoretical guarantee to live at the kink and since the location of the kink in \cite{Iliesiu:2017nrv} changed heavily with the imposed gap on $\Delta_{\sigma'}$, it is interesting to try to constrain the fixed point to lie within an island through which the line  $\Delta_{\epsilon} = \Delta_\psi+1/2 = \Delta_\sigma + 1$ would pass. In this paper we will indeed find such a bound by making some mild assumptions about the gaps in the various sectors of the theory, providing additional evidence for supersymmetry in $N=1$ case, see section \ref{sec:results} for details.
 
 By using assuming $\cN=1$ supersymmetry for several scaling dimensions and OPE coefficients and using a mixed system of scalar operators~\cite{Rong:2018okz, Atanasov:2018kqw, Atanasov:2022bpi} found highly accurate estimates for the CFT data in the $\cN=1$ super-Ising critical theory, the fixed point to which it was suggested that the $O(1)$ GNY model flows in the IR. For instance, the scaling dimension of $\Delta_\sigma$ was found to be \cite{ Atanasov:2022bpi}, $\Delta_\sigma = 0.5844435(83)$,
from which the scaling dimensions of  $\Delta_{\epsilon}$ and $ \Delta_\psi$ can also be found. By imposing relations between scaling dimensions within the same supermultiplet, \cite{Atanasov:2022bpi} also found accurate estimates for the scaling dimensions of $\Delta_{\psi'} = 3.3869(25)$, $\Delta_\chi = 4.88$, $\Delta_{\epsilon'} = 3.8869(25)$, and $\Delta_{\sigma'} = 2.8869(25)$. We will use these results to inform our gap assumptions for other values of $N$ as we describe below.

\subsection{Gap assumptions at $N=1,2,4,8$}
\label{sec:vals-of-N-of-focus}

\begin{figure}[th]
	\begin{center}
		\includegraphics[width=.95\textwidth]{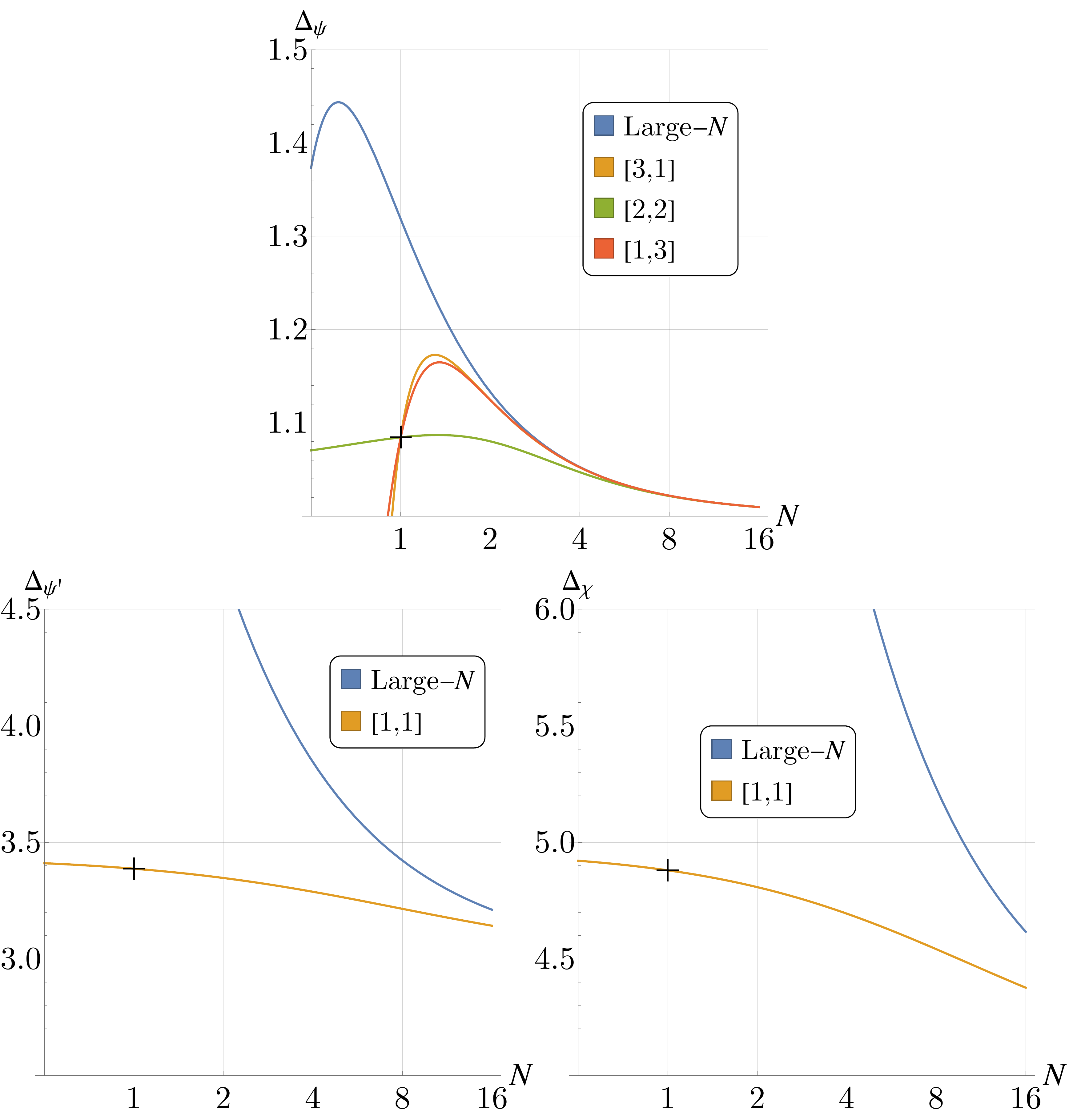}
	\end{center}
	\caption{Two sided-Pad\'e approximation for the scaling dimensions of low-lying fermionic operators, in the vector representation of the $O(N)$ global symmetry.  \label{fig:2-sided-Pade-for-fermions}}
\end{figure}

We now discuss what gaps we can reasonably assume in the spectrum when trying to find islands for the GNY fixed points. Given the abundance of evidence in favor of supersymmetry for $N=1$ it will be convenient to assume it when studying the theories with other values of $N$. Specifically, we can combine the highly accurate superconformal bootstrap results for the scaling dimensions of $N=1$ theory with the results from the large-$N$ expansion in a two-sided Pad\'e approximation as shown in figures~\ref{fig:2-sided-Pade-for-scalars} and~\ref{fig:2-sided-Pade-for-fermions}. This allows us to better estimate the scaling dimensions of the various low-lying operators in the theory and to better assess what gap assumptions we can make in the various sectors of the theory for the other values of $N$ that we study. We want to stress that while these approximations are by no means rigorous, we only use them to motivate our gap assumptions. Given the gap assumptions, which we state explicitly below, our bounds are rigorous.

The plots in figures~\ref{fig:2-sided-Pade-for-scalars} and \ref{fig:2-sided-Pade-for-fermions}, together with the $\epsilon$-expansion and other large-$N$ results, lead us to make the following gap assumptions:

\begin{figure}[th]
	\begin{center}
		\includegraphics[width=.95\textwidth]{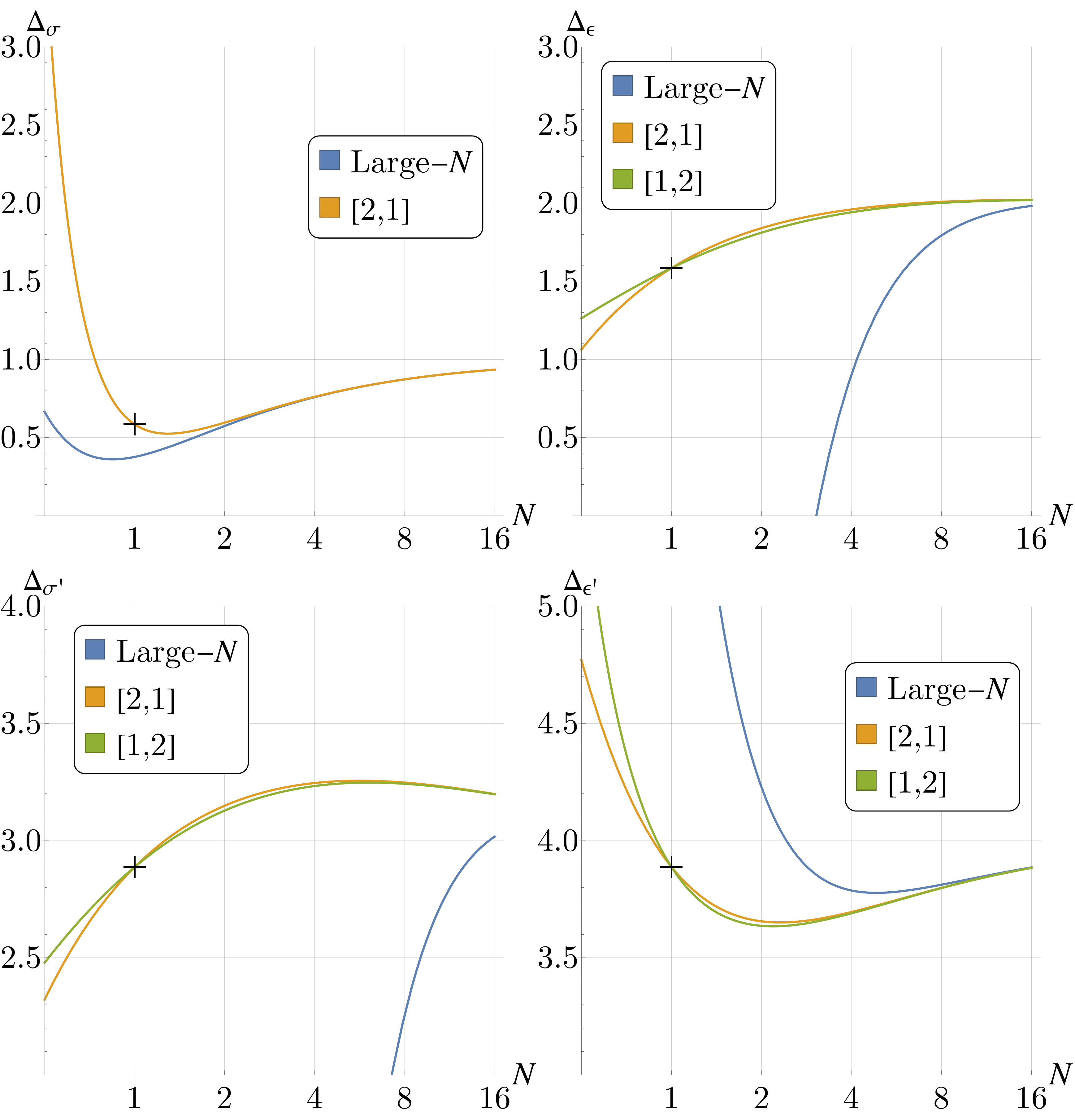}
	\end{center}
	\caption{Two sided-Pad\'e approximation for the scaling dimensions of low-lying scalars, singlets under the $O(N)$ global symmetry. The results are found using the results for the $\cN=1$ super-Ising model and the large-$N$ estimates for the GNY models. \label{fig:2-sided-Pade-for-scalars}}
\end{figure}

\begin{itemize}
\item Since $\epsilon$, $\sigma$ and $\psi_i$ will serve as the external operators in our bootstrap search, we will not make any assumption about their scaling dimensions.

\item We assume that $\epsilon$ is the only relevant neutral scalar operator in the theory. Consequently, we will assume $\Delta_{\epsilon'}>3$. This assumption is substantiated by the Pad\'e approximation obtained in the bottom-right plot in figure \ref{fig:2-sided-Pade-for-scalars} as well as by $\epsilon$-expansion estimates. 

\item Motivated by the large-$N$ equation of motion which removes the operator $\psi_i \psi_i$ from the spectrum of primaries, it will also be useful to impose bounds on $\Delta_{\sigma'}$. The Pad\'e approximation for this scaling dimension is shown in the bottom-left plot of figure~\ref{fig:2-sided-Pade-for-scalars}. It suggests that this operator is always irrelevant for $N\geq 2$. For most of our plots we will make the assumption of irrelevance $\Delta_{\sigma'} > 3$, but for \(N=2\) we will also study the more conservative assumption $\Delta_{\sigma'} > 2.5$.

\item For operators that are not part of the spectrum for $N=1$, such as $\sigma_T$, we can no longer rely on the two-sided Pad\'e approximations and will instead use the estimates from the large-$N$ expansion as well as the past bootstrap results \cite{Iliesiu:2017nrv}, which showed a kink in the $(\Delta_\psi, \Delta_{\sigma_T})$ at the expected location for the GNY critical point.  There it was  found that  $\Delta_{\sigma_T}>2$ for all $N$ which, emboldened by the large-$N$ estimates, we will take to be our gap value in this sector.

\item We will also assume gaps for the low-lying fermionic operators. In particular, due to the equation of motion $\slashed \partial \psi_i = - g \phi \psi_i$,\footnote{In mean field theory one would have $\chi_i\propto \phi\psi_i$. However, the equation of motion shows that $\phi\psi_i$ is removed from the primary spectrum, and therefore $\chi_i$ in interacting theory is expected to have a larger scaling dimension. } $\chi_i$ has a larger scaling dimension than could be na\"ively expected. We will conservatively assume that $\Delta_{\chi}>3.5$. This assumption is well within the expectations from the two-sided Pad\'e interpolation that is shown in the lower-right plot of figure~\ref{fig:2-sided-Pade-for-fermions}, which in fact estimates that $\Delta_\chi >4$ for all $N\geq 1$. 

\item We will also assume $\Delta_{\psi'}>2$, based on the two-sided Pad\'e approximation shown in the lower-left plot of figure~\ref{fig:2-sided-Pade-for-fermions}.

\item For the computations at $N=1$ we will use a similar set of gap assumptions, taking $\Delta_{\sigma'} > 2.5,\,\, \Delta_{\epsilon'} > 3,\,\, \Delta_{\psi'} > 2,$ and $\Delta_{\chi} > 3.5$. 

\item We also assume in all cases a small twist gap of $10^{-6}$ to improve numerical stability of the semidefinite programming algorithm. That is, we assume that all operators except the identity, the stress tensor, and the conserved $O(N)$ current have twist at least $10^{-6}$ higher than allowed by the unitarity bound. This a safe assumption because, based on the existing estimates of the scaling dimensions of $\sigma$ and $\psi$, we expect the smallest twists to be on the order of $10^{-2}$ or more above the unitarity bound~\cite{Fitzpatrick:2012yx,Komargodski:2012ek}. 
 \end{itemize}
 
Let us briefly comment on another interpretation of these gaps and why some of them may be helpful for isolating the GNY model. One can consider nonlocal deformations of the GNY fixed point, obtained by coupling its low-dimension operators to generalized free fields. Similar nonlocal deformations of the GNY models were discussed in~\cite{Chai:2021wac} and are generalizations of the description of the long-range Ising model developed in~\cite{Behan:2017dwr, Behan:2017emf, Behan:2018hfx}, e.g.~one can couple $\psi$ to a generalized free field $\chi_\text{GFF}$ of dimension $\sim 3-\Delta_{\psi} \approx 2$ and potentially flow to a nearby nonlocal fixed point. Our gap $\Delta_{\chi} > 3.5$ then excludes any such nearby solution to the bootstrap equations. Similarly, one could couple $\sigma$ to a generalized free field $\sigma_\text{GFF}$ of dimension $\sim 3-\Delta_{\sigma} \approx 2.2 - 2.4$. Our gaps $\Delta_{\sigma'} > 2.5 \text{ or } 3$ similarly exclude these possible solutions. One could also consider nonlocal deformations involving the $\epsilon$, $\sigma_T$, or $\slashed{\partial} \psi$ operators, which would also be excluded by our gap assumptions. It would be interesting to give a more systematic study of the nonlocal fixed points (and their dualities) that could be reached by deforming the GNY models.

\subsection{A distinction between two GNY fixed points}
\label{sec:two-GNYs}

Before diving into the detailed analysis of the condensed matter applications of the $O(N)$ GNY model in \eqref{eq:lagrangian}, we would like to compare this model to another theory, referred to in the literature as the ``chiral'' GNY model, which possesses an $O(N/2)^2 \rtimes \Z_2$ symmetry. The distinction between the two models and physical examples of each universality class is not always clearly described in the literature. Below we shall show that while the two models flow to different fixed points, many scaling dimensions and OPE coefficients at the two fixed points agree up to a high order in a perturbative expansion. To make the distinction clear, we will henceforth refer to the two theories by their global symmetry groups.\footnote{In discussing the related gauged QED\(_3\)-GN(Y) theories, there is some literature which describes a similar distinction with different nomenclature. The \(\frac{SU(N_f)\times U(1)_\text{top}}{\mathbb{Z}_{N_f}}\rtimes{Z^{\mathcal{C}}_2}\)-symmetric case is referred to as QED\(_3\)-GN(Y)\(_+\) and the \(\frac{(SU(N_f/2)^2\times U(1)_b\times U(1)_\text{top})\rtimes\Z_2^e}{\Z_{N_f}}\rtimes\Z_2^\mathcal{C}\)-symmetric case as QED\(_3\)-GN(Y)\(_-\)~\cite{Benvenuti:2018cwd,Boyack:2018zfx}. As far as we are aware, the analogous notation has not been used regularly in the gauge-free GN(Y) theories.}

In the $O(N/2)^2 \rtimes \Z_2$ GNY model there are two species of two-component Majorana fermions, \(\psi_i^L\) and \(\psi_i^R\), such that each species has \(N/2\) flavor components. Both species have a Yukawa coupling to a pseudoscalar \(\phi\), but with opposite signs:
\be\label{eq:chirallagrangian}
	\cL_{O(N/2)^2 \rtimes \Z_2\textrm{ GNY}} &= - \frac12 (\partial \phi)^2 - i \frac12 \psi^A_i\slashed{\ptl}\psi^A_i - \frac{1}{2}m^2\phi^2 - \frac{\lambda}{4}\phi^4 -i\frac{g}{2} \phi (\psi_i^L\psi_i^L - \psi_i^R\psi_i^R) \nn\\
	&\qquad \text{ (not what we study)}\,.
\ee 
Here \(i=1\dots \frac{N}{2}\) and \(A=L,R\). Each species of fermion has its own \(O(N/2)\) symmetry. Additionally, there is a discrete $\Z_2$ ``chiral'' symmetry of \(\psi_i^L\leftrightarrow\psi_i^R\), \(\phi\to-\phi\) which exchanges the fermion species. Note that this symmetry is not really chiral in (2+1)d since there is no notion of ``left'' or ``right'' fermions.\footnote{It is worth noting that the chiral symmetry is indeed related to the spacetime chiral symmetry of a (3+1)d fermionic theory with 4-component fermions. This connection is laid out explicitly in appendix~\ref{sec:more-on-diff-GNY-vs-chiral-GNY}.} In total, the flavor symmetry is \(O(N/2)^2\rtimes \Z_2\). When the fermions spontaneously generate mass due to $\phi$ getting a VEV, they preserve a parity and a time-reversal symmetry but break the $\Z_2$ symmetry. The $\Z_2$ symmetry breaking of this theory is characteristic of the so-called \emph{chiral Ising} universality classes, and it has been studied extensively \cite{rosenstein1993critical,Mihaila:2017ble,fei2016yukawa,Janssen:2014gea,Vacca:2015nta,Chandrasekharan:2013aya,Wang:2014cbw,Li:2014aoa,Huffman:2017swn,Hesselmann:2016tvh,Gracey:1990wi,Gracey:1992cp,Vasiliev:1992wr,Vasiliev:1993pi,Gracey:1993kb,Gracey:1993kc,Derkachov:1993uw,Gracey:2017fzu,Manashov:2017rrx,gracey1990three,Gracey:1991vy,Luperini:1991sv,Zerf:2017zqi,Ihrig:2018hho,Huffman:2019efk,Liu:2019xnb}. 

At the critical value of \(m^2\) perturbative calculations show that the model \eqref{eq:chirallagrangian} should also be described by a CFT whose scaling dimensions precisely agree at low perturbative orders with that of the critical model \eqref{eq:lagrangian}. Due to this seeming coincidence of the perturbative estimates of the CFT data, the distinction between the \(O(N)\) GNY universality class and the \(\ONc\) GNY a.k.a. chiral Ising universality class has been unresolved.\footnote{In fact it has been stated that the two models can be related through a field redefinition of $\psi^L$ and $\psi^R$ \cite{Herbut:2009qb}. However, while that redefinition makes the Yukawa interaction terms in \eqref{eq:chirallagrangian} to be the same as in \eqref{eq:lagrangian}, the fermionic kinetic terms then differ. } We will now clarify this ambiguity and show that the two models are distinct when computing observables to higher perturbative orders.

Let us thus compare the two models in the large-$N$ expansion. To compute correlators in the two models, we consider the Feynman diagrams with the leading order propagators of $\phi$ and $\psi_i$ (or  $\psi_i^{L,R}$) denoted by
\be 
&\text{$O(N)$ GNY: }\quad
&
 \<\psi_i(x) \psi_j(y)\>
 &= \delta_{ij}\,\, \raisebox{-0mm}{\includegraphics[scale=0.1]{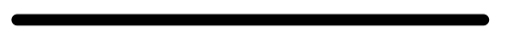}}  \,,\nn\\ 
&\text{$O(N/2)^2 \rtimes \Z_2$ GNY: }\quad
& \<\psi_i^L(x) \psi_j^L(y)\>
 &= \delta_{ij}\,\, \raisebox{-0mm}{\includegraphics[scale=0.1]{diagrams/prop1.jpg}}  \,,
  \quad \<\psi_i^R(x) \psi_j^R(y)\> = \delta_{ij}\,\, \raisebox{-0mm}{\includegraphics[scale=0.1]{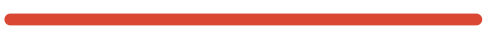}}  \,,\nn \\ 
&\text{In both:}  \quad 
& \<\phi(x) \phi(y)\> &= \,\, \raisebox{-0mm}{\includegraphics[scale=0.1]{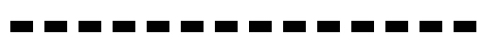}}  \,.
\ee
The only relevant interactions in the two models are of the form $\phi \psi \psi$ and $\phi ({ \psi}\,^L \psi^L -  \psi\,^R \psi^R)$, respectively, and corrections to the two-point function of $\phi^k$ are given by fermion loops whose vertices involve such interactions. Since in a fermionic loop, both the fermions $\psi^L$ and $\psi^R$ can always propagate, the only distinction between the large-$N$ Feynman diagrams of the two models can come from fermionic loops with an odd number of vertices.  
In the $O(N/2)^2 \rtimes \Z_2$ GNY fixed point, the contribution of such loops is always vanishing since for each loop in which $L$ fermions propagate, there is a loop in which  $R$ fermions propagate that has the opposite sign, 
\be
\label{eq:chiral-ising-cancellation}
\begin{tikzpicture}[baseline={([yshift=0ex]current bounding box.center)}, scale=0.10 ]
 \pgftext{\includegraphics[scale=1.0]{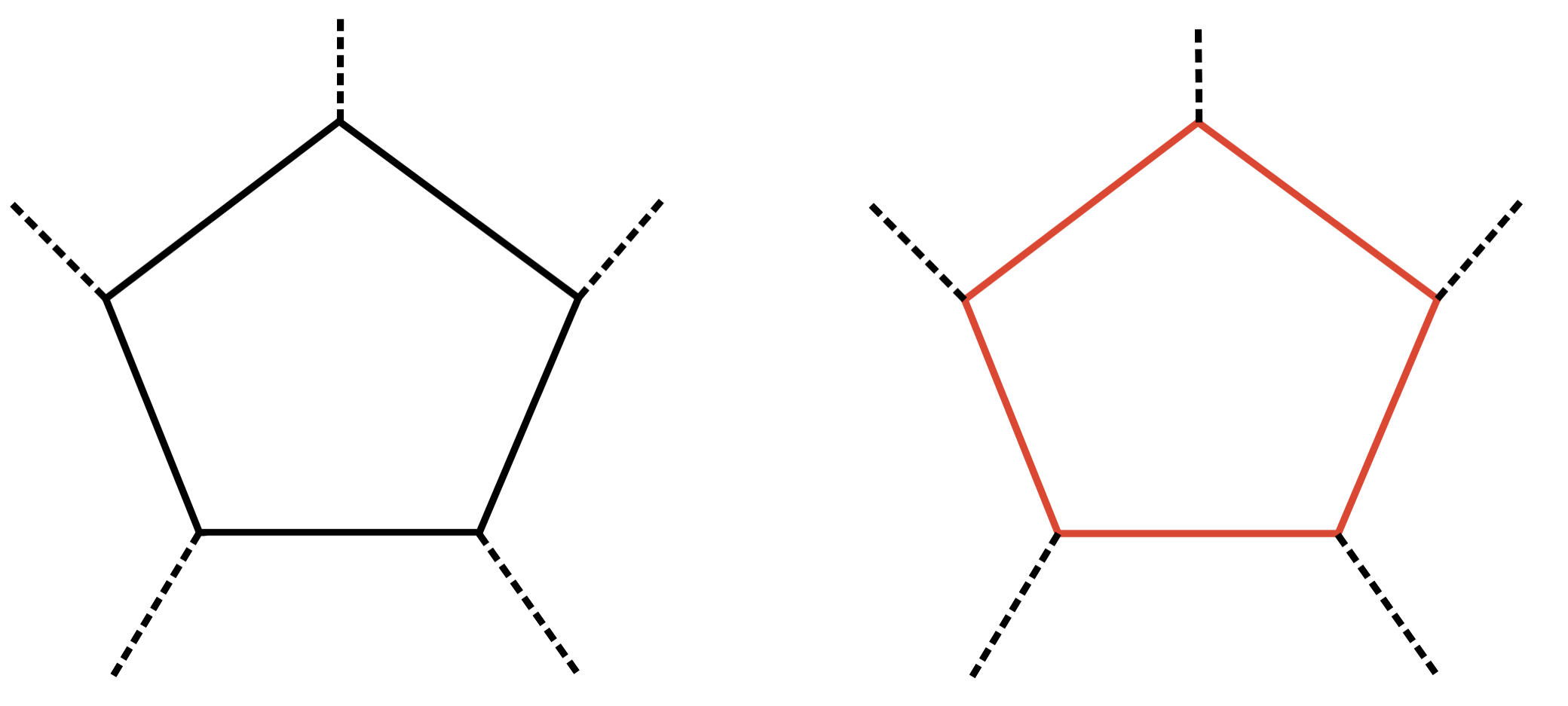}} at (0,0);
  \draw (0,-2.5) node {$+$};
    \draw (40.5,-2.5) node {$=0\,.$};
  \end{tikzpicture} 
\ee
This cancellation is due to the presence of the chiral $\Z_2$ symmetry under which the field $\phi$ is charged.
In the $O(N)$ model, however, such loops are not guaranteed to vanish when the odd number $k$ of fermions propagating through the loop is $k\geq 5$.\footnote{For $k=3$ such loops can explicitly be shown to vanish due to $\text{Tr}(\slashed{x}_{12} \slashed{x}_{23} \slashed{x}_{31} )= 0$. Alternatively one can use the fact that the three-point function of pseudo-scalars in 3d is always vanishing to arrive at the same conclusion.} This leads to a five-point point function for the pseudoscalar $\phi$
\be\label{eq:5pt} 
\< \phi(x_1) \phi(x_2)  \phi(x_3) \phi(x_4) \phi(x_5)\> = \begin{tikzpicture}[baseline={([yshift=-2ex]current bounding box.center)}, scale=0.10 ]
 \pgftext{\includegraphics[scale=1.0]{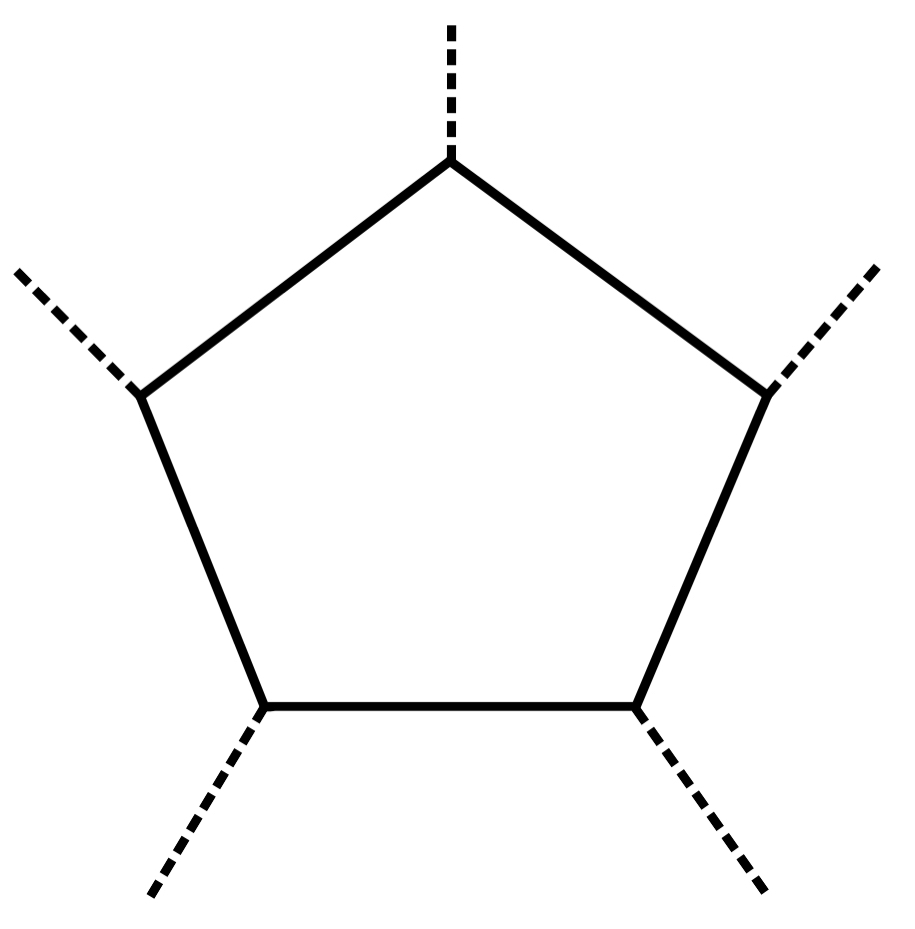}} at (0,0);
  \end{tikzpicture}   \,,
\ee 
that can be explicitly checked to be non-zero at order $1/N^{3/2}$.\footnote{We have checked this numerically, in momentum space, for fixed external momenta.}  Such higher $k$-point functions of pseudoscalars are in principle non-zero in 3d CFTs due to the existence of parity-odd $k$-point structures for $k\geq 5$ \cite{Kravchuk:2016qvl}. 
On the other hand, due to \eqref{eq:chiral-ising-cancellation} the five-point point function of $\phi$ in the $O(N/2)^2 \rtimes \Z_2$ GNY models vanishes. While this already proves that the fixed points in the two models are distinguishable, we would like to see how these differences are manifest in more commonly discussed observables such as the scaling dimensions of $\psi_i$, $\phi$, or $\phi^2$. For this we simply have to find the leading diagram that includes such loops with an odd number of fermion vertices. For instance, we find 
\be 
\Delta_\phi &\supset\begin{tikzpicture}[baseline={([yshift=-1ex]current bounding box.center)}, scale=0.12 ]
 \pgftext{\includegraphics[scale=1.0]{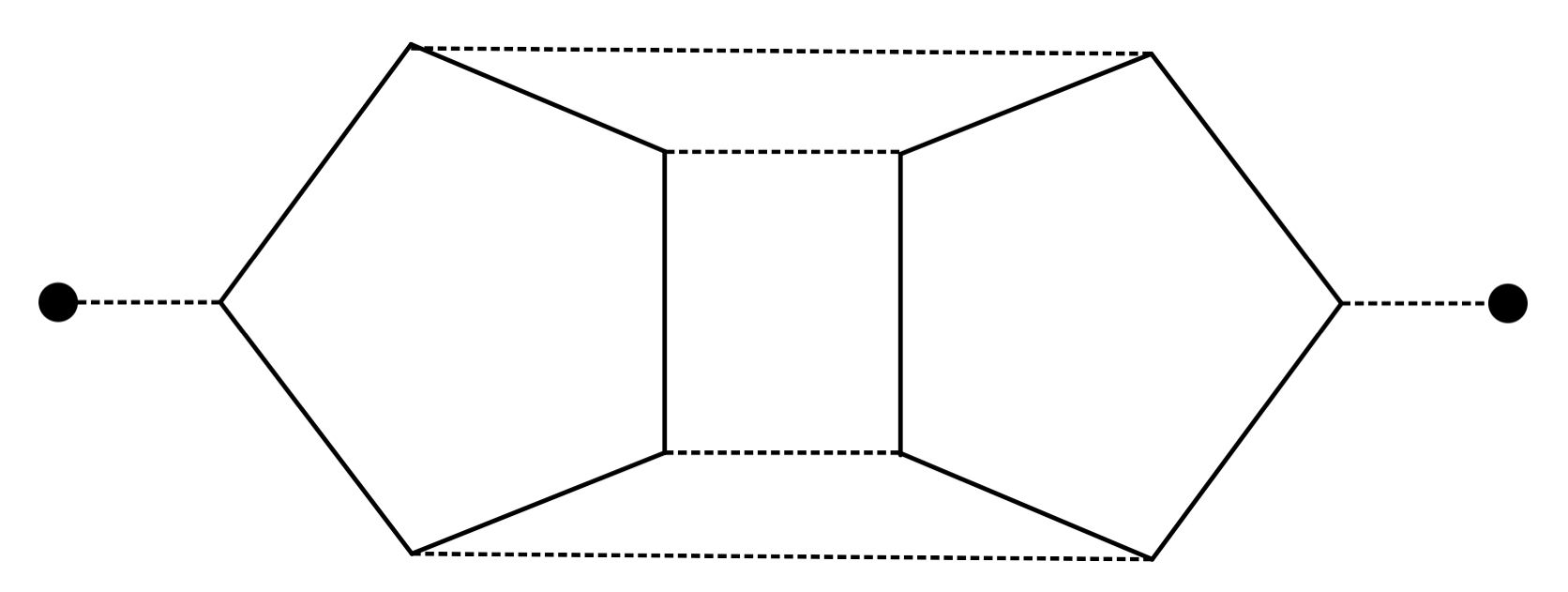}} at (0,0);
  \draw (-26.5,-5.0) node {$\phi(x)$};
    \draw (26.5,-5.0) node {$\phi(y)$};
  \end{tikzpicture}   \sim \frac{1}{N^3}\,, 
\ee
where we show an example diagram (rather than all the diagrams) contributing to the leading order at which the distinction between the two theories is present.

Consequently, even in these low-lying scaling dimensions the two models are different at a fairly high order in $1/N$.  We can similarly determine the large-$N$ expansion for all the other various operators in the $O(N/2)^2 \rtimes \Z_2$ GNY theory. Since these calculations require a lengthy discussion of the irreducible representation of $O(N/2)^2 \rtimes \Z_2$ --- which is not the symmetry for the theory we rigorously constrain in this paper using the bootstrap method~---~we discuss these calculations in appendix \ref{sec:more-on-diff-GNY-vs-chiral-GNY}.

Similar logic shows that the models differ only at high order in the $\epsilon$-expansion. For instance, it was noted in \cite{Zerf:2017zqi} for the $O(N/2)^2 \rtimes \Z_2$ GNY model at 4-loop order that the \(\e\)-expansion, continued to \(N=1\), is inconsistent with the expected emergence of supersymmetry; the authors of  \cite{Zerf:2017zqi} found that manually adding a 5-fermion loop diagram contribution restored the superscaling relation. As in the large-$N$ expansion, in the $O(N/2)^2 \rtimes \Z_2$ GNY model, loops with an odd number of propagating fermions have vanishing contributions owing to the sign difference in the Yukawa coupling between fermion species. Conversely, these odd-fermion loops should be generically nonzero for the \(O(N)\) theory starting at loops with 5-fermions (such as~(\ref{eq:5pt})). Such a prescription of adding back diagrams with an odd-number of propagating fermions can in principle be used when computing the scaling dimension of any operator in the theory and was named DREG\(_3\) by the authors of  \cite{Zerf:2017zqi}.   Since the scaling dimensions are only affected at high loop order, the differences between the estimates of $\{\Delta_{\psi}, \Delta_{\sigma}, \Delta_{\epsilon}\}$ for the two models are very small ($\lesssim 3 \times 10^{-6}$) for all values of $N$ that we consider.\footnote{We thank Michael Scherer for providing us with explicit calculations of the differences in the $\epsilon$-expansion at four loop order.} Nevertheless, when comparing our results to those from the $\epsilon$-expansion we will use the estimates from \cite{Ihrig:2018hho}, which rely on the DREG$_3$ prescription.

Materially for this paper, we see it fit to compare this work's results with previous results for the chiral Ising universality class, since the scaling dimensions are expected to be close to those of the \(O(N)\) GNY models even at low values of $N$. Nevertheless, our bounds will not be able to \emph{rigorously} constrain the chiral Ising GNY models. 
Rigorous bounds for these models, in which the \(O(N/2)^2\rtimes\Z_2\) global symmetry would be explicitly implemented in the bootstrap equations, represent a separate target for the bootstrap to be studied in the future.

\subsection{Condensed matter applications}

As mentioned in the introduction, the universality class of the GNY model is used to describe a variety of quantum phase transitions in condensed matter systems. In this subsection we will list some of the proposals in the literature. Since there can be confusion regarding the two universality classes associated to the $O(N)$ GNY critical point and the  $O(N/2)^2 \rtimes \Z_2$ GNY critical point, we will revisit this point for each quantum phase transition. 
 
\textbf{D-wave superconductors \cite{vojta2000quantum, vojta2003quantum}}:  In \cite{vojta2000quantum}, the possible quantum critical points in \(d\)-wave superconductors were classified according to their order parameter in an effort to describe anomalous behavior in cuprate superconductors.
The two transitions relevant to the discussion in this paper are the transition to $d_{x^2-y^2}+is$ pairing which is described by the universality class with symmetry \(O(4)^2\rtimes\mathbb{Z}_2\) (the chiral GNY model for \(N=8\)) and the transition to $d_{x^2-y^2}+id_{xy}$ pairing which can be seen to correspond to the universality class with symmetry \(O(8)\) (the non-chiral $N=8$ GNY model). As shown in~\cite{vojta2000quantum}, these are the only two transitions that have a nodal quasiparticle momentum distribution curve with a width proportional to $k_B T$. 
The additional requirement that the superconductor exhibits negligible scattering along the $(1,0)$ and $(0,1)$ directions uniquely isolates the transition $d_{x^2-y^2}+id_{xy}$ which thus underpins the importance of the \(O(8)\) GNY model for phase transitions in such superconductors.

\textbf{Chern insulators, and topological superconductors \cite{Grover:2013rc}}:
 There are more systems like the \(d\)-wave superconductors which have come into focus in the condensed matter community in recent years, which all possess the common factor of time-reversal symmetry breaking (TRSB). TRSB has been known to the condensed matter community for decades, and systems with a broken time-reversal symmetry are known to have the integer quantum anomalous Hall effect. Prototypical examples of these systems include the Haldane model (a Chern insulator in Cartan symmetry class A) \cite{Haldane:1988zza}. The \(O(N)\)-symmetric critical point
 represents a phase transition between the time-reversal-preserving class DIII and the time-reversal-breaking class D, which in the case \(N=1\) is discussed in \cite{Grover:2013rc}. More specifically, the universality class of the $O(1)$ GNY model is expected to describe a quantum phase transition, with emergent supersymmetry, at the boundary of a topological superconductor where the superconducting Majorana edge modes begin to gap out. Phases associated to this transition are expected to be found for a thin film of superfluid  He$_3$-B. 
 
In general, it is understood that the cases \(N>1\) also should have the same phase transition between symmetry classes DIII and D, though we are not aware of any proposed experimental design for empirical observation.\footnote{The notion of an interaction spontaneously modifying topological order is a matter of great interest, as it is believed that interactions break the class D free fermion classification of the topological invariant from \(\mathbb{Z}\) to \(\mathbb{Z}_{16}\) \cite{metlitski2014interaction}. In other words, the topological invariants of the \(O(N)\)-symmetric, time-reversal-breaking theories should have topological order defined by \(\nu = N\mod{16}\), since there is an adiabatic way to get from a nontrivial invariant with \(\nu=16\) to a trivial invariant with \(\nu=0\). However, this procedure requires breaking the \(O(N)\) symmetry while deforming the theory from the two topological phases.}

\textbf{Graphene \cite{herbut2006interactions,Herbut:2009qb, herbut2009relativistic,Mihaila:2017ble}}: The distinction between the \(O(N)\) and \(\ONc\)  theories is also apparent
in the case of graphene lattices. In \cite{Herbut:2009qb} the authors exhaustively enumerated the various order parameters that are expected to exist in graphene lattice theories. It's understood that the chiral Ising universality class, with symmetry \(\ONc\), describes the semimetal to charge density wave insulator transition on a honeycomb lattice \cite{herbut2006interactions}; specifically for \(N=8\) it describes a theory of spinful fermions and for \(N=4\) it describes spinless fermions. In  these two cases, time-reversal symmetry is preserved, but the fermions are gapped out, leading to an insulating phase. The corresponding time-reversal-breaking \(O(N)\) case is also expected to exist, but with a different order parameter which does break time-reversal symmetry.

\section{Numerical setup}
\label{sec:numerical-setup}

\subsection{SDP formulation of general crossing equations}
\label{sec:algo}

The system of crossing equations that we study in this paper is fairly complicated: it involves many correlation functions, several of which involve operators with non-trivial spin and/or flavor charges. On top of it, we are working with fermions and need to keep track of lots of minus signs associated with permutations. As a result, rewriting our system of equations in an SDP form suitable for numerical analysis is a rather non-trivial task and is especially prone to human error. 

Motivated by this, and also with a view towards future applications, we developed a computer code \texttt{bootstrap-bounds} (see appendix~\ref{app:software}) which handles most of the bookkeeping associated with passing from physically-transparent crossing equations to the SDP form. In this section we describe the basic algorithm that it uses, in the context of a general conformal bootstrap problem.

Let us consider a general conformal bootstrap problem for a set of external primary operators $\cO_1,\cdots, \cO_n$. We allow these operators to have arbitrary spins and flavor symmetry representations, and package their dependence on space-time coordinates and the various polarization indices into an abstract argument $\bp$, i.e.\ we write $\cO_i(\bp_i)$. In particular we assume that the action of all known symmetries on $\cO_i$ is expressed in terms of the action on $\bp_i$. For simplicity, we assume that all operators can be chosen to be Hermitian in Lorentzian signature (as is the case in our setup),
\be
	(\cO_i(\bp_i))^\dagger = \cO_i(\bp_i).
\ee
Furthermore, we introduce the notation $\cO_{\De,\r}$ for the primary operators that can appear in the OPE of $\cO_i$ but are not among the $\cO_i$, with $\De$ denoting the scaling dimension, and $\r$ all the other quantum numbers (such as spin, flavor symmetry representation, space parity, etc.).

First, we consider the three-point functions. We will need the following two types,
\be
	\<\cO_i(\bp_1)\cO_j(\bp_2)\cO_k(\bp_3)\>\qquad\text{and}\qquad\<\cO_i(\bp_1)\cO_j(\bp_2)\cO_{\De,\r}(\bp_3)\>.
\ee
Let us focus on the former. It can be written as
\be
	\<\cO_i(\bp_1)\cO_j(\bp_2)\cO_k(\bp_3)\>=\sum_a \l_{ijk}^a Q_{a,ijk}(\bp_1,\bp_2,\bp_3),
\ee
where $Q_{a,ijk}(\bp_1,\bp_2,\bp_3)$ is some basis of three-point tensor structures which we are free to choose. Intuitively, this statement is clear, but it turns out we need to formalize this a little in order to have a well-defined algorithm. Formally, choosing a basis of $Q_{a,ijk}$ amounts to the following:
\begin{itemize}
	\item \textit{For any choice of the ordered triple $i,j,k$ and the index $a$, specify a function $Q_{a,ijk}$ in three variables $\bp_1,\bp_2,\bp_3$, which is invariant under all the available symmetries when transformed with the quantum numbers of $\cO_i$ at $\bp_1$, of $\cO_j$ at $\bp_2$, and of $\cO_k$ at $\bp_3$.}
\end{itemize}
Note that the order of the arguments $\bp_1,\bp_2,\bp_3$ is fixed: for a function, we know what is the first, what is the second, and what is the third argument. This means that for fixed $i,j,k$ the functions $Q_{a,ijk},Q_{a,jik},Q_{a,kji},\cdots$ all require separate choices. But of course, the corresponding physical correlation functions 
\be
	\<\cO_i(\bp_1)\cO_j(\bp_2)\cO_k(\bp_3)\>,\quad \<\cO_j(\bp_1)\cO_i(\bp_2)\cO_k(\bp_3)\>,\quad \<\cO_k(\bp_1)\cO_j(\bp_2)\cO_i(\bp_3)\>, \cdots
\ee
are all related to each other in the obvious way (taking into the account fermionic permutation signs), and this induces a relation between the corresponding OPE coefficients $\l_{ijk}^a, \l_{jik}^a, \l_{kji}^a, \cdots$. We then demand, as is always possible to do, that the functions $Q_{a,ijk}$ are chosen so that the OPE coefficients $\l_{ijk}^a$ are functions of the \textit{unordered triple} $(i,j,k)$. In other words, so that
\be\label{eq:lambdasymmetry}
	\l_{ijk}^a=\l_{jik}^a=\l_{kji}^a=\cdots.
\ee
In terms of $Q_{a,ijk}$ this means
\be
	Q_{a,ijk}(\bp_1,\bp_2,\bp_3)=\pm Q_{a,jik}(\bp_2,\bp_1,\bp_3)=\cdots,
\ee
where $\pm$ account for fermionic permutation signs. Since these signs are included here,~\eqref{eq:lambdasymmetry} is true even when some $\cO_i$ are fermions. This choice of $Q_{a,ijk}$ is provided to the algorithm by the user; in this way, the algorithm does not have to know about the permutation properties of the operators, and can reason in terms of the simple coefficients $\l^{a}_{ijk}$. Furthermore, we require $Q_{a,ijk}$ to be chosen so that $\l^{a}_{ijk}\in \R$, which is again always possible to ensure. 

Next we consider the three-point functions
\be
	\<\cO_i(\bp_1)\cO_j(\bp_2)\cO_{\De,\r}(\bp_3)\>,
\ee
where the convention is the same. Concretely, we write
\be
	\<\cO_i(\bp_1)\cO_j(\bp_2)\cO_{\De,\r}(\bp_3)\>=\sum_a \l^a_{ij;\De,\r} Q_{a,ij;\De,\r}.
\ee
Now we only have to worry about the ordering in the pair $i,j$: we agree to always keep the generic operator $\cO_{\De,\r}$ at $\bp_3$.\footnote{Recall that we have agreed not to use the label $\cO_{\De,\r}$ for any of the $\cO_i$.} So in this case we need to provide only two structures for each pair $i,j$: $Q_{a,ij;\De,\r}$ and $Q_{a,ji;\De,\r}$. We again choose them in a way such that
\be\label{eq:lambdasymmetry2}
	\l^a_{ij;\De,\r}=\l^a_{ji;\De,\r}\in \R.
\ee

The above convention allows us to interface the general conformal block code \texttt{blocks\_3d}~\cite{Erramilli:2020rlr} from our algorithm. In order to produce a conformal block for the four-point function $\<\cO_i(\bp_1)\cO_j(\bp_2)\cO_k(\bp_3)\cO_l(\bp_4)\>$ for the exchange of $\cO_{\De,\r}$,\footnote{The code \texttt{blocks\_3d} can only compute the 3d conformal blocks; the flavor structure is then added by an additional layer of code.} this code requires the user to specify the three-point structures for
\be
	\<\cO_i(\bp_1)\cO_j(\bp_2)\cO_{\De,j}(\bp_3)\>\qquad\text{and}\qquad\<\cO_l(\bp_1)\cO_k(\bp_2)\cO_{\De,j}(\bp_3)\>.
\ee
Our algorithm can simply look up the functions $Q_{a,ij;\De,\r}$ and $Q_{b,lk;\De,\r}$ and pass them to \texttt{blocks\_3d}. 

As a result of our conventions, the expansion of the four-point function becomes
\be
	\<\cO_i(\bp_1)\cO_j(\bp_2)\cO_k(\bp_3)\cO_l(\bp_4)\>={\sum_{\De,\r}}'\sum_{a,b}\l^a_{ij;\De,\r}\l^b_{kl;\De,\r}G_{ab,ijkl,\De,\r}(\bp_1,\cdots\bp_4),
\ee
where $G_{ab,ijkl,\De,\r}$ is the block returned by \texttt{blocks\_3d}, and $\sum_{\De,\r}'=\sum_{\De,\r}+\sum_{n}$ denotes the sum over $\cO_{\De,\r}$ appearing in $\cO_i\times\cO_j$ OPE plus the sum over the $\cO_n$ appearing in the same OPE.

The above describes the conventions for the three-point functions, but we also need to specify the crossing equations. Any crossing equation involves only one four-point function, expanded in different channels. For a given four-point function $\<\cO_i(\bp_1)\cO_j(\bp_2)\cO_k(\bp_3)\cO_l(\bp_4)\>$, there are at most 3 distinct channels, depending on which operator out of $\cO_j,\cO_k,\cO_l$ we take the OPE of $\cO_i$ with. In any case, an equality of two channels in a four-point function takes the form
\be\label{eq:crossing-four-point}
	\<\cO_i(\bp_1)\cO_j(\bp_2)|\cO_k(\bp_3)\cO_l(\bp_4)\>\pm \<\cO_k(\bp_3)\cO_j(\bp_2)|\cO_i(\bp_1)\cO_l(\bp_4)\>=0,
\ee
where we used $|$ to separate the groups of operators between which we take the OPE. The $\pm$ sign is chosen based on the statistics of the operators. Note that 
\be
	\<\cO_j(\bp_1)\cO_i(\bp_2)|\cO_k(\bp_3)\cO_l(\bp_4)\>\pm \<\cO_k(\bp_3)\cO_j(\bp_1)|\cO_i(\bp_2)\cO_l(\bp_4)\>=0
\ee
expresses the same equality. We will use the convention in which we order operators in such a way that the two terms in the crossing equation differ by swapping the operators at $\bp_1$ and $\bp_3$, like in~\eqref{eq:crossing-four-point}. We choose some complete and independent set of crossing equations, written in this convention.

We can expand each four-point function in a basis of four-point tensor structures, 
\be
	\<\cO_i(\bp_1)\cO_j(\bp_2)|\cO_k(\bp_3)\cO_l(\bp_4)\>&=\sum_I g^I_{ijkl}(z,\bar z) T_{I,ijkl}(\bp_1,\bp_2,\bp_3,\bp_4),\label{eq:four-point1}\\
	\<\cO_k(\bp_1)\cO_j(\bp_2)|\cO_i(\bp_3)\cO_l(\bp_4)\>&=\sum_I g^I_{kjil}(z,\bar z) T_{I,kjil}(\bp_1,\bp_2,\bp_3,\bp_4).\label{eq:four-point2}
\ee
The conformal block expansions take the form
\be
g^I_{ijkl}(z,\bar z)&={\sum_{\De,\r}}'\sum_{a,b}\l^a_{ij;\De,\r}\l^b_{kl;\De,\r}G_{ab,ijkl,\De,\r}^I(z,\bar z),\\
g^I_{kjil}(z,\bar z)&={\sum_{\De,\r}}'\sum_{a,b}\l^a_{kj;\De,\r}\l^b_{il;\De,\r}G_{ab,ijkl,\De,\r}^I(z,\bar z),
\ee
where $G$ denotes the blocks computed by \texttt{blocks\_3d}. 

Again, we have a choice to make for the functions $T_{I,ijkl}$. We don't constrain it in any particular way. However, since the functions above differ only by the order of operators, we must have
\be
	T_{I,kjil}(\bp_3,\bp_2,\bp_1,\bp_4)=\sum_J M^J_I T_{J,ijkl}(\bp_1,\bp_2,\bp_3,\bp_4)
\ee
for some matrix $M^J_I$ (which depends on the choice of $i,j,k,l$). The crossing equation then takes the form
\be\label{eq:crossing}
	g^I_{ijkl}(z,\bar z)\pm \sum_{J} M^I_J g^J_{kjil}(1-z,1-\bar z)=0.
\ee
The appearance of $1-z$ is due to our convention for the crossing equation and the choice of cross-ratios.

We can furthermore expand the crossing equations~\eqref{eq:crossing} in a power series around $z=\bar z=\half$. In general, not all the Taylor coefficients are going to be linearly-independent~\cite{Kravchuk:2016qvl,Dymarsky:2017yzx,Dymarsky:2017xzb}. We therefore choose some independent set, with a cut-off $n_{max}$ on the order. We then introduce a combined label $\bI$ which runs over all crossing equations, all four-point tensor structures $I$ and the independent Taylor coefficients up to the cutoff $n_{max}$. We write $\hat g^\bI_{ijkl}$ for the contribution of the four point function OPE channel $\<\cO_i\cO_j|\cO_k\cO_l\>$, in this specific ordering of indices, to the scalar crossing equation labeled by $\bI$.\footnote{Note that for some orderings of $ijkl$ it might be that $\hat g^\bI_{ijkl}$ vanishes for all $\bI$ due to the choice of the crossing equations. Furthermore, some four-point functions vanish by symmetry and the corresponding $\hat g^\bI_{ijkl}$ are also 0.} For example, if $\bI$ corresponds to $I=0$ and the $(z-\thalf)^m (\bar z-\thalf)^n$ term in the equation~\eqref{eq:crossing}, then we have
\be
	\hat g^\bI_{ijkl}&=\frac{1}{m!n!}\ptl_z^m\ptl_{\bar z}^n g^0_{ijkl}(z,\bar z)\Big\vert_{z=\bar z=\half},\\
	\hat g^\bI_{kjil}&=\frac{(-1)^{m+n}}{m!n!}\sum_{J}M^0_J\ptl_z^m\ptl_{\bar z}^n g^J_{kjil}(z,\bar z)\Big\vert_{z=\bar z=\half},
\ee
with all other $\hat g^\bI_{\cdots}$ vanishing. With this notation, the full set of crossing equations takes the form
\be
	\forall\, \bI: \quad\sum_{ijkl}\hat g^\bI_{ijkl}=0,
\ee
where we sum over all choices of $ijkl$. Ultimately, the coefficients $\hat g^\bI_{ijkl}$ are provided to the algorithm by the user. In our implementation the user provides the equations~\eqref{eq:crossing} and the list of Taylor coefficients that need to be considered, from which the code reads off the $\hat g^\bI_{ijkl}$.

We can extend this notation in the obvious way to $G$, so that
\be
	\hat g^\bI_{ijkl}={\sum_{\De,\r}}'\sum_{a,b}\l^a_{ij;\De,\r}\l^b_{kl;\De,\r}\hat G_{ab,ijkl,\De,\r}^\bI.
\ee
This leads to the following expansion of the crossing equations,
\be\label{eq:crossingexpanded}
	0=\sum_{ijkl}{\sum_{\De,\r}}'\sum_{a,b}\l^a_{ij;\De,\r}\l^b_{kl;\De,\r}\hat G_{ab,ijkl,\De,\r}^\bI.
\ee
We now consider three types of contributions to~\eqref{eq:crossingexpanded}

\paragraph{Generic contributions}
We start with the contributions of generic operators $\cO_{\De,\r}$. After imposing a cut-off on the spin of $\cO_{\De,\r}$ there are finitely many distinct $\r$ appearing in~\eqref{eq:crossingexpanded}. We call such $\r$ ``operator channels.'' To each operator channel we associate the set
\be
	\cI_\r=\{(a,(i,j))|Q_{a,ij;\De,\r}\neq 0\},
\ee
where $(i,j)$ denotes an unordered pair. In other words, the elements of $\cI_\r$ label the OPE coefficients $\l_{ij;\De,\r}^a$ which are allowed by symmetries. 

We now introduce the PSD matrix $(\cP_{\De,\r})^{\a\b}$ with $\a,\b\in \cI_\r$
\be
	(\cP_{\De,\r})^{\a\b}=\sum_{\text{degeneracies}} \l^a_{ij;\De,\r}\l^b_{kl;\De,\r}\succeq 0,
\ee
where $\a=(a,(i,j))$, $\b=(b,(k,l))$, and we sum over all the contributions with given $\De,\r$ (accounting for possible degeneracies in the spectrum). Similarly, we introduce
\be
	(\hat\cG^\bI_{\De,\r})_{\a\b}=\half\p{\hat G_{ab,ijkl,\De,\r}^\bI+\hat G_{ba,klij,\De,\r}^\bI}+\text{permutations of $(ij)$ and $(kl)$},
\ee
with the same $\a,\b$ as above.

With this notation and using~\eqref{eq:lambdasymmetry}, the contribution of generic operators to the crossing equations is
\be
	0=\sum_{\r}\sum_{\De}\mathrm{Tr}\,\p{\cP_{\De,\r}\hat\cG^\bI_{\De,\r}}+\cdots.
\ee

\paragraph{External contributions}

We now focus on the contribution to the crossing equations of $\cO_i$ themselves. We define the index set
\be
	\cE=\{(a,(i,j,k))|Q_{a,ijk}\neq 0\},
\ee
where $(i,j,k)$ denotes unordered triples. That is, $\cE$ labels the OPE coefficients $\l_{ijk}^a$ which are not forced to be 0 by symmetries. 
We now define the rank-1 PSD matrix, assuming no degeneracies among the quantum numbers of $\cO_i$,
\be
	(\cP_{ext})^{\a\b}\equiv \l_{ijn}^a\l_{klm}^b,
\ee
where $\a=(a,(i,j,n))$ and $\b=(b,(k,l,m))$.

We define the symmetric matrix $\hat\cG^\bI_{ext}$ by the requirement that the contribution of the external operators to the crossing equation~\eqref{eq:crossingexpanded} is (recall~\eqref{eq:lambdasymmetry})
\be
0=\mathrm{Tr}\,\p{\cP_{ext}\hat\cG^\bI_{ext}}+\cdots.
\ee
The matrix $\hat\cG^\bI_{ext}$ has a straightforward expression in terms of $\hat G_{ab,ijkl,\De,\r}^\bI$ which is however awkward to describe.

\paragraph{Special operators}

There are sometimes special exchanged operators such as the stress-tensor $T^{\mu\nu}$ or the identity operator $\mathds{1}$. These operators contribute to the crossing equations as
\be
	0=\hat \cG^\bI+\cdots,
\ee
where $\hat \cG^\bI$ is determined by the special block for the given operator. It may or may not depend on parameters such at OPE coefficients or some scaling dimensions. For example, for the identity operator the contribution 
\be
	0=\hat \cG^\bI_{\mathds{1}}+\cdots
\ee 
is completely fixed by the normalization of two-point functions of $\cO_i$.

\paragraph{Final crossing equation}

The final crossing equation then takes the form
\be
	0=\hat\cG^\bI_{\mathds{1}}+\mathrm{Tr}\,\p{\cP_{ext}\hat\cG^\bI_{ext}}+\sum_{\r}\sum_{\De}\mathrm{Tr}\,\p{P_{\De,\r}\hat\cG^\bI_{\De,\r}},
\ee
and potentially additional contributions from $T$ or other special operators. Here, $\cP_{ext},\cP_{\De,\r}\succeq0$. So, for example, in a feasibility study we look for functionals $F_\bI$ such that
\be
	\sum_{\bI}F_{\bI}\hat\cG^\bI_{\mathds{1}}&=1,\label{eq:crossing-id}\\
	\sum_{\bI}F_{\bI}\hat\cG^\bI_{ext}&\succeq 0,\label{eq:crossing-ext}\\
	\sum_{\bI}F_{\bI}\hat\cG^\bI_{\De,\r}&\succeq 0.\label{eq:crossing-general-channel}
\ee 
Crucially, these conditions can be formed automatically once the user provides the three-point structures $Q$ and the crossing equations in the form~\eqref{eq:crossing}.

\subsection{Three-point functions}
Returning to the bootstrap problem for the GNY model, each local primary operator is characterized by the scaling dimension and three other quantum numbers, namely, spin $j$, space parity $P$, and an $O(N)$ irreducible representation $\mu$.\footnote{Note that the parity makes sense both for integer and half-integer $j$ (we use the transformation defined in~\cite{Kravchuk:2016qvl}), although for the latter the notions of “even” and “odd” parities can be exchanged by redefinition of the parity transformation by $(-1)^F$. This allows us to choose the parity for one fermionic operator at will, so we choose $\psi$ to be parity-even. } We consider the mixed system of the lowest dimension operators $\psi$, $\eps$ and $\s$, where their quantum numbers $(j,P,\mu)$ are 
\be
	\s&:\quad (0,\text{odd},\bullet)\\
	\eps&: \quad(0,\text{even},\bullet)\\ 
	\psi&:\quad (\thalf,\text{even},\myng{(1)}). 
\ee

When $N=1$, the global symmetry becomes $O(1) = \mathbb{Z}_2$. Since the non-trivial element sends $\psi\to-\psi$ and $\phi\to\phi$, it coincides with $(-1)^F$ and should not be considered as a separate global symmetry. Hence, in this case the operators are labeled by $(j,P)$ only.\footnote{In our discussion, which is valid for generic $N$, one can take \(\bullet\to\bullet\) and \(\myng{(1)}\to\bullet\) while all other representations are omitted.} Consequently, there are no flavor structures to consider. There is a corresponding reduction in the number of equations and OPE channels.

The crossing equations under study involve the following set of four point functions 
\be
\{\< \psi \psi \psi \psi\>,  \< \eps \eps \eps \eps\>, \< \s\s\s\s\> , \< \psi \psi \eps \eps\>, \<\psi \psi \s\s \>,   \< \s \eps \psi \psi \>,\< \s\s\eps \eps \>\}.
\ee
The conformal block expansions of these correlation functions involve OPEs between all combinations of $\psi,\e,\s$.  

We list the tensor structures appearing in relevant three-point functions in table \ref{tab:three-point-so3-mixed} and table \ref{tab:three-point-so3-fermion}. We choose to represent the conformal structures in the $\SO(3)_r$ basis following the convention in~\cite{Erramilli:2019njx,Erramilli:2020rlr},\footnote{In~\cite{Erramilli:2020rlr} it was called the $\SO(3)$ basis.} and we define the $O(N)$ flavor structures as follows,  
\be
	T^{\bullet,\bullet}_\bullet &= 1,\\
	T^{i,j}_{\myng{(1)}} &=\de^{ij},\\
	T^{ij,(kl)}_{\myng{(2)}} &= \half\p{\de^{ik}\de^{jl}+\de^{il}\de^{jk}}-\frac{1}{N}\de^{ij}\de^{kl},\\
	T^{ij,[kl]}_{\myng{(1,1)}} &=\half\p{\de^{ik}\de^{jl}-\de^{il}\de^{jk}}.
\ee
Some conformal structures might disappear for low spin of the exchanged operator $\cO$. The selection rule is that $|j_{12},j_{123}\>$ is present in $\<\cO_1\cO_2\cO\>$ if $j_{12}\in j_1\otimes j_2$ and $j_{123}\in j_{12}\otimes l$, where $j_1$ and $j_2$ are the spins of $\cO_1$ and $\cO_2$. In practice, to obtain the conformal structures, we did calculations in the $q$-basis~\cite{Kravchuk:2016qvl} and then converted them to the $\SO(3)_r$ basis as discussed in~\cite{Erramilli:2020rlr}.

Tensor structures for three external operators are obtained from those in tables~\ref{tab:three-point-so3-mixed} and~\ref{tab:three-point-so3-fermion} by restricting $\cO$ to the relevant special case. The only exception is the structure for $\<\psi\psi\sigma\>$ which we take to be $-|1,1\>$ (differs by a factor of $-1$ from table~\ref{tab:three-point-so3-fermion}).

Our choice of the tensor structure basis ensures the equality of the following OPE coefficients, as required by the algorithm in section~\ref{sec:algo}:
\be
\lambda_{\psi \psi \s} &=\lambda_{\psi \s \psi } = \lambda_{\s \psi \psi }, \\
\lambda_{\psi \psi \eps} &=\lambda_{\psi \eps \psi }\, = \lambda_{\eps \psi \psi }, \\
\lambda_{\s \s \eps} &=\lambda_{\s \eps \s }\,\, = \lambda_{\eps \s \s },
\ee
and similarly $\l_{\psi\sigma\cO}=\l_{\sigma\psi\cO}$ etc. A slight difference from~\ref{sec:algo} is that the OPE coefficients for our structures are purely imaginary if they involve fermionic operators. This difference is accounted for in the software.

\begin{table}[h]
\begin{center}
  \begin{tabular}{c|l|l|l}
     \hline\hline 
	  OPE & $\cO \in (l, P, \mu)$        &         $\<\cO_a\cO_b\cO_c\>$&\multicolumn{1}{c}{Structures}\\
     \hline
     $\s\times\s$                             &  \multirow{2}*{$(l\in2\Z,\text{even},\bullet)$}   & $ \<\s \s \cO\>$  & \multirow{2}*{$ \phantom{\de^{ij}(-1)^{l+\half}}|0,l\>$}\\
     
     $\eps\times\eps$                   &                                                                    &   \( \<\eps\eps \cO\> \) & \\
     \hline
     \multirow{2}*{$\s\times\eps$}                        & \multirow{2}*{$ (l\in2\Z,\text{odd},\bullet) $  }                         &    $ \<\s \eps \cO\>$ & $\phantom{\de^{ij}(-1)^{l+\half}}|0,l\>$ \\
			                                                                                                               & & \(\<\eps\s\cO\>\) & \(\phantom{\de^{ij+\half}}(-1)^l |0,l\> \) \\ 
     \hline
     \multirow{4}*{$\s\times\psi $} & \multirow{2}*{$(l\in\Z+\thalf,\text{even},\myng{(1)})$} & $ \<\s\psi^i\cO^j\> $&$(-1)^{l+\half} \de^{ij}|\thalf,l+\thalf\>$\\
	      			                                                                                                        & & $\<\psi^i\s\cO^j\>$&$\phantom{(-1)^{l+\half}}\de^{ij}|\thalf,l-\thalf\>$\\
      \cline{2-4}
	                                        	& \multirow{2}*{$(l\in\Z+\thalf,\text{odd},\myng{(1)})$}    & $
				                                                                                	 	\<\s\psi^i\cO^j\>$&$(-1)^{l-\half}\de^{ij}|\thalf,l-\thalf\>$\\
			                                                                             	     & &    $\<\psi^i\s\cO^j\>$&$\phantom{(-1)^{l+\half}}\de^{ij}|\thalf,l+\thalf\> $\\
     \hline
     \multirow{4}*{$\eps\times\psi $} & \multirow{2}*{$(l\in\Z+\thalf,\text{even},\myng{(1)})$} & $ \<\eps\psi^i\cO^j\>$&$(-1)^{l-\half}\de^{ij}|\thalf,l-\thalf\>$ \\		  
		                                                                                                  	         & &       $ \<\psi^i\eps\cO^j\>$&$\phantom{(-1)^{l+\half}}\de^{ij}|\thalf,l+\thalf\>   $\\
      \cline{2-4}
	                                        	&\multirow{2}*{ $(l\in2\Z+\thalf,\text{odd},\myng{(1)})$}    & $ \<\eps\psi^i\cO^j\>$&$(-1)^{l+\half}\de^{ij}|\thalf,l+\thalf\>$\\	                                                           
		                                                              	                               	   	  & &           $\<\psi^i\eps\cO^j\>$&$\phantom{(-1)^{l+\half}}\de^{ij}|\thalf,l-\thalf\> $\\ 
      \hline\hline
  \end{tabular}
   \caption{\label{tab:three-point-so3-mixed}Tensor structures appearing in the OPEs of mixed scalar-fermion operators.} 
   \end{center}
\end{table}

\begin{table}[h]
\begin{center}
	\begin{tabular}{ l | l | l }
		\hline\hline
		  OPE & $\cO \in (l, P, \mu)$        & $\<\psi^i\psi^j\cO^a\>$ Structures\\
		  \cline{1-3}
		\multirow{6}*{ $\psi\times\psi $} 
		& $(l\in 2\Z,\, \text{even},\, \mu\in\{\bullet,\myng{(2)}\})$ & \(\begin{aligned} T_\mu^{ija} & |0,l\> \\ T_\mu^{ija} & |1,l\> \end{aligned}\) \\
		\cline{2-3}
		& $(l\in 2\Z+1,\,\text{even},\,\mu=\myng{(1,1)}) $  &$\begin{aligned}T_\mu^{ija}&|0,l\> \\ T_\mu^{ija}&|1,l\>\end{aligned}$ \\
		\cline{2-3}
		& $(l\in 2\Z,\,\text{odd},\,\mu\in \{\bullet,\myng{(2)}\})$ & $T_\mu^{ija}(\sqrt{l+1}|1,l+1\>-\sqrt{l}|1,l-1\>)$ \\
		\cline{2-3}
		& $(l\in 2\Z+1,\,\text{odd},\,\mu \in \{\bullet,\myng{(2)}\})$ &$T_\mu^{ija}(\sqrt{l+1}|1,l-1\>+\sqrt{l}|1,l+1\>)$ \\
		\cline{2-3}
		& $(l\in (2\Z)_{\geq 2},\,\text{odd},\,\mu=\myng{(1,1)})$ & $ T_\mu^{ija}(\sqrt{l+1}|1,l-1\>+\sqrt{l}|1,l+1\>)$ \\
		\cline{2-3}
		& $(l\in 2\Z+1,\,\text{odd},\,\mu=\myng{(1,1)})$ & $T_\mu^{ija}(\sqrt{l+1}|1,l+1\>-\sqrt{l}|1,l-1\>)$ \\
		\hline\hline
	\end{tabular}
	\caption{\label{tab:three-point-so3-fermion}Tensor structures appearing in the OPE of the fermionic operators.}
\end{center}
\end{table}

The stress-tensor $T$ and the conserved $O(N)$ current $J$ transform with $(l,P,\mu)$ equal to $(2,\text{even},\bullet)$ and $(1,\text{even},\myng{(1,1)})$. Their OPE coefficients are constrained by Ward identities as
\be
	&f_{\s\s\hat T}^1 = -\sqrt{\frac{3}{2}}\frac{\De_\s}{4\pi\sqrt{C_T}},\quad f_{\e\e\hat T}^1 = -\sqrt{\frac{3}{2}}\frac{\De_\e}{4\pi\sqrt{C_T}},\\
	&f_{\psi\psi\hat T}^1 = i\frac{\sqrt 3}{4\pi}\frac{\De_\psi}{\sqrt{C_T}},\quad f_{\psi\psi\hat T}^2 = -i\frac{3}{4\sqrt 2\pi}\frac{1}{\sqrt{C_T}},\\
	&f_{\psi\psi\hat J}^1 = i\frac{1}{\sqrt 2\pi}\frac{1}{\sqrt{C_J}},\quad f_{\psi\psi\hat J}^2\,\,\,\,\text{not constrained},
\ee
where $\hat T=C_T^{-\thalf} T$ and $\hat J=C_J^{-\thalf} J$ are canonically normalized in the conventions of~\cite{Erramilli:2019njx,Erramilli:2020rlr}, which are
\be
	\<\hat T\hat T\>=\frac{H_{12}^2}{X_{12}^5},\quad \<\hat J^{[ij]}\hat J^{[kl]}\>=\frac{1}{2}\p{\de^{ik}\de^{jl}-\de^{il}\de^{jk}}\frac{H_{12}}{X_{12}^{3}}.
\ee
In practice, the Ward identity for $J$ does not affect the numerics because it only gives an interpretation of an OPE coefficient in terms of $C_J$. The Ward identity for $T$ doesn't affect the numerics unless a gap above $T$ is assumed. As explained in section~\ref{sec:vals-of-N-of-focus}, we do assume a gap of $10^{-6}$ above $T$. However, since this gap is small, it is likely that imposing the $T$ Ward identity doesn't have a significant effect on our results.

\subsection{Four-point functions and crossing equations}
We now study the four-point functions of our mixed system. Following the procedures outlined by \eqref{eq:crossing-four-point}, \eqref{eq:four-point1}, and \eqref{eq:four-point2}, we construct the four-point tensor structures $T_{I, ijkl}$ and find the crossing equations in the form of \eqref{eq:crossing}. 

In general, one single four-point structure is the product of a flavor and a conformal structure. The full structure should be invariant under kinematic permutations, which is the group that preserves all conformal cross-ratios formed by the coordinates of the four operators. We refer to \cite{Kravchuk:2016qvl} for the details on this group. 

When the four operators are identical, $\rho_1 =\rho_2 =  \rho_3 =  \rho_4$ and the kinematic permutation group is $\Z_2\times \Z_2 = \{ e, (12)(34), (13)(24),(14)(23) \}$. The irreducible representations of $\Z_2\times \Z_2$ can be labelled by $(++),(--),(+-),(-+)$ where the signs stand for the eigenvalues under $(12)(34)$ and $(13)(24)$ respectively. On the other hand, for four-point functions with two pairs of identical operators the kinematic permutation group is  $\Z_2$. For example, if $\rho_1 =\rho_2$ and $\rho_3 =  \rho_4$ then the kinematic permutation group is just $\Z_2= \{ e, (12)(34)\}$. The irreducible representations of $\Z_2$ can be labelled simply by $+$ and $-$. 

To ensure that the full structure is invariant under the kinematic permutations, the flavor and conformal structures should transform in the same irrep of the kinematic permutation group. 

There are two additional requirements for the  four-point tensor structures. Firstly,  the structures must have space parity consistent with the parity of the operators in the four-point function. And secondly, they must have definite parity under the transformation $z\leftrightarrow \bar z$, which we refer to as the t-parity transformation.\footnote{Not to be confused with time reversal.} This is needed to simplify the corresponding symmetry of the coefficient functions $g(z,\bar z)$, see~\cite{Kravchuk:2016qvl}.

\subsubsection{$ \<\psi \psi \psi \psi\>$ }

We first consider the four point function of four fermions
\be
\<\psi^i\psi^j\psi^k\psi^l\>  = \sum_{I, a} {{t}}_{I}\,  {T}^{ijkl}_{a}  {g}_{\psi \psi\psi\psi} ^{I, a}(z,\bar z), 
\ee
where ${T}_a$ stands for the flavor structures and ${t}_{I}$ for the conformal structures. It participates in the crossing equations of the form
\be
	\<\psi\psi|\psi\psi\>=-\<\psi\psi|\psi\psi\>,
\ee
understood in the sense of equation~\eqref{eq:crossing-four-point}.

There are three possible flavor structures for $\<\psi^i\psi^j\psi^k\psi^l\>$, namely $\de^{ij}\de^{kl}$, $\de^{ik}\de^{jl}$, and $\de^{il}\de^{jk}$. They all are invariant under the kinematic permutation group $\Z_2\x\Z_2$, i.e.\ are all in the $(++)$ representation. For future convenience, we define the following linear combinations of these structures,
\be
	T^{ijkl}_+=\de^{ij}\de^{kl}+\de^{il}\de^{jk},\quad T^{ijkl}_3=\de^{ik}\de^{jl},\quad T^{ijkl}_-=\de^{ij}\de^{kl}-\de^{il}\de^{jk}.
\ee
We also notice that under crossing symmetry permutation $(13)$,  $T_+$ and $T_3$ are symmetric while $T_-$ is anti-symmetric. 

In order to construct the conformal structures, we use the $q$-basis defined in~\cite{Kravchuk:2016qvl}. We look for conformal structures that are $\Z_2\x\Z_2$ invariant, parity-even, and have definite parity under the t-parity transformation:\footnote{This example has been worked out in detail in~\cite{Kravchuk:2016qvl} and~\cite{Erramilli:2020rlr}.}
\be
\<\up \up\up\up\>^\pm &= \<\up \up\up\up\>\pm \<\dn \dn \dn\dn\>,\\
\<\up \up\dn\dn\>^+ &= \<\up \up\dn\dn\> + \<\dn\dn\up \up\>,\\
\<\up \dn\up\dn\>^+ &= \<\up \dn\up\dn\> + \<\dn\up \dn\up\>,\\
\<\dn \up \up\dn\>^+ &= \<\dn \up \up\dn\> + \< \up\dn\dn \up\>. 
 \ee
Here $\pm$ denotes the t-parity, which acts on the individual fermionic spins by $[\up] \to i [\dn]$ and $[\dn]\to i [\up]$.  We use $\up$ to denote $+\thalf$ and $\dn$ to denote $-\thalf$.

From the results of \cite{Kravchuk:2016qvl} we can determine the phase factor picked up by the conformal structures under crossing symmetry\footnote{Equation (4.44) in~\cite{Kravchuk:2016qvl} assumes $\rho_1=\rho_3$ and includes an extra $(-1)$ for fermionic permutations. The general result~\eqref{eq:phase-13} can be derived from appendix B of~\cite{Kravchuk:2016qvl}.}
\be\label{eq:phase-13}
(13): \<q_1 q_2 q_3 q_4\>\rightarrow (-1)^{q_1+q_2+q_3-q_4} \<q_3 q_2 q_1 q_4\>. 
\ee
Multiplied by the extra factor of $(-1)$ coming from the fermion exchange, we find that all the conformal structures are invariant under the $(13)$ permutation. 

 Combining conformal and flavor structures, we form the following two sets of linear combinations that are crossing-symmetric and  crossing-antisymmetric respectively, 
 \be
 \text{symmetric: } &\mathlarger g_{\<\up\up\up\up\>^\pm, T_{+}}\,, \quad \mathlarger{g}_{\<\up\up\dn\dn\>^+, T_{+}} + \mathlarger g_{\<\dn\up\up\dn\>^+,  T_{+}}\,, \quad \mathlarger g_{\<\up\dn\up\dn\>^+,  T_{+}}\,,\\ 
 &\mathlarger g_{\<\up\up\up\up\>^\pm, T_{3}}\,, \quad \mathlarger g_{\<\up\up\dn\dn\>^+, T_{3}} + \mathlarger g_{\<\dn\up\up\dn\>^+,  T_{3}}\,, \quad \mathlarger g_{\<\up\dn\up\dn\>^+,  T_{3}}\,,\\ 
 &\mathlarger g_{\<\up\up\dn\dn\>^+, T_{-}} - \mathlarger g_{\<\dn\up\up\dn\>^+,  T_{-}}\, ,\\
  \text{anti-symmetric: } &\mathlarger g_{\<\up\up\dn\dn\>^+, T_{+}} - \mathlarger g_{\<\dn\up\up\dn\>^+,  T_{+}}\, ,\\
  &\mathlarger g_{\<\up\up\dn\dn\>^+, T_{3}} - \mathlarger g_{\<\dn\up\up\dn\>^+,  T_{3}}\, ,\\
  &\mathlarger g_{\<\up\up\up\up\>^\pm, T_{-}}\,, \quad \mathlarger g_{\<\up\up\dn\dn\>^+, T_{-}} + \mathlarger g_{\<\dn\up\up\dn\>^+,  T_{-}}\,, \quad \mathlarger g_{\<\up\dn\up\dn\>^+,  T_{-}}\, ,
\ee
where we used the simplified notation
\be
	g_{I,a}\equiv g^{I,a}_{\psi \psi\psi\psi} .
\ee
The above functions satisfy crossing equations  with the sign determined by whether they fall in the ``symmetric'' or ``anti-symmetric'' category above. For example, we have
\be
	g_{\<\up\up\up\up\>^-, T_{+}}(z,\bar z)&=g_{\<\up\up\up\up\>^-, T_{+}}(1-z,1-\bar z),
\ee
while
\be
	&g_{\<\up\up\dn\dn\>^+, T_{+}}(z,\bar z) - \mathlarger g_{\<\dn\up\up\dn\>^+,  T_{+}}(z,\bar z)\nn\\
	&=-\p{g_{\<\up\up\dn\dn\>^+, T_{+}}(1-z,1-\bar z) - \mathlarger g_{\<\dn\up\up\dn\>^+,  T_{+}}(1-z,1-\bar z)}.
\ee
When Taylor-expanding these equations, one should keep in mind the t-parity of the conformal structures which determines the parity of the above functions under $z\leftrightarrow \bar z$.

Finally, the crossing equations for $g_{\<\up\up\up\up\>^+, T_{a}}(z,\bar z)$  at $z=\bar z$ are redundant with other crossing equations and should not be imposed in order to avoid numerical instabilities (see appendix A of~\cite{Kravchuk:2016qvl}). In practice this is done by requiring that $n>0$ derivatives are taken in the direction orthogonal to $z=\bar z$ for this structure.
 
\subsubsection{$ \<\psi \psi \s \s\>$ and $ \<\psi \psi \eps \eps\>$  }
 Mixed four-point functions containing both $\sigma$ and $\psi$ give rise to two crossing channels in the sense of~\eqref{eq:crossing-four-point},
 \be
 \< \sigma \sigma| \psi \psi \>&=\< \psi \sigma |\sigma \psi\>, \\
  \< \psi \sigma |\psi\sigma \>&= -\< \psi \sigma| \psi \sigma \>. 
 \ee
We expand each ordering of the four point function as
\be
 \<\sigma \sigma \psi^i \psi^j \>  = \sum_{I, a} {{t}}_{I}\,  {T}^{ij}_{a} \, {g}_{\sigma\sigma\psi\psi} ^{I, a}(z,\bar z),\nn\\
\<\psi^i \sigma \sigma \psi^j \>  = \sum_{I, a} {{t}}_{I}\,  {T}^{ij}_{a} \, {g}_{\psi\sigma\sigma\psi} ^{I, a}(z,\bar z),\nn\\
\<\psi^i \sigma \psi^j \sigma\>  = \sum_{I, a} {{t}}_{I}\,  {T}^{ij}_{a} \, {g}_{\psi\sigma\psi\sigma} ^{I, a}(z,\bar z).
\ee
We are slightly abusing the notation since the conformal structures are different for each of the three orderings (since different operators are inserted at different points $x_i$).

Only one flavor structure ${T}^{ij}_{a}  = \delta^{ij}$ exists for each of these four-point functions, and it is kinematic-symmetric. It is also mapped to itself under the crossing permutation $(13)$. Hence, we will ignore the flavor structure in the following discussion. Furthermore, since the products $\sigma\sigma$ and $\eps\eps$ have the same parity, the analysis below is the same for the correlation functions  $\<\psi \psi \s \s\>$ and $ \<\psi \psi \eps \eps\>$ . We will focus on the correlator $\<\psi \psi \s \s\>$ for concreteness.

The two parity-even conformal structures for the ordering $\<\sigma \sigma \psi \psi \>$ are $[0,0, \frac12, \frac12 ]$ and $[0,0,  \text{-}\frac12,\text{-}\frac12]$, and similar expressions apply for the other two orderings. These structures are automatically kinematically symmetric.  We will simplify the notation and write them as $[\up\up]$ and $[\dn\dn]$ for each of the orderings (i.e.\ omitting the 0 charges), and define the structures with definite t-parity as
\be
[\up\up]^{\pm} = [\up\up]\mp [\dn\dn].
\ee

Applying \eqref{eq:phase-13} and factors from fermion permutations, we find that conformal structures with the same label are mapped into each other under the $(13)$ permutation. Hence, we can form crossing-symmetric and anti-symmetric combinations as 
 \be
\text{symmetric: } &\mathlarger g^{\s\s\psi\psi}_{[\up\up]^\pm} + \mathlarger g^{\psi\s\s\psi}_{[\up\up]^\pm},\,\quad \mathlarger g^{\psi\s\psi\s}_{[\up\up]^\pm}, \\
\text{anitsymmetric: } &\mathlarger g^{\s\s\psi\psi}_{[\up\up]^\pm} - \mathlarger g^{\psi\s\s\psi}_{[\up\up]^\pm}\,. 
 \ee
The convention for the crossing equations is the same as in the previous subsection. This time, there are no redundancies between the crossing equations.

 \subsubsection{$\<\s\e\psi\psi\>$}
The independent crossing equations in this case are
 \be
\< \sigma \e| \psi\psi\>&=\< \psi \e |\sigma \psi\>, \\
\< \psi \sigma |\psi \e \>&=-\< \psi \sigma| \psi\e \>. 
\ee
Similar to the case above, we have a trivial flavor structure that can be ignored. However, the overall parity $\<\s\e\psi\psi\>$ is now odd and so a separate analysis of the conformal structures is required. 

For all of the orderings of the operators, the parity-odd conformal structures are $[\up\dn]$ and $[\dn\up]$, using  the same notation as in the previous subsection. There is no kinematic symmetry to consider. To form structures with definite t-parity, we write
\be\label{eq:updnstruct}
	[\up\dn]^\pm = [\up\dn]\mp [\dn\up] = \mp [\dn\up]^\pm.
\ee

Applying \eqref{eq:phase-13} and taking into account the factors from fermion permutation, we can form the crossing-symmetric and crossing-antisymmetric linear combinations as before, 
 \be
\text{symmetric: } &\mathlarger g^{\s\e\psi\psi}_{[\up\dn]^\pm} -\mathlarger g^{\psi\e\s\psi}_{[\up\dn]^\pm}\,,\quad \mathlarger g^{\psi\s\psi\e}_{[\up\dn]^+}\,, \\
\text{anitsymmetric: } &\mathlarger g^{\s\e\psi\psi}_{[\up\dn]^\pm} + \mathlarger g^{\psi\e\s\psi}_{[\up\dn]^\pm}\, ,\quad \mathlarger g^{\psi\s\psi\e}_{[\up\dn]^-}\,.
\ee
A slight subtlety in this case is that the structure $[\up\dn]^\pm$ for $\<\psi\s\psi\e\>$ ordering is mapped by (13), up to a phase, to $[\dn\up]^\pm$, and so we need to use~\eqref{eq:updnstruct} to reduce it back to the $[\up\dn]^\pm$ basis.
 
\subsubsection{Scalar four-point functions}
Since both flavor and conformal structures are trivial, it is a straightforward exercise to form crossing-symmetric and crossing-antisymmetric functions, 
 \be
\text{symmetric: } &g^{\s\s\s\s}, \, g^{\e\e\e\e}, \, g^{\s\e\s\e}, \, g^{\s\s\e\e} + g^{\e\s\s\e} , \\
\text{anitsymmetric: } &g^{\s\s\e\e} - g^{\e\s\s\e} . 
 \ee
The t-parity of all these functions is $+1$.

In total,  we end up with $38$ crossing equations for $N\ge2$ and $28$ equations for $N=1$.

\subsection{Numerical computations}
\label{sec:software-algorithms}
After setting up the crossing equations and OPE channels, our workflow is automated by softwares described in appendix~\ref{app:software} to search for the functionals $F_{\bI}$ to satisfy (\ref{eq:crossing-id}--\ref{eq:crossing-general-channel}).

The space of CFT data that we searched over is 6-dimensional, but is better understood as two separate searches done together. We have three scaling dimensions and three OPE coefficient ratios that we search over; the latter is done by a cutting-surface search algorithm, and the former is done by a Delaunay mesh search algorithm. We will now briefly go over these search algorithms. 

\subsubsection{OPE scan}
\label{sec:ope-algorithm}
As noted in \cite{Kos:2016ysd}, by including assumptions of OPE coefficient ratios in our bootstrap computations, the allowed region in the space of scaling dimensions can be improved at the cost of also having to search over those OPE coefficient ratios. To that end, we employ the algorithm described in \cite{Chester:2019ifh} to include constraints imposed by the OPE coefficients involving the only the external operators, $\lambda_{\text{ext}}$. The non-vanishing OPE coefficients are 
\be
\vec{ \lambda}_{\text{ext}} = 
\begin{pmatrix}
\lambda_{\psi \psi \sigma }\\
\lambda_{\psi \psi \epsilon }\\
\lambda_{\sigma\sigma\epsilon}\\
\lambda_{\epsilon\epsilon\epsilon}
\end{pmatrix}. 
\ee

With \eqref{eq:crossing-ext}, we require that the contribution of external scalar OPE coefficients to the crossing equation has a definite sign after applying the functional, independent of the values of those coefficients.
\be
\sum_{\bI}F_{\bI}\hat\cG^\bI_{ext}&\succeq 0\quad \Rightarrow \quad\sum_{\bI} \mathrm{Tr}\,\p{M_{ext}F_{\bI}\hat\cG^\bI_{ext}} \ge 0. 
\ee
However, \eqref{eq:crossing-ext} is strong enough that the conclusion above stands for any matrix with the decomposition $M_{ext}= A^\dag A $. 
We, on the other hand, are only interested in the case $M_{ext} = \cP_{ext} =\vec\lambda_{ext}  \vec\lambda_{ext}^T $, which is a rank-1 matrix. Therefore, we instead look for functionals $F_{\bI}$'s that satisfy
\be
\vec \lambda_{ext}^T \left(\sum_{\bI}F_{\bI}\hat\cG^\bI_{ext}\right) \vec\lambda_{ext}   \ge 0
 \ee
 for each $ [\lambda_{ext}] \in \mathbb{RP}^3$ (independent of the magnitude or sign of the vector),  along with \eqref{eq:crossing-id}  and \eqref{eq:crossing-general-channel}. A point in the dimensional space $(\Delta_\psi, \Delta_\s, \Delta_\eps)$ is ruled out if such $F_{\bI}$'s exist for all $[\lambda_{ext}]$, and is allowed otherwise. Hence, by scanning over the OPE space, we obtain a union of allowed regions in dimension space, each permitted by some OPE direction: 
 \be
 \bigcup_{ [\lambda_{ext}] \in \mathbb{RP}^3} \mathcal{D}_{\lambda_{ext}}. 
 \ee
Note that this union of allowed regions is contained in the allowed region obtained by imposing the stronger \eqref{eq:crossing-ext}, as this is the special case when imposing the condition that $M_{ext} $ is rank-1. We should note that for more than one OPE coefficient ratio, the cutting surface algorithm is non-rigorous \cite{Chester:2019ifh}. 

For each point in the dimension that the cutting surface algorithm doesn't exclude, it outputs a direction in the OPE space that could not be excluded by our constraints. Combined over all the allowed points in the dimension space, this produces a list of OPE coefficient ratios with a relatively small variance. We report the full range of values of the OPE coefficient ratios that we found in our searches in section~\ref{sec:results} as estimates of the real ratios. It should be noted however that these estimates are not rigorous. In particular, we cannot exclude a systematic error that could arise from the way the cutting surface algorithm samples the OPE ratios. Similarly, the error bars are not rigorous.

\subsubsection{Delaunay mesh search}
\label{sec:delaunay}
The islands that we show were computed by the Delaunay mesh search algorithm; we will refer for details on the algorithm to the original paper \cite{Chester:2019ifh}. However, in order to properly interpret how we have chosen to represent our results, we will provide a brief qualitative review of this method.

The principle of the algorithm is to divide the search space into a suitable simplicial complex, known as a Delaunay mesh, where each vertex is a point that has already been computed as being allowed or disallowed. A simplex that has entirely allowed or disallowed vertices is assumed to be completely on the interior or exterior of our island. If a simplex has both an allowed and disallowed vertex, then the simplex is deemed to be on the boundary of the island; we can think of the volume of the simplex as the region of uncertainty between allowed and disallowed. This implied boundary can be further refined by computing the feasibility of the point at the simplex's centroid.\footnote{The Delaunay search algorithm selects the centroid, which is also the mean of the vertices, as the next point to be computed in the numerical bootstrap \cite{Chester:2019ifh}. Interpreting the simplex as the uncertain region of the boundary, the centroid is the mean of that uncertain region.} Thus, with each iteration we get an increasingly refined mesh.

After a search has finished, we can determine the island's boundary in one of two ways. The first is to take the centroids of the boundary simplicies and compute their convex hull. This method produces fairly smooth islands, but whose bounds are not strictly rigorous. The second, more conservative, approach is to take the convex hull of all boundary simplices, which is equivalent to taking the convex hull of all vertices that neighbor allowed vertices. This includes the interiors of all boundary simplices into our island so we can be sure that the area outside this hull is strictly disallowed by our bootstrap constraints, up to assumptions of convexity. Thus, the latter method produces islands and bounds that are rigorous. However, it should be noted that the latter method's islands tend to be more jagged than islands computed with the centroid method given the same set of points. This distinction is heightened when working with relatively small numbers of points computed. In this work, we have opted to take the more conservative approach.
\section{Results}

\label{sec:results}

In this section, we present the results of our numerical bootstrap computations for various values of $N$. While the space of CFT data that we searched over is 6-dimensional (as discussed in section \ref{sec:numerical-setup}), in most cases, we have projected the results into the planes $(\Delta_\psi, \Delta_\sigma)$ and $(\Delta_\sigma, \Delta_\epsilon)$ for ease of visualization. 

We first discuss the bounds obtained for $N = 2, 4$, and $8$ and compare them with the existing studies from $\epsilon$-expansions (after Borel resummation) and from Monte Carlo simulations \cite{Ihrig:2018hho,Tabatabaei:2021tqv,Huffman:2019efk,Liu:2019xnb}. We then focus on the $N=1$ theory and discuss how our results, without assuming supersymmetry a priori, are strong evidence for the emergence of supersymmetry in the IR in the $N=1$ critical GNY model.

\subsection{$N = 2,\,4,$ and $8$}

\begin{figure}[t!]
\begin{center}
\includegraphics[width=0.98\textwidth]{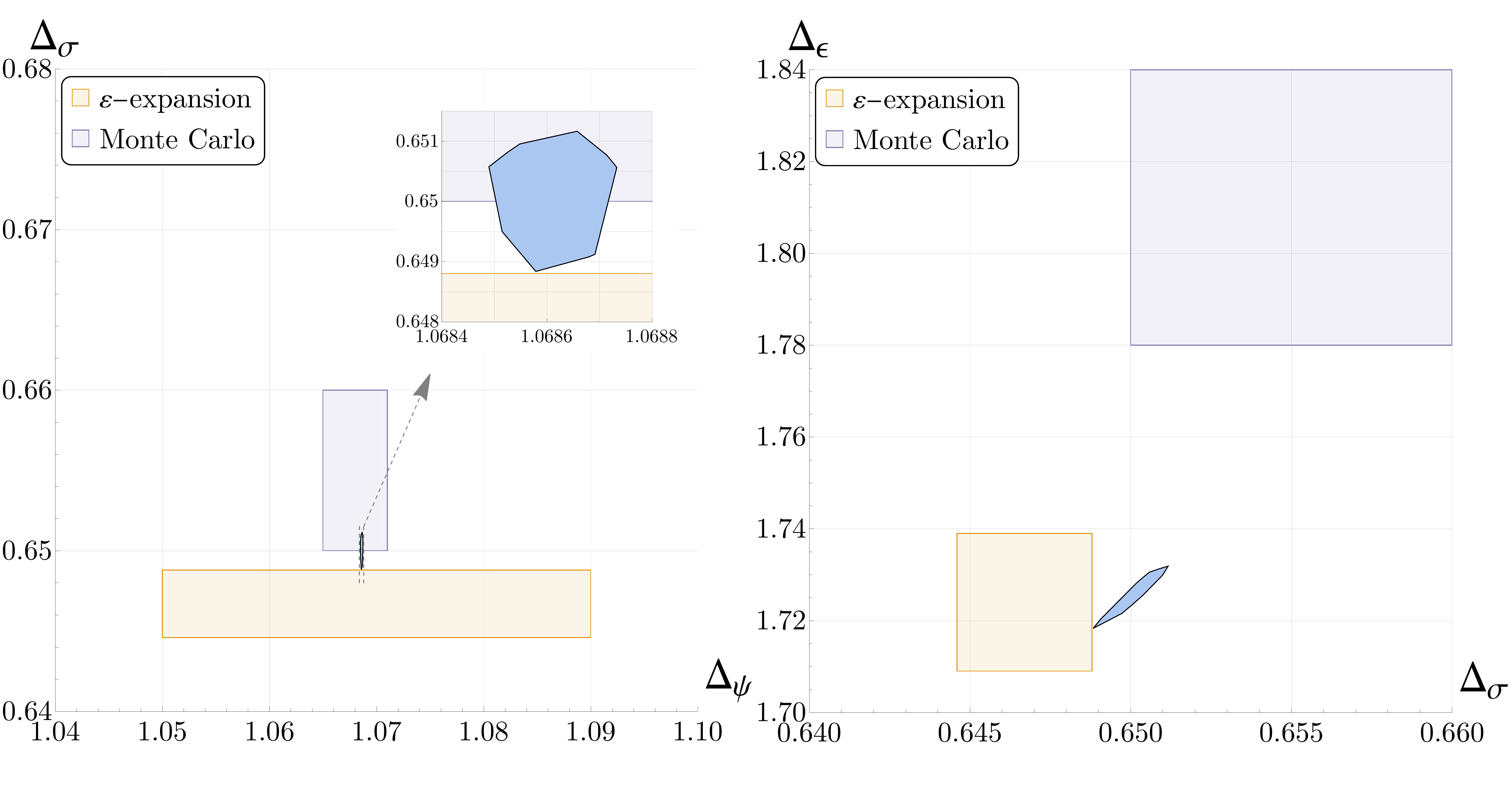}
\end{center}
\caption{The allowed region for the $N=2$ critical GNY model when imposing that $\Delta_{\sigma'}>3$ computed at $\nmax=18$, projected to the $(\Delta_\sigma,\, \Delta_\epsilon)$ and $(\Delta_\psi,\, \Delta_\sigma)$ planes. This should be compared to the Borel re-summed result obtained from the $\epsilon$-expansion \cite{Ihrig:2018hho} (shown in orange) and with the Monte Carlo results from  \cite{Tabatabaei:2021tqv} (shown in purple).}\label{fig: N2_combined}
\end{figure}

\begin{figure}[t!]
\begin{center}
\includegraphics[width=0.98\textwidth]{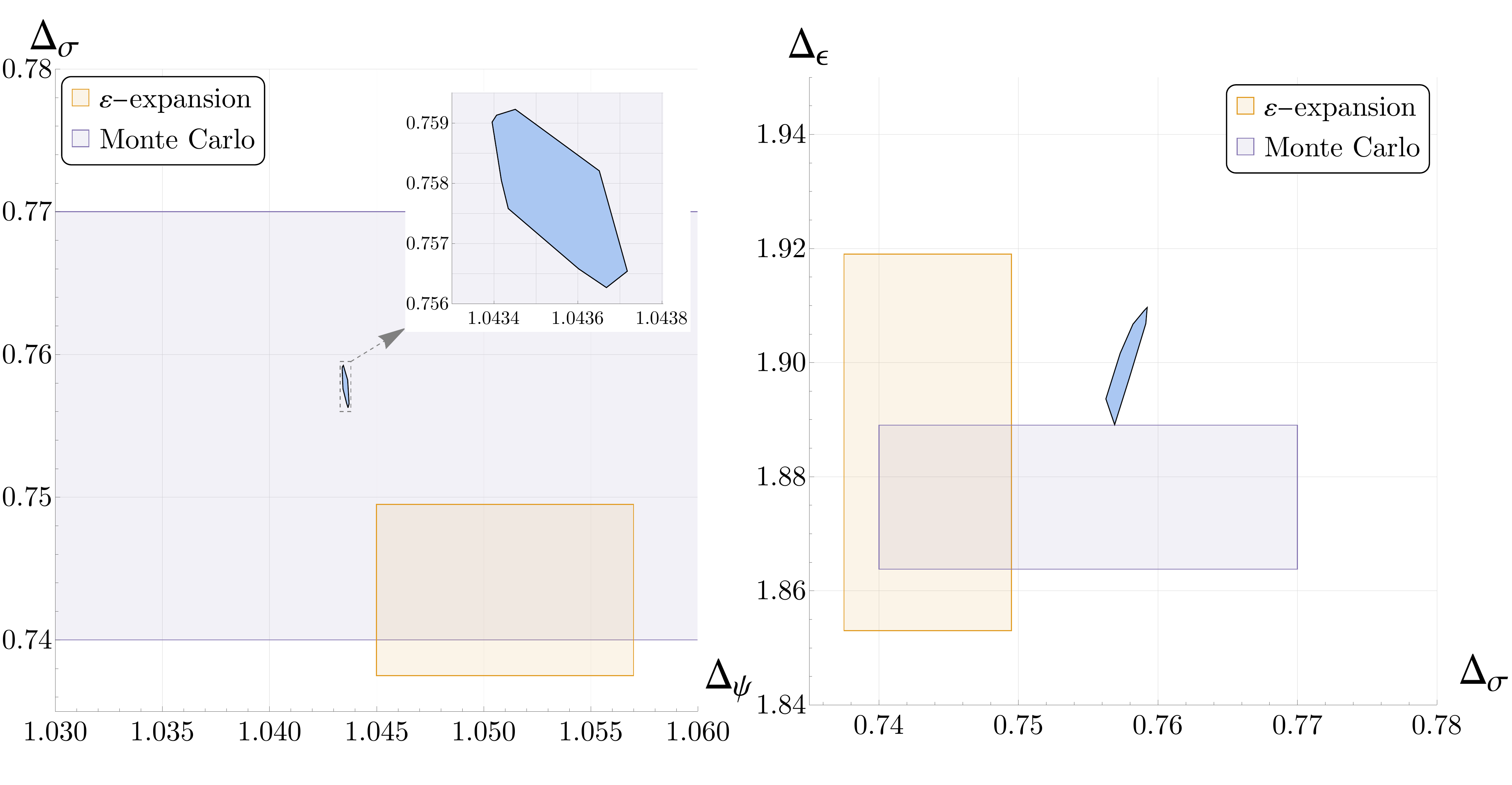}
\end{center}
\caption{The allowed region for the $N=4$ critical GNY model when imposing that $\Delta_{\sigma'}>3$ computed at \(\nmax=18\), projected to the $(\Delta_\psi,\, \Delta_\sigma)$ and $(\Delta_\sigma,\, \Delta_\epsilon)$ planes. This should be compared to the Borel re-summed result obtained from the $\epsilon$-expansion \cite{Ihrig:2018hho} obtained using the DREG$_3$ regularization scheme (shown in orange) and with the Monte Carlo results from \cite{Huffman:2019efk} (shown in purple). Results for $\Delta_\psi$ are not available from the Monte Carlo study cited.}\label{fig: N4_combined}
\end{figure}

\begin{figure}[t!]
\begin{center}
\includegraphics[width=0.98\textwidth]{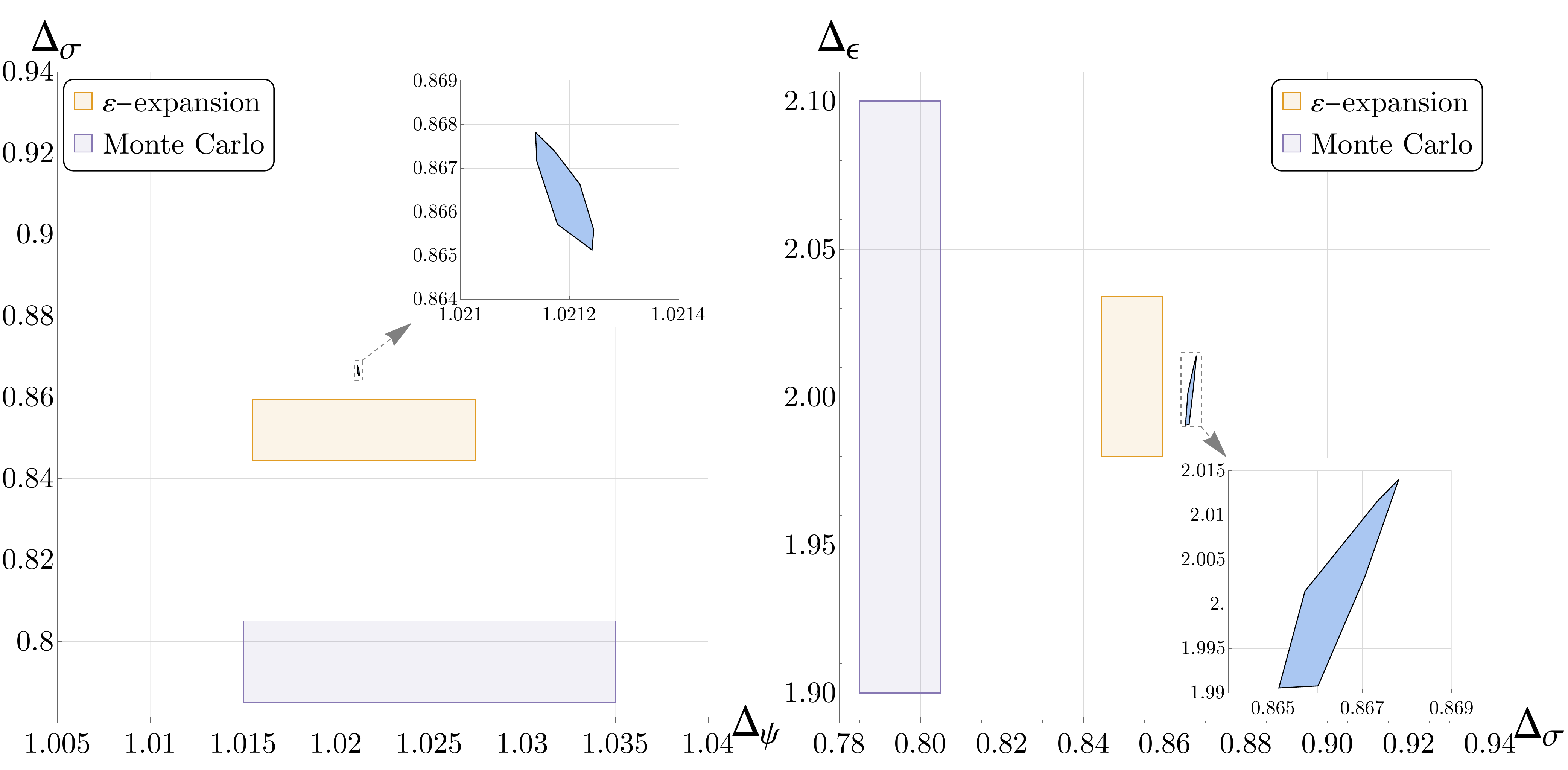}
\end{center}
\caption{The allowed region for the $N=8$ critical GNY model when imposing that $\Delta_{\sigma'}>3$ computed at \(\nmax=18\), projected to the $(\Delta_\sigma,\, \Delta_\epsilon)$ and $(\Delta_\psi,\, \Delta_\sigma)$ planes. This should be compared to the Borel re-summed result obtained from the $\epsilon$-expansion \cite{Ihrig:2018hho}, once again obtained using the DREG$_3$ regularization scheme (and shown in orange), and with the Monte Carlo results from \cite{Liu:2019xnb} (shown in purple).}\label{fig: N8_combined}
\end{figure}

For \(N=2\) at \(\nmax=18\) we can report rigorous estimates of our external scaling dimensions of \(\Delta_\psi=1.06861\mathbf{(12)}\), \(\Delta_\sigma=0.6500\mathbf{(12)}\), \(\Delta_\epsilon=1.725\mathbf{(7)}\), given the assumptions as discussed in section \ref{sec:vals-of-N-of-focus} of \(\Delta_{\epsilon'}>3,\,\Delta_{\sigma'}>3,\,\Delta_{\sigma_T}>2,\,\Delta_{\psi'}>2,\,\Delta_{\chi}>3.5\). Figure \ref{fig: N2_combined} shows the allowed regions\footnote{Readers familiar with bootstrap results may be concerned as to the jagged nature of the islands reported. The \(\nmax=18\) computations are very computationally expensive, so we only have a relatively small number of allowed points in each island. However, because we have a lot of information as to the disallowed points from lower \(\nmax\), and because the precision of these islands are still improved, we have elected to report the superficially more jagged islands as they represent our best results. We have taken pains to ensure that these results are rigorous, as outlined in section \ref{sec:delaunay}.} after projections to the $(\Delta_\sigma,\, \Delta_\epsilon)$ and $(\Delta_\psi,\, \Delta_\sigma)$ planes. As shown in the $(\Delta_\sigma,\, \Delta_\epsilon)$-plane, the bootstrap results exclude the reported error bars from earlier Monte Carlo studies from  \cite{Tabatabaei:2021tqv}; as can be seen in both plots, the bootstrap results also marginally exclude the reported error bars from the \(\e\)-expansion results after Borel resummation \cite{Ihrig:2018hho}. For both studies, the bootstrap results improve the precision of some of these estimates by orders of magnitude.

We also report (nonrigorous) estimates at \(N=2\) at \(\nmax=14\)\footnote{We report our \(\nmax=14\) estimates because we ran out of computing resources while computing \(\nmax=18\). While we have enough data to feel confident in our rigorous scaling dimension estimates, we deferred to our lower \(\nmax\) results for the OPE coefficient ratio estimates.} of the OPE coefficient ratios of \(\frac{\lambda_{\psi\psi\sigma}}{\lambda_{\sigma\sigma\psi}} = 0.5087(10)\),  \(\frac{\lambda_{\psi\psi\e}}{\lambda_{\sigma\sigma\psi}} = 0.2392(6) \), and \(\frac{\lambda_{\e\e\e}}{\lambda_{\sigma\sigma\psi}} = 1.629(13)\).  Note that these estimates are nonrigorous as discussed in section \ref{sec:ope-algorithm}; for the reader's convenience they are also reported in table \ref{tab:bootstrap-GNY-OPE-summary}. 

As noted in section \ref{sec:vals-of-N-of-focus}, the assumption of \(\Delta_{\sigma'}>3\) for \(N=2\) is perhaps at risk of being too strong because of the \(N=1\) value which violates this assumption. We note that despite this assumption we are still able to find feasible points. Moreover, the two-sided Pad\'e interpolations shown in figure~\ref{fig:2-sided-Pade-for-scalars} support this assumption, so we think this gap is justified. In future work we plan to study the CFT data in the $\sigma'$ sector more rigorously using the navigator method~\cite{Reehorst:2021ykw}. For now, we report our estimates using both \(\Delta_{\sigma'}>3\) as well as the more conservative assumption of \(\Delta_{\sigma'}>2.5\) in tables \ref{tab:bootstrap-GNY-summary} and \ref{tab:bootstrap-GNY-OPE-summary}.

For \(N=4\) at \(\nmax=18\), we can report rigorous estimates of our external scaling dimensions of \(\Delta_\psi=1.04356\mathbf{(16)}\), \(\Delta_\sigma=0.7578\mathbf{(15)}\), \(\Delta_\epsilon=1.899\mathbf{(10)}\), given the assumptions as discussed in section \ref{sec:vals-of-N-of-focus} of \(\Delta_{\epsilon'}>3,\,\Delta_{\sigma'}>3,\,\Delta_{\sigma_T}>2,\,\Delta_{\psi'}>2,\,\Delta_{\chi}>3.5\). Figure \ref{fig: N4_combined} shows the allowed regions after projections. In this case, the $\epsilon$-expansion estimates \cite{Ihrig:2018hho} are excluded by the conformal bootstrap.\footnote{It's worth noting that the \(\epsilon\)-expansion results encounter a pole for \(N\simeq 2\) which appears to distort some of their resummations~\cite{Ihrig:2018hho}.} On the other hand, the existing Monte Carlo results  \cite{Huffman:2019efk} give no estimates on $\Delta_\psi$, and the reported error bars are excluded by the conformal bootstrap in the $(\Delta_\sigma,\, \Delta_\epsilon)$ plane. A subtlety to be noticed is that the MC estimates shown in the plots were obtained based on the $O(2)^2 \rtimes \Z_2$ ``chiral" GNY model as discussed in section \ref{sec:two-GNYs}, whose CFT data for $\{\Delta_{\psi}, \Delta_{\sigma}, \Delta_{\epsilon}\}$ is expected to be slightly different from that of the $O(4)$ GNY model that we implemented in the conformal bootstrap.  

We can also report estimates at \(N=4\) of OPE coefficient ratios of \(\frac{\lambda_{\psi\psi\sigma}}{\lambda_{\sigma\sigma\psi}} = 0.4386(6)\),  \(\frac{\lambda_{\psi\psi\e}}{\lambda_{\sigma\sigma\psi}} = 0.15530(19)\), and \(\frac{\lambda_{\e\e\e}}{\lambda_{\sigma\sigma\psi}} = 1.682(18)\). Note that these estimates are nonrigorous as discussed in section \ref{sec:ope-algorithm}; for the reader's convenience they are also reported in table \ref{tab:bootstrap-GNY-OPE-summary}.

For \(N=8\) and \(\nmax=18\), we can report rigorous estimates of our external scaling dimensions of \(\Delta_\psi=1.02119\mathbf{(5)}\), \(\Delta_\sigma=0.8665\mathbf{(13)}\), \(\Delta_\epsilon=2.002\mathbf{(12)}\), given the assumptions as discussed in section \ref{sec:vals-of-N-of-focus} of \(\Delta_{\epsilon'}>3,\,\Delta_{\sigma'}>3,\,\Delta_{\sigma_T}>2,\,\Delta_{\psi'}>2,\,\Delta_{\chi}>3.5\). Figure \ref{fig: N8_combined} shows the allowed regions after projections.
The reported error bars for the \(\Delta_\sigma\) estimates of the $\epsilon$-expansion \cite{Ihrig:2018hho} and the Monte Carlo results \cite{Liu:2019xnb} are excluded by the conformal bootstrap, while the other estimates show good agreement.
Precision is considerably improved for all scaling dimensions, especially for \(\Delta_\psi\).
It should be noted again, the MC estimates were obtained by studying the chiral theory, discussed in section \ref{sec:two-GNYs}.

We can also report estimates at \(N=8\) of OPE coefficient ratios of \(\frac{\lambda_{\psi\psi\sigma}}{\lambda_{\sigma\sigma\psi}} = 0.3322(8)\),  \(\frac{\lambda_{\psi\psi\e}}{\lambda_{\sigma\sigma\psi}} = 0.08082(12)\), and \(\frac{\lambda_{\e\e\e}}{\lambda_{\sigma\sigma\psi}} = 1.71(4)\). Note that these estimates are nonrigorous as discussed in section \ref{sec:ope-algorithm}; for the reader's convenience they are also reported in table \ref{tab:bootstrap-GNY-OPE-summary}. 

\begin{table}[t!]
	\begin{center}
		\resizebox{\columnwidth}{!}{
		\begin{tabular}{l | L L L | L L L}
			\hline\hline
			 & \hspace{0.5cm} \Delta_\psi & \hspace{0.5cm} \Delta_\sigma &\hspace{0.5cm}  \Delta_\epsilon&  \lambda_{\psi\psi\sigma}/\lambda_{\sigma\sigma\epsilon} &  \lambda_{\psi\psi\epsilon}/\lambda_{\sigma\sigma\epsilon} & \lambda_{\e\e\epsilon}/\lambda_{\sigma\sigma\epsilon} \\
			\hline
			\(\mathbf{N=2}\) & &  &  \\
			\(n_\text{max}=18, \Delta_{\sigma'}>2.5\)  
			 & 1.0672\mathbf{(25)} & 0.657\mathbf{(13)} & 1.74\mathbf{(4)} & 0.5071(15) & 0.2347(35) & 1.636(17) \\
        	\(n_\text{max}=14, \Delta_{\sigma'}>3\)  
			 &1.06860\mathbf{(16)}  & 0.6498\mathbf{(14)} & 1.724\mathbf{(8)} & 0.5087(10) &0.2392(6)  &1.629(13)\\
			\(n_\text{max}=18, \Delta_{\sigma'}>3\)  
			 & 1.06861\mathbf{(12)} & 0.6500\mathbf{(12)} & 1.725\mathbf{(7)} & -  & - & - \\
			
			\hline 
			\(\mathbf{N=4}\) & & & \\ 
			\(n_\text{max}=18, \Delta_{\sigma'}>3\)  
			  & 1.04356\mathbf{(16)}& 0.7578\mathbf{(15)} & 1.899\mathbf{(10)} & 0.4386(6) & 0.15530(19) & 1.682(18) \\

			\hline
			\(\mathbf{N=8}\) & & & \\
			\(n_\text{max}=18, \Delta_{\sigma'}>3\)  
			 & 1.02119\mathbf{(5)} & 0.8665\mathbf{(13)} & 2.002\mathbf{(12)} & 0.3322(8) & 0.08082(12) & 1.71(4) \\
			
			\hline\hline
		\end{tabular}}
 	\caption{A summary of the results of this work with estimates of all six search parameters, compiled here for the reader's convenience. We do not report the \(\nmax=18\) OPE coefficient ratio estimates, as we did not have sufficient statistics. Note that we have reported the scaling dimension estimates as only scaling dimensions, and we have only included this work's results. For critical exponents and comparisons, see table \ref{tab:bootstrap-GNY-summary}.
\label{tab:bootstrap-GNY-OPE-summary}}
\end{center}
\end{table}

\subsection{$N = 1$ and emergent supersymmetry}

\begin{figure}[t!]
	\begin{center}
		\includegraphics[width=0.98\textwidth]{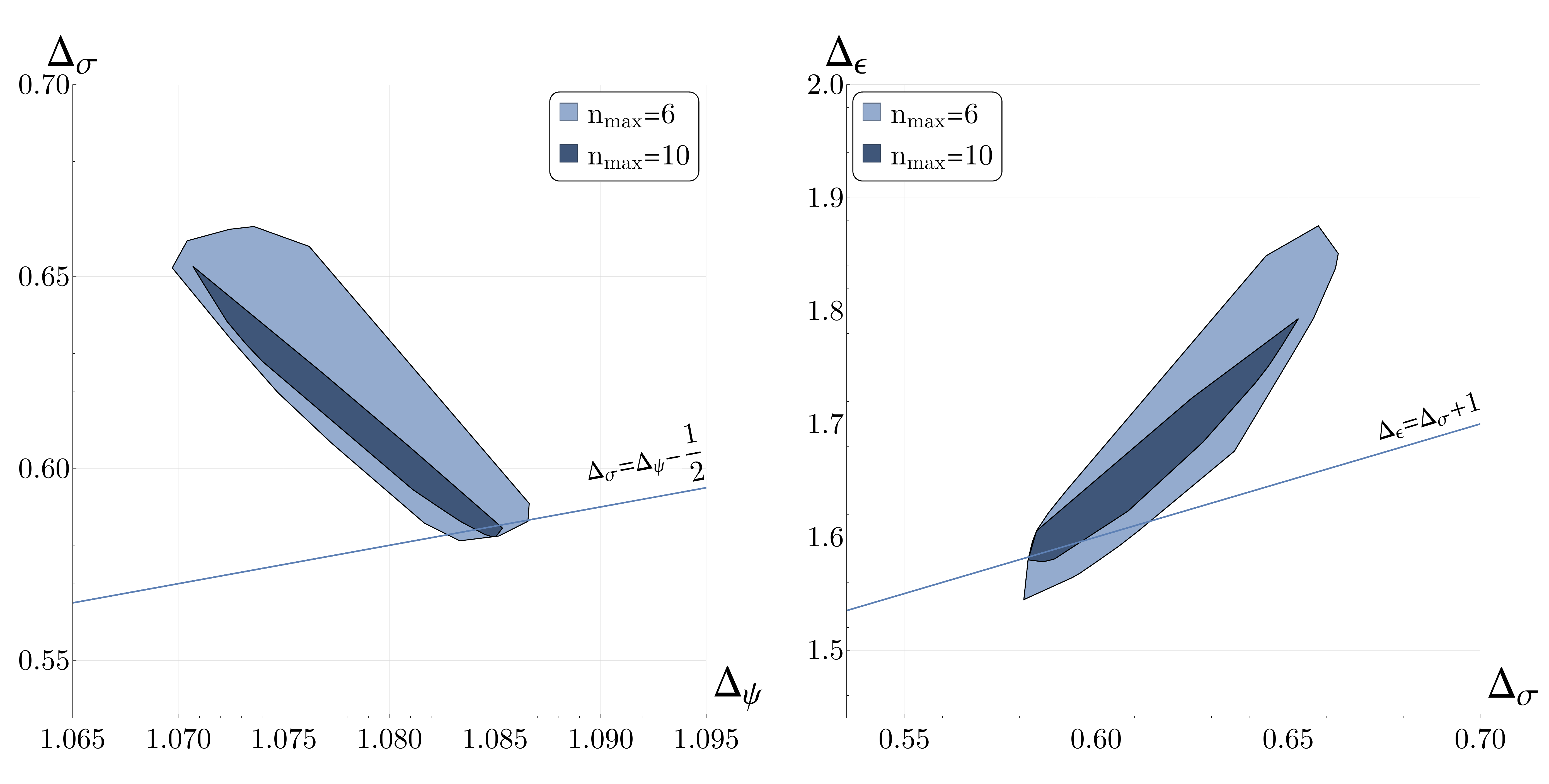}
	\end{center}
	\caption{The allowed region for the $N=1$ critical GNY model when imposing that $\Delta_{\sigma
	'} > 2.5$, projected to the $(\Delta_\psi,\, \Delta_\sigma)$ and $(\Delta_\sigma,\, \Delta_\epsilon)$ planes. The solid lines capture the expected relation between scaling dimensions in an  $\mathcal  N=1$  SCFT. The tip intersects with the supersymmetric constraint on scaling dimensions, though note that due to the projection and visualization method (described in section~\ref{sec:delaunay}) that the extent of that intersection is exaggerated, as the line and island are actually in 3-dimensional space. We separately computed a binary search along the supersymmetric line in all three external dimensions and found a rigorous estimate of \(\Delta_{\sigma'} = 0.58444(\mathbf{8})\). This agrees exactly with those found for the $\cN=1$ super-Ising model in~\cite{Atanasov:2022bpi}.  }
	\label{fig:susy-emergence-plot}
\end{figure}

We also computed islands at \(\nmax=6,10\) for the \(N=1\) case, shown in figure \ref{fig:susy-emergence-plot}. Shown also in the plot is the expected relation between scaling dimensions for an \(\cN=1\) SCFT; our three external operators are expected to all be in a supermultiplet with each other. We can see that the very tip of the island indeed is consistent with the assumption of supersymmetry. There is a long tail, however, which seems to get cut away as \(\nmax\) is increased. The intersection of the tip of the island with the supersymmetric line is very narrow, and at \(\nmax=10\) we can report a rigorous estimate of \(\Delta_\sigma\) of \(0.58444(\mathbf{8})\), which is both completely consistent though roughly an order of magnitude less precise than the \(\nmax=30\) superconformal bootstrap results reported in \cite{Atanasov:2022bpi}.\footnote{While the \(\nmax\) values reported here are quite different, we should note that the system of crossing equations studied in \cite{Atanasov:2022bpi} has only 4 crossing equations while our \(N=1\) system has 28.}

One advantage of our mixed fermion bootstrap setup is that we now have access to half-integer spin exchange channels. In this case, the parity-odd \(\ell=3/2\) channel would be expected to have a supercurrent at the unitarity bound of \(\Delta = 2.5\) if the solution to crossing corresponds to a supersymmetric CFT. We can therefore investigate whether a given solution to crossing in our \(N=1\) island \emph{must} have supersymmetry by imposing gaps in the supercurrent channel and seeing what the upper bound of that gap is. If the feasibility of a solution to crossing is sensitive to this gap assumption, we can say that in that solution there must be a low-lying operator in that spectrum. 

\begin{figure}[t!]
\begin{center}
\includegraphics[width=0.48\textwidth]{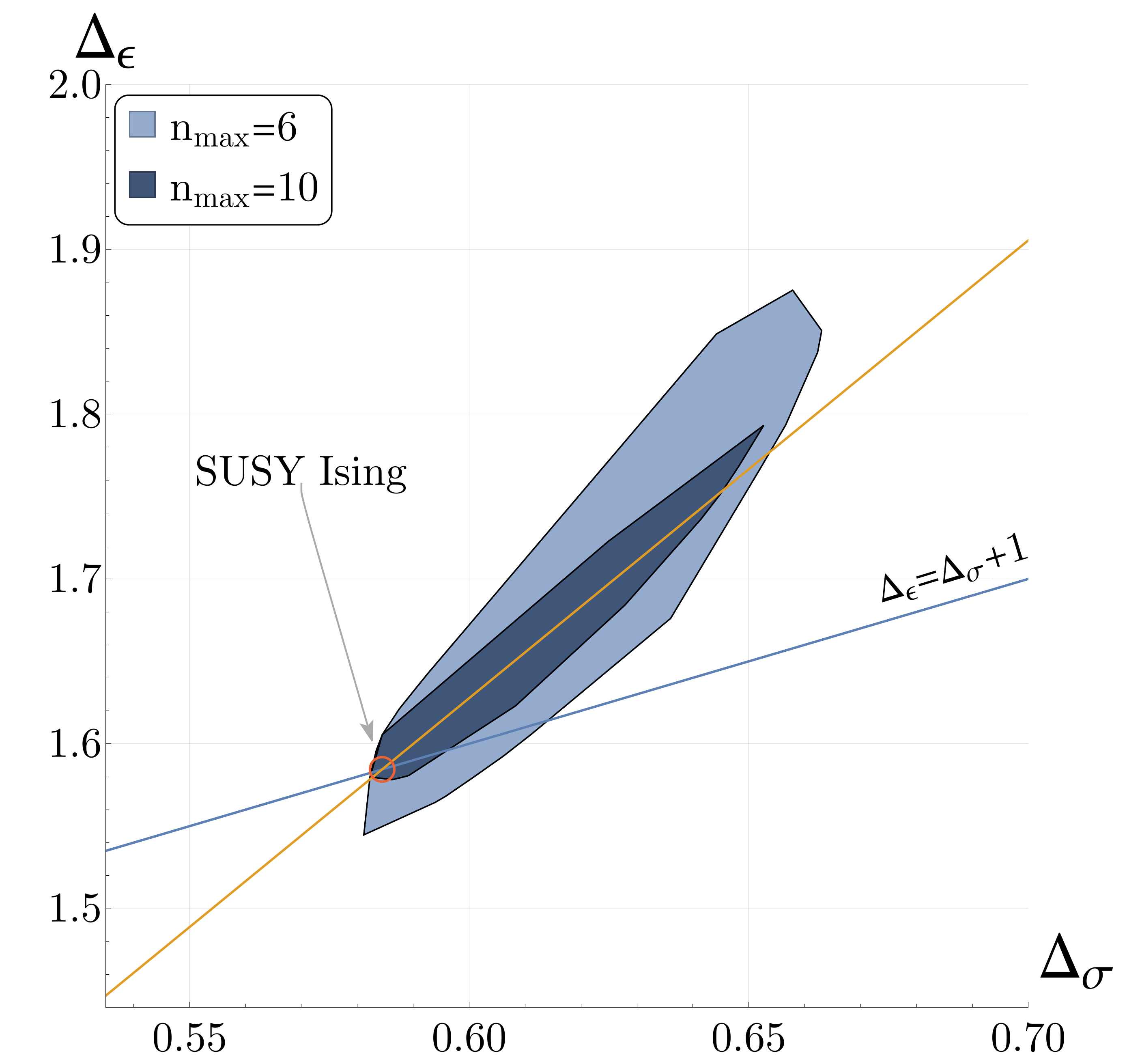}
\includegraphics[width=0.48\textwidth]{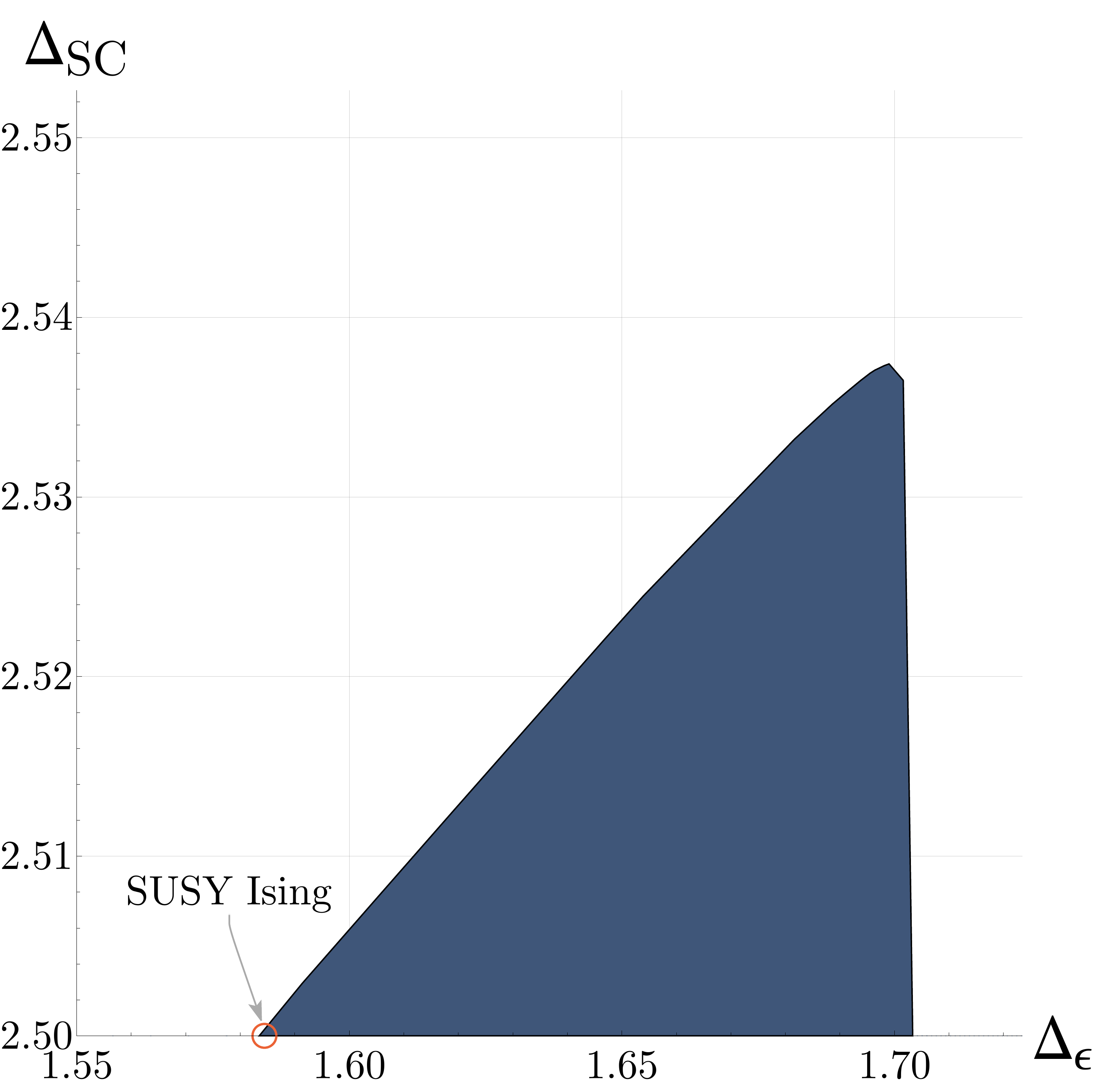}
\end{center}
\caption{\label{fig:susy-supercurrent-bound}On the right we show the upper-bound at \(\nmax=10\) on the scaling dimension of the lowest spin-$3/2$ operator as a function of $\Delta_\epsilon$ obtained along an axis of the $N=1$ island when imposing $\Delta_{\sigma'}>2.5$. The axis was selected such that it interpolates between the location of the \(\cN=1\) super-Ising model as determined in previous literature \cite{Rong:2018okz,Atanasov:2018kqw,Atanasov:2022bpi} and the center of the \(\nmax=10\) island. We show this axis projected into \((\Delta_{\sigma},\Delta_\epsilon)\) space in orange in the figure on the left. We see that all points along this axis are forced to have a spin-$3/2$ operator that is close to the unitarity bound, where such operator becomes a supercurrent. The red circle in the plot on the right gives the location of the critical $\cN=1$ super-ising model as determined from the $\cN=1$ superconformal bootstrap \cite{Atanasov:2022bpi}.}
\end{figure}

Specifically, we can perform a binary search in the supercurrent gap to find precisely what is the upper bound of the scaling dimension of the leading operator in that channel. Scanning along an axis of our \(\nmax=10\) island, which goes from the tip through the center (shown in figure \ref{fig:susy-supercurrent-bound}), we find that for the entire scan \(\Delta_\text{SC} < 2.54\). In particular, at the tip of the island (corresponding to the \(\cN=1\) super-Ising model), the upper bound drops to \(\Delta_\text{SC}<2.5003219\). This implies that any CFT with these parameters must be, to a high degree of precision, supersymmetric.\footnote{We also did a preliminary exploration of how the bound on lowest spin-$3/2$ operator changes as a function of the gap in $\Delta_{\sigma'}$. As this gap is increased towards the value that it takes in the $\cN=1$ super-critical Ising model, we saw that the upper-bound on the dimension of the spin-$3/2$ operator becomes even stronger, as the right part of the plot (the tail of the island) shrinks away.}

\section{Discussion}
\label{sec:discussion}

In this work we have obtained the first rigorous and precise islands for the conformal data of the 3d $O(N)$ Gross-Neveu-Yukawa fixed points from the conformal bootstrap. Much like the 3d Ising and $O(N)$ vector models, these theories appear to be readily amenable to bootstrap methods. In particular, we have shown that one can obtain small islands in the parameter space of CFT data after studying the crossing relations for the operators $\{\psi, \sigma, \epsilon\}$ and  imposing gaps in the spectrum which isolate the leading scalar and spin-$1/2$ operators. The gaps we have chosen are motivated by perturbative calculations in the large-$N$ and $\epsilon$-expansions. They can also be viewed as being necessary in order to exclude potential nearby nonlocal fixed points obtained by coupling the leading operators in the GNY theories to generalized free fields. In the $\chi$ and $\sigma'$ sectors the gaps can be viewed as imposing a consequence of the equations of motion for the fundamental fields. We saw a particular sensitivity of our results to the gap in $\Delta_{\sigma'}$, and establishing its irrelevance down to $N=2$ appears to be an important open problem. In future work we plan to study the allowed values of this gap more rigorously using navigator methods~\cite{Reehorst:2021ykw} and use the results to further improve our islands.

In the present study, we focused on the $O(N)$-invariant GNY theories with $N=1,2,4,8$, but it is clear that the results can be extended to any $N$. In the case of $N=1$, the fixed point is believed to have emergent 3d $\mathcal{N}=1$ supersymmetry. If one assumes supersymmetry then one can perform a precise bootstrap of this model using only external scalar fields as was demonstrated in~\cite{Atanasov:2022bpi}. If one does not assume supersymmetry, then we have seen in the present work that the numerical bootstrap with external fermions still forces the solution to be approximately supersymmetric, requiring a spin-$3/2$ ``supercurrent'' operator in the spectrum that is near the unitarity bound.

In this paper we also highlighted the distinction between the $O(N)$ GNY fixed points and the $O(N/2)^2 \rtimes \Z_2$ GNY fixed points. For the leading operators $\{\psi,\sigma,\epsilon\}$, differences in their scaling dimensions only show up at 4-loop order in the $\epsilon$ and large-$N$ expansions and they are expected to be extremely close to each other. On the other hand, the models have more significant differences in other parts of the spectrum, e.g.~in the number of conserved currents and spectrum of fermion bilinear operators. In the future it will be interesting to study the  $O(N/2)^2 \rtimes \Z_2$ GNY fixed points using numerical bootstrap methods and see if we can clearly resolve the difference between these models. It will additionally be interesting to study bounds on OPE coefficients and central charges in both models. These fixed points can also readily be generalized to models with multiple scalar fields, e.g.~the chiral-XY and chiral-Heisenberg GNY fixed points, which will also be interesting targets for the bootstrap.

Our work was the first significant application of the new software tool \texttt{blocks\_3d}~\cite{Erramilli:2020rlr}, which enables systematic and efficient calculations of 3d conformal blocks with arbitary spinning operators. Along with it, we have developed an extensible and modular software stack, as described in appendix \ref{app:software}. The success of our study makes it clear that this approach can be used in other studies of interest, e.g.~mixed correlators containing various combinations of scalars, fermions, currents, and stress tensors. We hope that 
future systematic studies of the bootstrap constraints for such correlators will lead to the discovery of exciting new islands in the vast  ocean of possible CFTs.

\section*{Acknowledgments}

We thank Meng Cheng, Simone Giombi, John Gracey, Yin-Chen He, Filip Kos, Igor Klebanov, Yakov Landau, Walter Landry, Zhijin Li, Marco Meineri, Max Metlitski, Matthew Mitchell, Silviu Pufu, Slava Rychkov, Michael Scherer, Ning Su, and Yuan Xin for discussions. DSD and AL were supported by Simons Foundation grant 488657 (Simons Collaboration on the Nonperturbative Bootstrap) and by a DOE Early Career Award under grant DESC0019085. DP and RSE were supported by Simons Foundation grant 488651 (Simons Collaboration on the Nonperturbative Bootstrap) and DOE grant DE-SC0017660. LVI was supported in part by the Simons Collaboration on Ultra-Quantum Matter, a Simons Foundation Grant with No. 651440. PK was supported by DOE grant DESC0009988 and the Adler Family Fund at the Institute for Advanced Study. This work used the Extreme Science and Engineering Discovery Environment (XSEDE) Expanse Cluster at the San Diego Supercomputing Center (SDSC) through allocation PHY190023, which is supported by National Science Foundation grant number ACI-1548562. Computations in this work were also performed on the Caltech High Performance Cluster, partially supported by a grant from the Gordon and Betty Moore Foundation, on the Yale Grace computing cluster, supported by the facilities and staff of the Yale University Faculty of Sciences High Performance Computing Center, and on the Institute for Advanced Study Helios cluster. This work was performed in part at the Aspen Center for Physics, which is supported by National Science Foundation grant PHY-1607611.

\appendix

\section{Software}
\label{app:software}

For this work, we implemented and used several software packages, which we briefly describe in this appendix.\footnote{Note: many of these packages are works in progress, and their names and APIs are subject to change.} We indicate with a ``$\star$" the packages that were newly written for this work. Other packages have been re-used (and in some cases modified) from other projects. The profusion of libraries is because we have made an effort to split them up into an orthogonal set of features.
\begin{itemize}
\item {\tt SDPB} (\url{https://github.com/davidsd/sdpb}): a C++ program for solving semidefinite programs \cite{Simmons-Duffin:2015qma,Landry:2019qug}.
\item {\tt blocks\_3d} (\url{https://gitlab.com/bootstrapcollaboration/blocks_3d}): a C++ program for computing spinning conformal blocks in 3d \cite{Erramilli:2019njx,Erramilli:2020rlr}.
\item {\tt scalar\_blocks} (\url{https://gitlab.com/bootstrapcollaboration/scalar_blocks}): a C++ program for computing scalar conformal blocks in general $d$.
\item {\tt hyperion} (\url{https://github.com/davidsd/hyperion}): a Haskell framework for concurrent computations on an HPC cluster.
\item {\tt hyperion-bootstrap} (\url{https://gitlab.com/davidsd/hyperion-bootstrap}): A Haskell library for computing numerical bootstrap bounds on an HPC cluster using {\tt hyperion}.
\item {\tt sdpb-haskell} (\url{https://gitlab.com/davidsd/sdpb-haskell}): A Haskell interface to {\tt SDPB}.
\item[$\star$] {\tt bootstrap-build} (\url{https://gitlab.com/davidsd/bootstrap-build}): A Haskell ``build system" for building objects and their dependencies.
\item {\tt bootstrap-math} (\url{https://gitlab.com/davidsd/bootstrap-math}): A Haskell math library containing datatypes and algorithms useful for bootstrap computations.
\item[$\star$] {\tt blocks-core} (\url{https://gitlab.com/davidsd/blocks-core}): Core Haskell datatypes and functions for conformal blocks in bootstrap
computations.
\item[$\star$] {\tt blocks-3d} (\url{https://gitlab.com/davidsd/blocks-3d}): A Haskell interface to the C++ program {\tt blocks\_3d}, built on {\tt blocks-core}.
\item[$\star$] {\tt scalar-blocks} (\url{https://gitlab.com/davidsd/scalar-blocks}): A Haskell interface to the C++ program {\tt scalar\_blocks}, built on {\tt blocks-core}.
\item[$\star$] {\tt bootstrap-bounds} (\url{https://gitlab.com/davidsd/bootstrap-bounds}): A Haskell library for setting up crossing equations using information about three- and four-point structures. This library implements the algorithm described in section~\ref{sec:numerical-setup}.
\item {\tt quadratic-net} (\url{https://gitlab.com/davidsd/quadratic-net}): A search algorithm for solving non-convex quadratic constraints in low numbers of dimensions, used in the OPE scan algorithm of \cite{Chester:2019ifh}.
\item[$\star$] {\tt scalars-3d} (\url{https://gitlab.com/davidsd/scalars-3d}): An implementation of several bootstrap bounds on scalar theories in 3d, using the libraries listed here.
\item[$\star$] {\tt fermions-3d} (\url{https://gitlab.com/davidsd/fermions-3d}): An implementation of several bootstrap bounds on GNY models in 3d, using the libraries listed here. This is the main ``umbrella" package for the computations in this work.
\end{itemize}

\section{$\De_\chi$ and $\De_{\psi'}$ at large $N$}
\label{sec:fermion-op-in-large-N-exp}

To better isolate the $O(N)$ GNY-model using the conformal bootstrap it is useful to get a better estimate for the scaling dimension of various operators in the theory, especially for those that are involved in the equation of motion or in  the new fermionic channels that are involved in a fermionic-scalar OPE. In particular, we would like to compute the $1/N$ correction to the dimension of the fermionic operators $\phi^2 \psi_i$ (the lowest dimensional operator above $\psi_i$) and $\phi^3 \psi_i$ (the lowest dimensional primary in its parity sector and $O(N)$ representation).  We will determine these scaling dimension by extracting the logarithmic divergence coefficient in the propagator of the more generic operator $\cO_i  =\phi^k \psi_i$. The bare dimension of this operator is $\Delta_{\phi^k \psi_i}=  k+1 + O(1/N)$. We will follow a similar logic to that used to determine the dimension of $\phi^k$ in appendix B of \cite{Iliesiu:2015qra}. 

Up to order  $1/N$, this propagator (in the concrete, but generalizable, example with $k=2$) is given by the following diagrams: 
\be 
 &D_{\cO_i}(p)= 
  \begin{tikzpicture}[baseline={([yshift=-.5ex]current bounding box.center)}, scale=0.22]
 \pgftext{\includegraphics[scale=1.0]{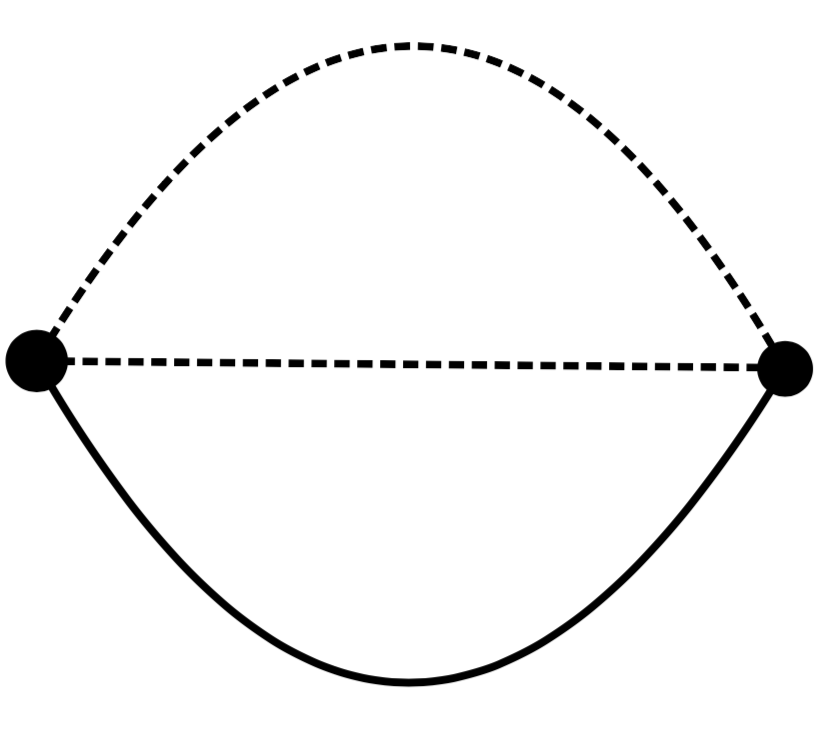}} at (0,0);
 \draw (-9.8,0) node  {$\mathcal O_{i}(p)$};
  \draw (9.9,0) node  {$\mathcal O_{i}(-p)$};
  \end{tikzpicture} + k \left[
    \begin{tikzpicture}[baseline={([yshift=-.35cm]current bounding box.center)}, scale=0.22]
   \pgftext{\includegraphics[scale=0.65]{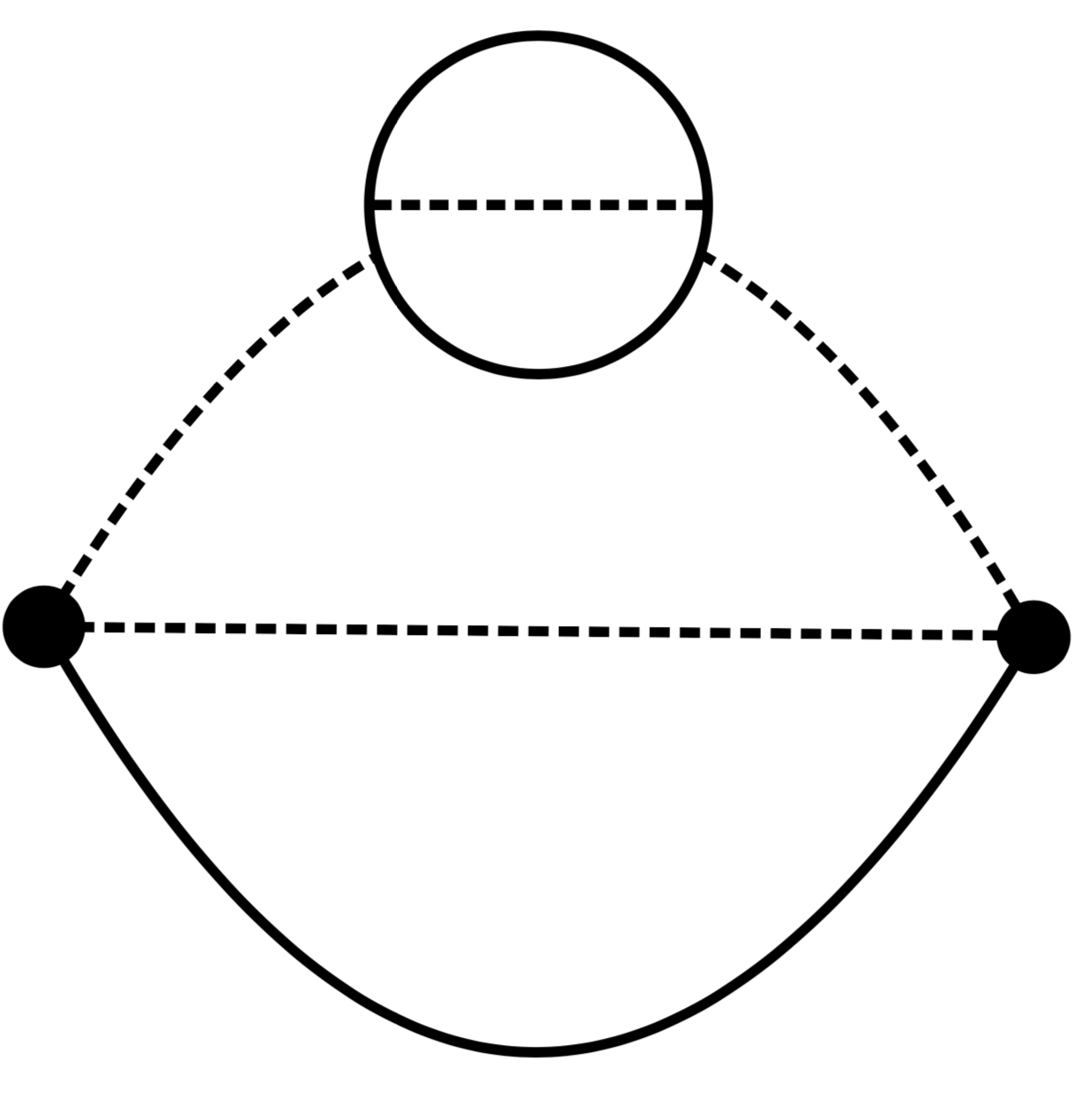}} at (0,0);
  \end{tikzpicture} +   
    \begin{tikzpicture}[baseline={([yshift=-.35cm]current bounding box.center)}, scale=0.22]
   \pgftext{\includegraphics[scale=0.65]{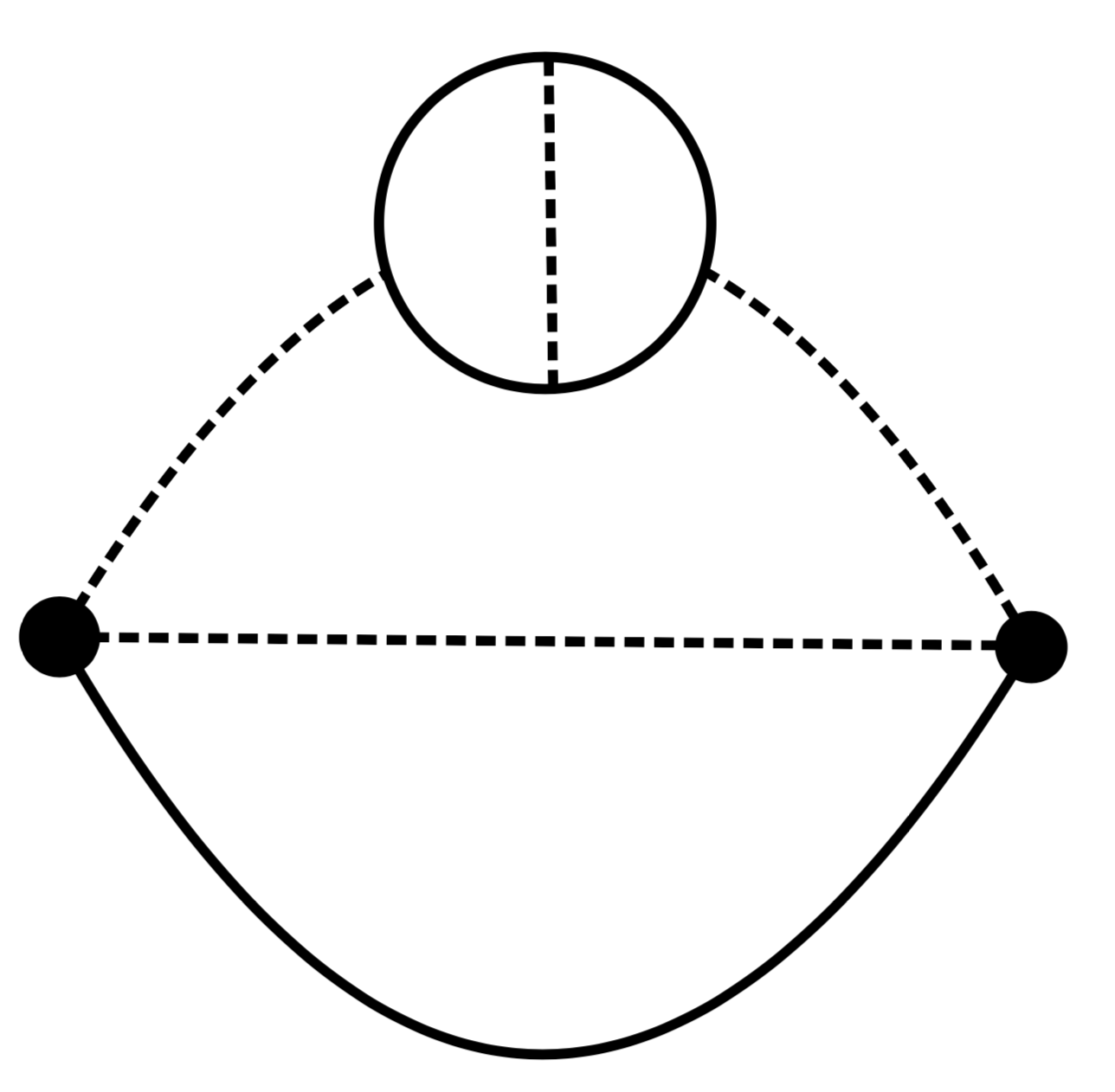}} at (0,0);
  \end{tikzpicture} 
  \right] \nn \\ 
  &+\frac{k(k-1)}2 \left[  \begin{tikzpicture}[baseline={([yshift=-.35cm]current bounding box.center)}, scale=0.22]
   \pgftext{\includegraphics[scale=0.65]{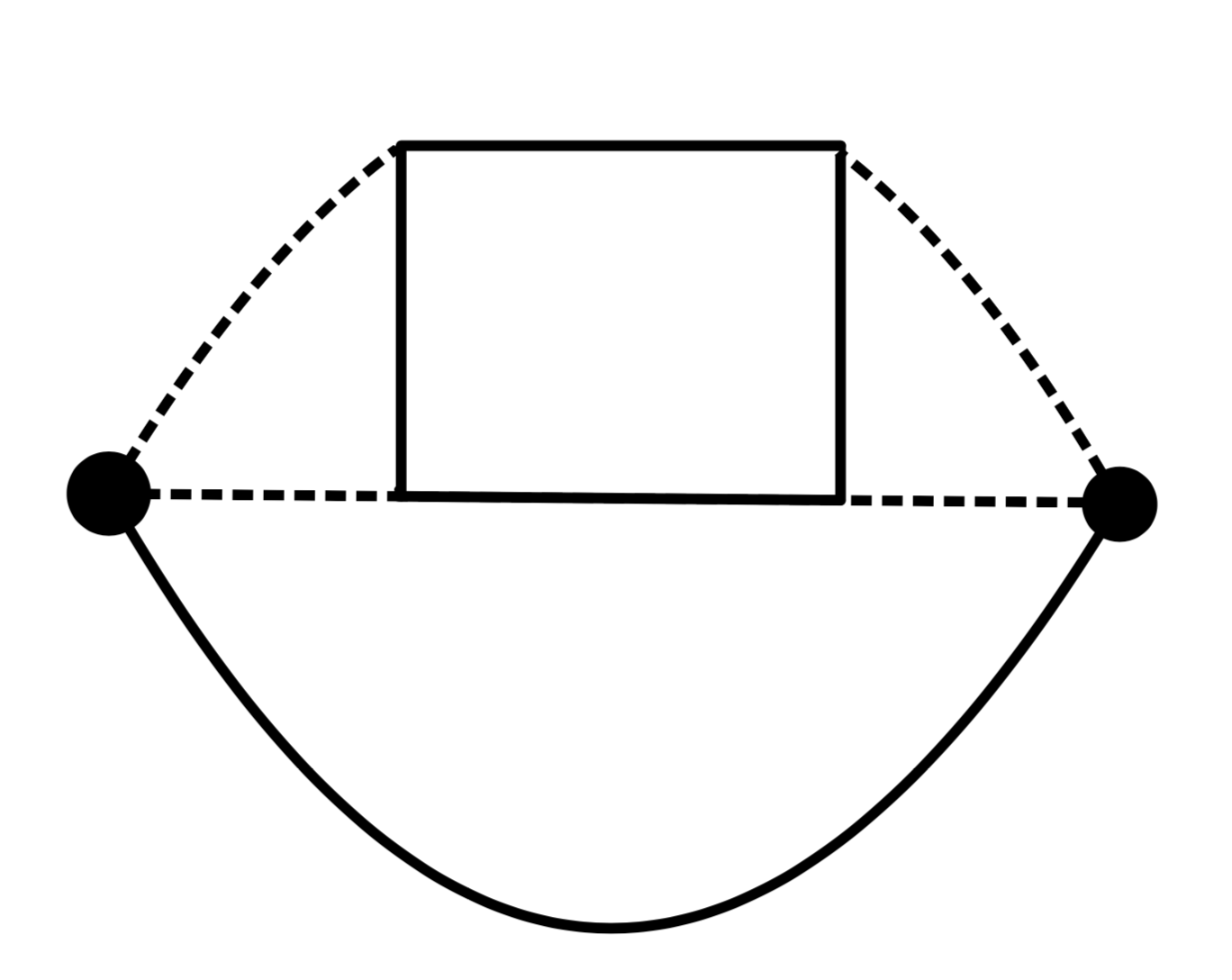}} at (0,0);
  \end{tikzpicture} \right] +  \begin{tikzpicture}[baseline={([yshift=-.35cm]current bounding box.center)}, scale=0.22]
   \pgftext{\includegraphics[scale=0.65]{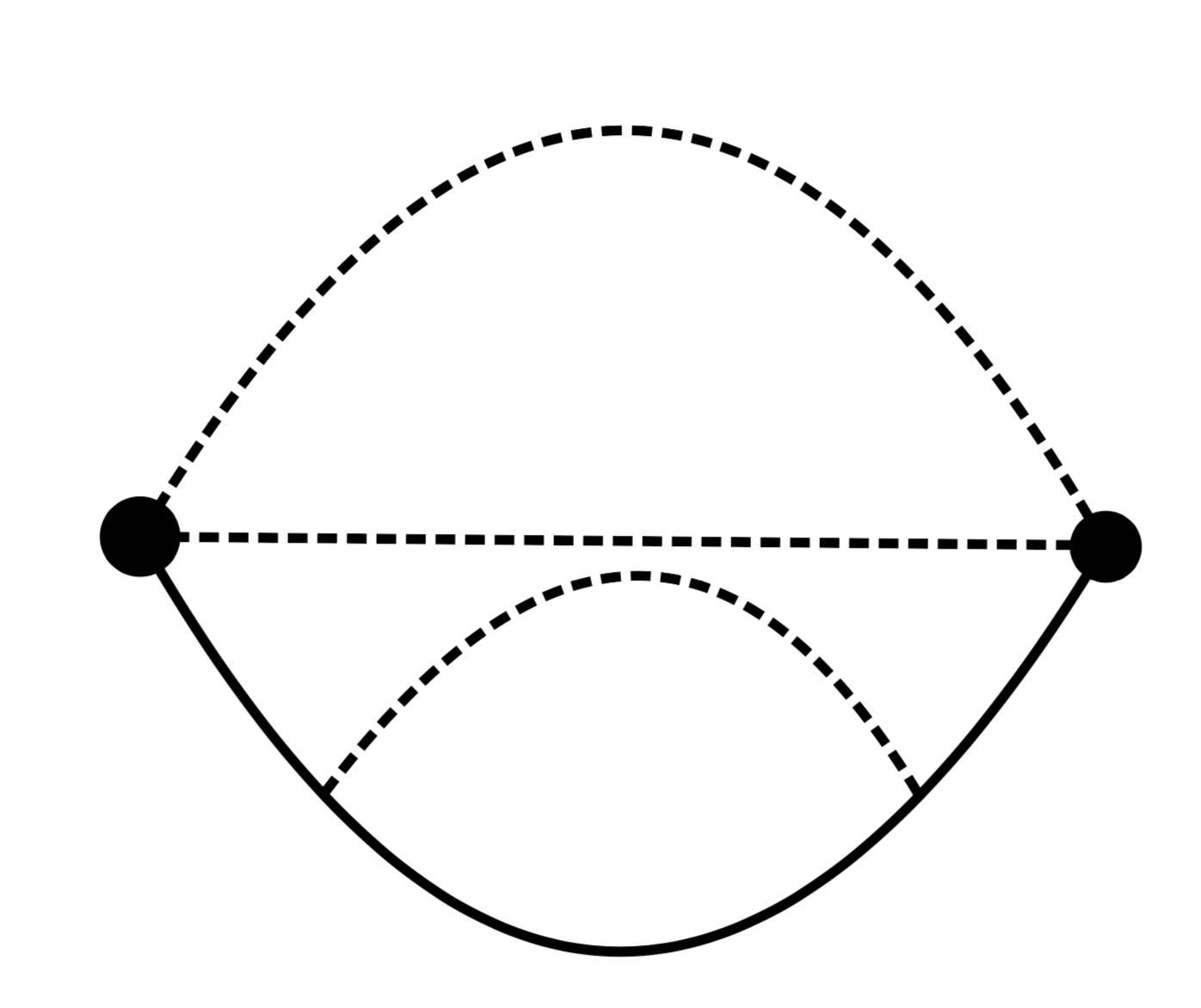}} at (0,0);
  \end{tikzpicture} + k\left[  \begin{tikzpicture}[baseline={([yshift=-.35cm]current bounding box.center)}, scale=0.22]
   \pgftext{\includegraphics[scale=0.65]{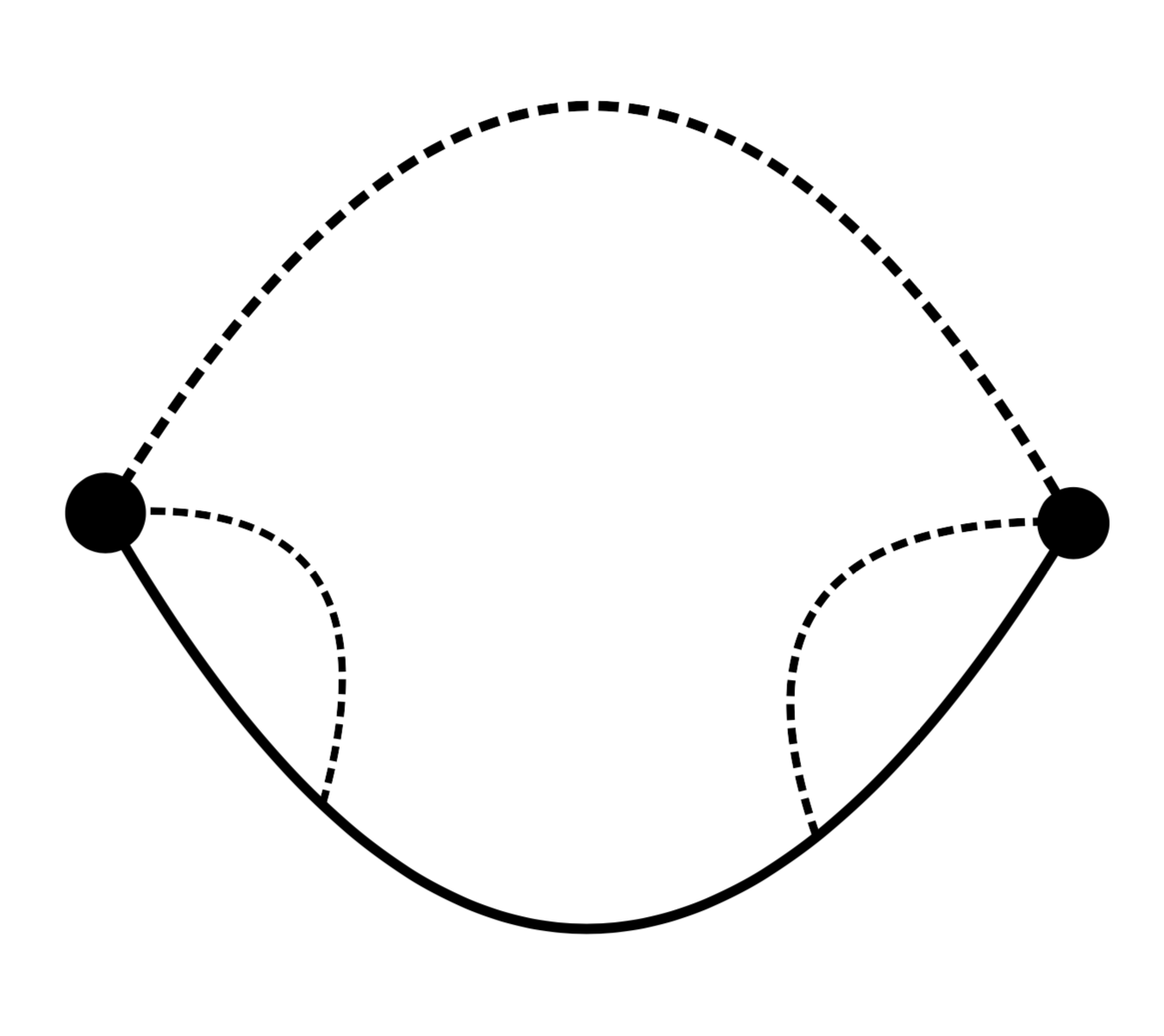}} at (0,0);
  \end{tikzpicture} \right] \,.
\ee
The first diagram gives the leading bare propagator. The next two subleading terms (i.e.~the one proportional to $k$ and the one proportional to $k(k-1)/2$) capture the anomalous dimension of $\phi^k$, which was found to be $\delta_{\phi^k} = \frac{1}N \frac{16k(3k-5)}{3\pi^2}$ \cite{Iliesiu:2015qra}. The next term yields the $1/N$ correction to the fermionic propagator and gives the contribution of the anomalous dimension of the fermionic field, $\delta_{\psi_i} =  \frac{1}{N} \frac{4}{3\pi^2}$. Finally, the last term is a new diagram which yields a correction $k\eta$ to the anomalous dimension. Thus, the scaling dimension of the operator is given by,
\be
\Delta_{\phi^k \psi_i} = k+1+\delta_{\phi^k} + \delta_{\psi_i} + k\eta\,. 
 \ee  
To determine $\eta$ we note that in the special case when $k=1$, following from the equations of motion,  the operator $\cO_i$ is a descendant and therefore, $\Delta_{\phi\psi_i}=1+\Delta_{\psi_i}$. This implies that the logarithmic divergences for the following diagrams should cancel each other:
\be 
&
  \left[
    \begin{tikzpicture}[baseline={([yshift=-.35cm]current bounding box.center)}, scale=0.22]
   \pgftext{\includegraphics[scale=0.65]{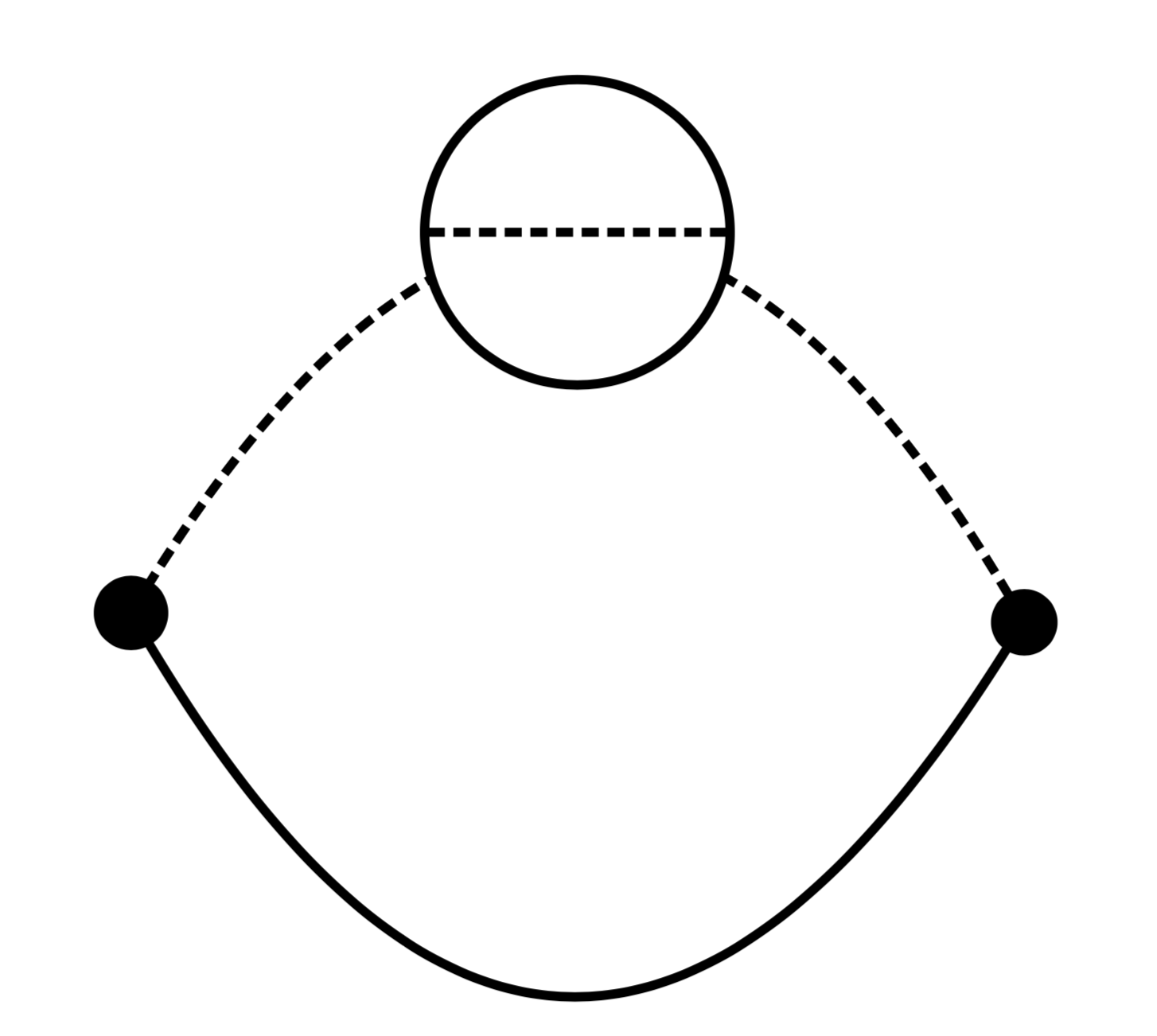}} at (0,0);
  \end{tikzpicture} +   
    \begin{tikzpicture}[baseline={([yshift=-.35cm]current bounding box.center)}, scale=0.22]
   \pgftext{\includegraphics[scale=0.65]{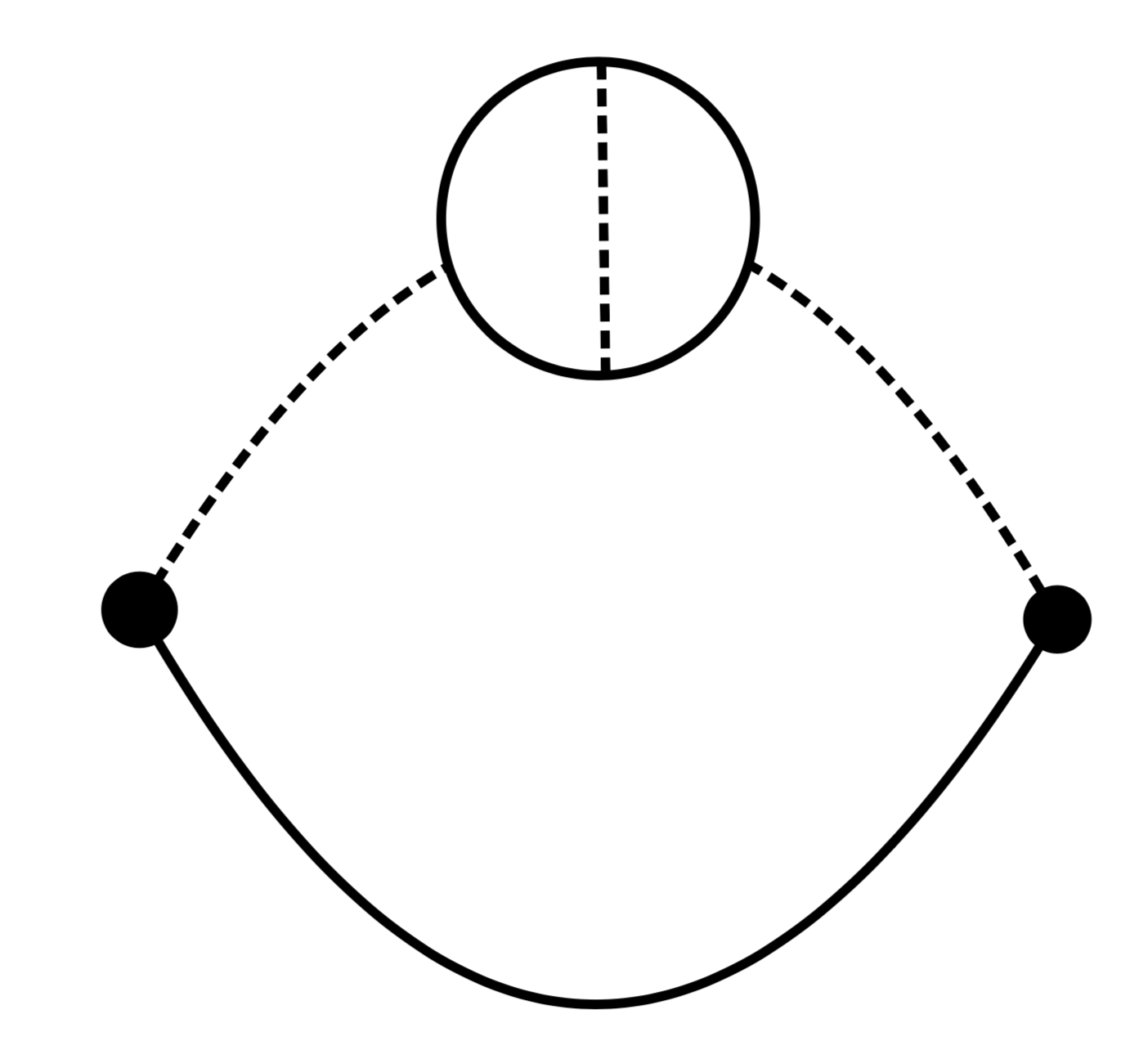}} at (0,0);
  \end{tikzpicture} 
  \right]+ \left[  \begin{tikzpicture}[baseline={([yshift=-.35cm]current bounding box.center)}, scale=0.22]
   \pgftext{\includegraphics[scale=0.65]{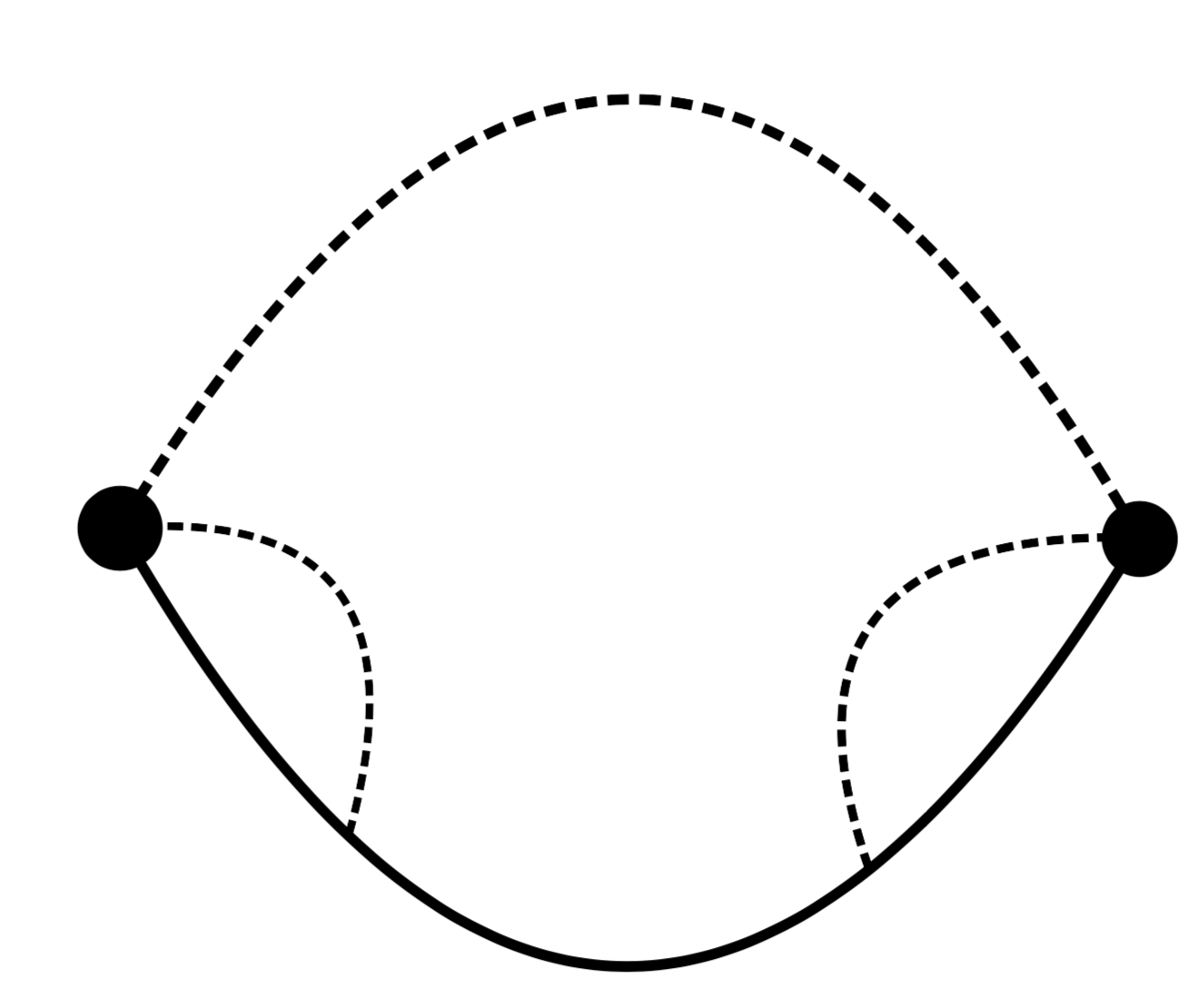}} at (0,0);
  \end{tikzpicture} \right] = 0 \log \Lambda + \dots \,,
\ee 
from which we conclude that $\eta =- \delta_\phi = \frac{32}{3\pi^2 N}$. Consequently, we find:
\be 
\Delta_{\phi^k \psi_i} = k+1+\frac{1}N \left(\frac{16k(3k-5)}{3\pi^2} + \frac{4}{3\pi^2} + \frac{32 k }{3\pi^2} \right) = k+1+\frac{48k(k-1) +4}{3\pi^2 N } + O\left(\frac{1}{N^2}\right)\,. 
\ee
In particular, we find 
\be 
\Delta_{\psi_i'}  &=  3+\frac{100}{3\pi^2 N} + O\left(\frac{1}{N^2}\right)\,,\\ 
\Delta_{\chi_i}  &=  4+\frac{292}{3\pi^2 N} + O\left(\frac{1}{N^2}\right)\,.
\ee

\section{More on the differences between the $O(N)$ and $O(N/2)^2 \rtimes \Z_2$ GNY models}
\label{sec:more-on-diff-GNY-vs-chiral-GNY}

\subsection{2- vs 4-component}
We will note that in the literature sometimes authors will describe the theories discussed in this paper using the language of 4-component fermions as opposed to the natively 3-dimensional 2-component language which we use in this paper. This is especially true in the case of \(4-\epsilon\) calculations. A Lagrangian may look like
\be\label{eq:chiral4lagrangian}
	\cL_\chi = -\frac12 (\partial \phi)^2 - i\bar{\Psi}_i\slashed{\ptl}\Psi_i - \frac{1}{2}m^2\phi^2 - \frac{\lambda}{4}\phi^4 - i g \phi \bar{\Psi}_i\Psi_i ,
\ee
where \(\Psi_i\) are 4-component Dirac spinors\footnote{Here we define \(\bar\Psi_i\equiv \Psi_i^\dagger \widetilde\gamma^0\), where \(\widetilde\gamma^0\) is defined in \ref{eq:4-gammas}. The Clifford algebra satisfies \(\{\widetilde\gamma^\mu,\widetilde\gamma^\nu\}= 2\eta^{\mu\nu}\), just like the \(2\times 2\) gamma matrices defined in \cite{Iliesiu:2015qra} that we use in this paper.} and \(i=1\dots N_4,~N_4\equiv N/4\). Notably, the Lagrangian has an additional discrete ``chiral" symmetry of \(\Psi_i\to \gamma^5\Psi_i\), \(\phi\to -\phi\), inhereted from the four-dimensional theory.\footnote{We denote the ``chiral'' symmetry in quotes as there is no inherent spacetime notion of chirality in three (or indeed, any odd number of) dimensions. This symmetry amounts to an internal flavor symmetry of the fermions.}
Models of this form have been studied extensively to understand the chiral Ising universality class, which for \(N=4,8\) (i.e.~\(N_4=1,2\)) describe spinless/spinful critical points for the semimetal-to-CDW transition in graphene. The critical points describe breaking of the symmetry while preserving time-reversal symmetry \cite{Ihrig:2018hho,Zerf:2017zqi}.

To make the distinction between this theory more clear from the theory that is the focus of our paper, it is helpful to consider decomposing our 4-component Dirac spinors into 2-component Majorana spinors. In a suitably convenient \(4\times 4\) basis of the gamma matrices, such as the following defined in \cite{Boyack:2018zfx,Kubota:2001kk} in terms of the \(2\times 2\) \(\gamma^\mu\) we have used throughout this paper (defined in \cite{Iliesiu:2015qra}), we have:
\be
\label{eq:4-gammas}
\widetilde{\gamma}^\mu = \begin{pmatrix}
	\gamma^\mu & 0 \\ 0 & -\gamma^\mu
\end{pmatrix},~\mu=0,1,2;\quad \widetilde\gamma^3 = \begin{pmatrix}
	0 & -i\\i & 0
\end{pmatrix}.
\ee

We should note that the \(\widetilde\gamma^3\) will not play a role in the 3d theory; thus this gamma matrix basis reduces to a block-diagonal form in 3d. Therefore, the 4-component spinors can be broken down into 2-component spinors as
\be
\Psi_i = \begin{pmatrix}
	\psi^L_i\\
	\psi^R_i
\end{pmatrix} = \begin{pmatrix}
	\psi_i\\
	\psi_{i+N_4}
\end{pmatrix}.
\ee 

We get
\be
\cL_\chi =  -\frac12 (\partial \phi)^2 - i\bar{\psi}_i\slashed{\ptl}\psi_i - \frac{1}{2}m^2\phi^2 - \frac{\lambda}{4}\phi^4 - i g \phi (\bar{\psi}_i^L\psi_i^L - \bar{\psi}_i^R\psi_i^R).
\ee

(This can be further decomposed into the requisite Majorana spinors without significant difference in form, save for the index \(i\) going from 1 to \(2N_4\).) In this form, we see that we have two distinct fermion species with opposite signs on their Yukawa couplings; the chiral symmetry becomes \(\psi_i^L \leftrightarrow \psi_i^R\), \(\phi \to -\phi\). We can see that, if there are \(N\) Majorana fermions total, the symmetry of this Lagrangian is \(O(N/2)^2\rtimes\Z_2\), where \(\Z_2\) is the chiral symmetry. For completeness, we will note that the 4-component spinor notation for the fermion bilinear which appears in (\ref{eq:lagrangian}) is \(i\bar\Psi_i\gamma^3\gamma^5\Psi_i\).

\subsection{Irreps of $O(N/2)^2\rtimes \Z_2$}

In the next subsection we will classify the various low-lying primary operators of the  $O(N/2)^2 \rtimes \Z_2$ GNY theory. To attempt such a classification, we should first discuss the irreps of $O(N/2)^2 \rtimes \Z_2$. We will start with a more general discussion. Given a compact simple Lie group $G$, let $H=(G\times G)\rtimes \Z_2$. The group $H$ is generated by $(g_L,g_R)\in G\x G$, together with an element $s$ such that $s^2=1$ and
\be
s(g_L,g_R)=(g_R,g_L)s.
\ee

For each irrep $\rho$ of $H$, we can consider its restriction to $G\x G$. The irreps of $G\x G$ have the form $\rho_1\boxtimes \rho_2$ where $\rho_1,\rho_2$ are irreps of $G$. The symbol $\boxtimes$ means we take a tensor product as vector spaces, but not as $G$ representations. The first $G$ acts on the left-tensor factor and the second $G$ acts on the right tensor factor.

Suppose the restriction of $\rho$ contains the $G\x G$ irrep $\rho_1\boxtimes \rho_2$. That is, we have a $G\x G$ homomorphism
\be
\phi \in \mathrm{Hom}_{G\x G}(\rho_1\boxtimes \rho_2,\rho).
\ee
Now consider $\phi':\rho_2\boxtimes \rho_1 \to \rho$ given by
\be
\phi'(v_2\otimes v_1) &\equiv s \phi(v_1\otimes v_2),
\ee
where $s\f(v_1\otimes v_2)$ denotes the action of $s\in H$ on the vector $\f(v_1\otimes v_2) \in \rho$.
We have
\be
\phi'((g_L,g_R)(v_2\otimes v_1))
&= \phi'(g_Lv_2\otimes g_R v_1) \nn\\
&= s \phi(g_R v_1\otimes g_L v_2) \nn\\
&= s(g_R,g_L) \phi(v_1\otimes v_2) \nn\\
&= (g_L,g_R) \f'(v_2\otimes v_1),
\ee
so we see that $\phi'$ is a $G\x G$ homomorphism from $\rho_2\boxtimes \rho_1$ to $\rho$:
\be
\phi':\mathrm{Hom}_{G\x G}(\rho_2\boxtimes \rho_1,\rho).
\ee
We have two cases to consider.

\begin{itemize}

\item Suppose that $\rho_1,\rho_2$ are distinct. Clearly $\f(\rho_1\boxtimes \rho_2)+\f'(\rho_2\boxtimes \rho_1)$ is an $H$-invariant subspace of $\rho$, and thus must be all of $\rho$. Since each of $\f,\f'$ is a $G\x G$ isomorphism onto its image, this furnishes an isomorphism 
\be
\rho \cong (\rho_1\boxtimes \rho_2) \oplus (\rho_2\boxtimes \rho_1)
\ee
as $G\x G$ representations. Furthermore, the two summands are swapped by the action of $s$. We denote the corresponding $H$-representation by $\rho=\<\rho_1,\rho_2\>$.

\item  Suppose that $\rho_1=\rho_2$. In this case, we claim that $\rho_\pm = (\f\pm \f')(\rho_1\boxtimes \rho_1)$ are $H$-invariant subspaces of $\rho$. Clearly they are $G\x G$-invariant subspaces. To prove they are also $H$-invariant, note that
\be
s(\f \pm \f')(u\otimes v) &=s(\f(u\otimes v) \pm s \f(v\otimes u))  \nn\\
&= s\f(u\otimes v) \pm \f(v\otimes u)\nn\\
&= \pm (\f\pm \f')(v\otimes u) \in \rho_\pm.
\ee
By irreducibility, one of the $\rho_\pm$ must be all of $\rho$, and the other must vanish. When $\rho=\rho_+$ or $\rho=\rho_-$, we denote the $H$-representation by $\<\rho_1,\rho_1\>_\pm$. A basis of $\<\rho_1,\rho_1\>_\pm$ is given by $u\otimes v$ with $u,v\in \rho_1$, with the $s$-action
\be
s(u\otimes v) &= \pm v\otimes u.
\ee
\end{itemize}

Below, when listing the perturbative estimates for the scaling dimensions for some of the low-lying operators in the theory we will be interested in $\rho_i \in \{\bullet,\,\myng{(1)},  \myng{(2)},  \myng{(1,1)}\}$ which corresponds to the singlet, vector, symmetric traceless tensor and antisymmetric representations. 

\subsection{Large-$N$ computations in the  $O(N/2)^2 \rtimes \Z_2$ GNY model}
\label{sec:largeNcGN}

As explained in section \ref{sec:two-GNYs}, the large-$N$ Feynman diagrams in the $O(N)$ GNY model and  $O(N/2)^2 \rtimes \Z_2$ GNY model only differ by diagrams that contain fermionic loops with five or more fermion propagators. For instance, in the GNY model we have that 
\be
\Delta_{\psi_i} &\supset \begin{tikzpicture}[baseline={([yshift=-1ex]current bounding box.center)}, scale=0.12 ]
 \pgftext{\includegraphics[scale=1.0]{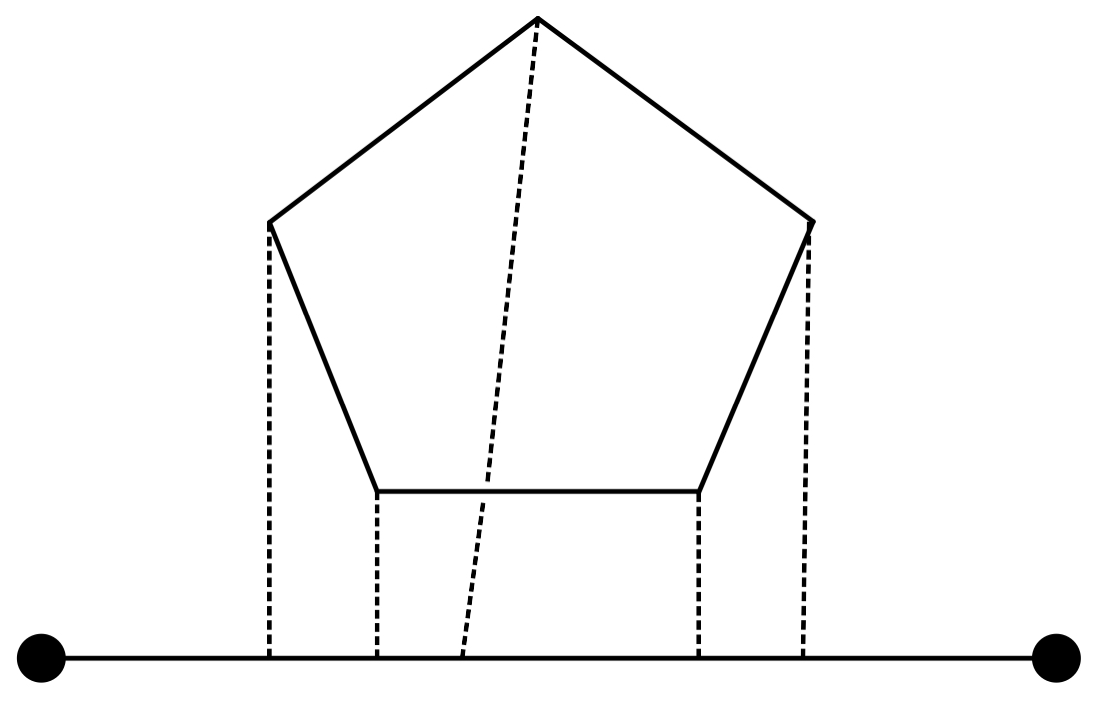}} at (0,0);
  \draw (-18.5,-6.0) node {$\psi_i(x)$};
    \draw (18.5,-6.0) node {$\psi_i(y)$};
  \end{tikzpicture}  \sim \frac{1}{N^4} \,, \nn \\
\Delta_{\phi^2} &\supset \begin{tikzpicture}[baseline={([yshift=0ex]current bounding box.center)}, scale=0.12 ]
 \pgftext{\includegraphics[scale=1.0]{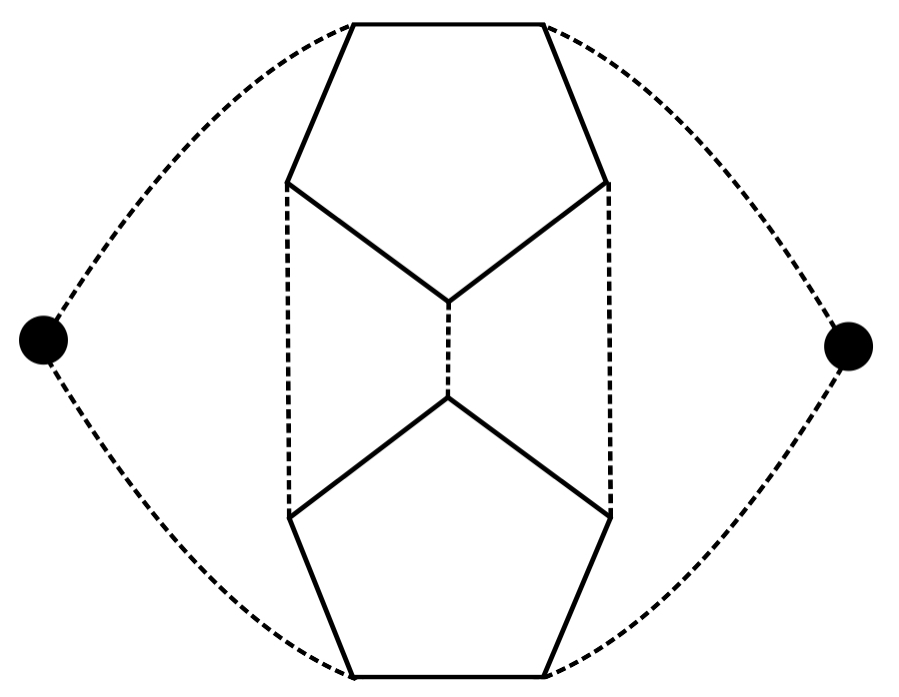}} at (0,0);
  \draw (-15.5,-6.0) node {$\phi^2(x)$};
    \draw (15.5,-6.0) node {$\phi^2(y)$};
  \end{tikzpicture}     \sim \frac{1}{N^3} \,,
 \ee
 where we have again listed examples of diagrams (rather than all the diagrams) contributing to the leading non-vanishing order. As explained in section \ref{sec:theoretical-background-GNY}, these diagrams cancel in the $O(N/2)^2 \rtimes \Z_2$ GNY models. More generally, since such loops only contribute at higher order we see that at leading order 
\be
\Delta^{O(N)}_{\cO_\rho} = \Delta^{\ONc}_{\cO_{\<\rho, \bullet\>}} + O\left(\frac{1}{N^3}\right)\,,
\ee
where $\cO$ can represent an operator with any spin and parity representation, $\rho$ can be one of the $\{{\myng{(1)}},\myng{(1,1)}, \myng{(2)}\}$ irreps of $O(N)$ or $O(N/2)$, and we have added the $O(N)$ and $\ONc$ superscripts to differentiate between the two models.

If $\rho = \bullet$, then we have to consider the $\ONc$ irreps $\<\bullet, \bullet\>_\pm$. For operators that have the same representation under parity as under the chiral $\Z_2$, their scaling dimensions match that of the operator with the same parity in the $O(N)$ GNY model for low enough order in the large-$N$ expansion: for instance, $\Delta_{\sigma_{\<\bullet, \bullet\>_-}}^{\ONc} \approx  \Delta_{\sigma}^{O(N)}$ and  $\Delta_{\epsilon{\<\bullet, \bullet\>_+}}^{\ONc} \approx  \Delta_{\epsilon}^{O(N)}$. If that is not the case, a more elaborate analysis is needed. For example, consider the operator $\sigma_{\<\bullet, \bullet\>_+} \sim \psi\,^L_i \psi^L_i + \psi\,^R_i \psi^R_i$.\footnote{This is in a different irrep than $\phi$ itself which as mentioned above lies in $\<\bullet, \bullet\>_-$, but transforms in the same way under parity.} The diagrams contributing to the two-point function of this operator are precisely the same as those contributing to the two-point function of $\sigma_{\symtensor} $. The fact that the $O(N/2)$ indices are contracted differently does not matter at low enough order when computing the scaling dimension of the operator and only affects the overall normalization of the two-point function. Therefore, we find that 
\be 
\Delta_{\sigma_{{\tiny \myng{(2)}}}}^{O(N)} \approx \Delta_{\sigma_{\symtensor}}^{\ONc} \approx \Delta_{\sigma_{\<\bullet, \bullet\>_+}}^{\ONc}\,.
\ee
This approximate relation between the scaling dimensions in the singlet sector and those in the $\symtensor$ irrep can be extended to the parity even sector. For instance, consider the operator $\epsilon_{\sigma_{{\tiny \myng{(2)}}}} \sim \phi  \psi_{(i} \psi_{j)} $ in the $O(N)$ model and the operators $\epsilon_{\symtensor} \sim \phi  \psi_{(i}\,^{L, R} \psi_{j)}^{L, R} $ and $\epsilon_{\<\bullet, \bullet\>_-} \sim \phi( \psi\,^L_i \psi^L_i + \psi\,^R_i \psi^R_i)$ in the  $O(N/2)^2 \rtimes \Z_2$ model. Once again the large-$N$ Feynman diagrams of the above operators are identical and we find 
\be 
\Delta_{\epsilon_{{\tiny \myng{(2)}}}}^{O(N)} \approx \Delta_{\epsilon_{\symtensor}}^{\ONc} \approx \Delta_{\epsilon_{\<\bullet, \bullet\>_-}}^{\ONc}\,.\ee

\begin{table}[t!]
\begin{center}
\renewcommand{\arraystretch}{1.2}
\resizebox{\columnwidth}{!}{
\begin{tabular}{l |c c c|l |c}
\hline \hline Operator~ & Spin &Parity & Global irrep. & $\Delta$ at large $N$&  GNY op.\\
\hline  $\psi_{\<{\myng{(1)},\bullet}\>} \sim \psi_i^{L,R}$ & $\frac{1}2$ & $+$ & $\<{\myng{(1)},\bullet}\>$ & $1+ \frac{4}{3\pi^2 N} + \dots$ & ${\psi_{\tiny \myng{(1)}}}$ \\
 $\psi_{\<{\myng{(1)},\bullet}\>}' \sim \phi^2 \psi_i^{L,R}$& $\frac{1}2$  & $+$ & $\<{\myng{(1)},\bullet}\> $ & $3+ \frac{100}{3\pi^2 N} +  \dots$ & ${\psi'_{\tiny \myng{(1)}}}$  \\
  $\chi_{\<{\myng{(1)},\bullet}\>} \sim \phi^3 \psi_i$ & $\frac{1}2$ & $-$ & $\<{\myng{(1)},\bullet}\>$ & $4+ \frac{292}{3\pi^2 N}+  \dots$ & ${\chi_{\tiny \myng{(1)}}}$ \\
 $ \sigma_{\<\bullet, \bullet\>_-}\sim \phi$ & $0$ & $-$  & $\<\bullet, \bullet\>_-$ & $1- \frac{32}{3\pi^2 N} + \dots$ &  $\sigma_\bullet$\\
  $\epsilon_{\<\bullet, \bullet\>_+}\sim \phi^2$ & $0$ & $+$ &  ${\<\bullet, \bullet\>_+}$ & $2+\frac{32}{3\pi^2 N}  + \dots$ & $\epsilon_\bullet$ \\ 
  $\sigma_{\<\bullet, \bullet\>_-}' \sim \phi^3 $  & $0$ & $-$ &  ${\<\bullet, \bullet\>_-}$ & $3+\frac{64}{\pi^2 N}  + \dots$ & $\sigma'_\bullet$ \\
  $\epsilon_{\<\bullet, \bullet\>_+}' \sim \phi^4 $ & $0$ & $+$ &  ${\<\bullet, \bullet\>_+}$ & $4+\frac{448}{3\pi^2 N}  + \dots$  & $\epsilon'_\bullet$ \\
  $\phi^k $ & $0$ & $(-1)^k$ & ${\<\bullet, \bullet\>_{(-1)^k}}$ & $k+\frac{16k(3k-5)}{3\pi^2 N}  + \dots$  & ${\phi^k}$\\
  $\sigma_{\symtensor} \sim \psi_{(i}^{L,R}\psi_{j)}^{L,R} $ & $0$ & $-$ & $\symtensor$ & $2+\frac{32}{3\pi^2 N} + \dots$ & ${\sigma_{\tiny \myng{(2)}}}$ \\
  $\sigma_{\<\bullet, \bullet\>_+} \sim \psi_{i}\,^{L}\psi_{i}^{L} + \psi_i\,^R \psi_i^R$ & $0$ & $-$ & $\<\bullet, \bullet\>_+$ & $2+\frac{32}{3\pi^2 N} + \dots$ & ${\sigma_{\tiny \myng{(2)}}}$ \\ 
 $\sigma_{\bifundamental_+} \sim  \psi\,^L_i \psi^R_j + \psi\,^R_i\psi_j^L $  & $0$ & $-$ & $\bifundamental_+$ & $2-\frac{16}{3\pi^2 N} + \dots$ & $-$ \\
  $\sigma_{\<\tiny \myng{(1,1)}, \bullet\>} \sim j^\mu_{\<\tiny \myng{(1,1)}, \bullet\>}\phi^2 \partial_\mu \partial^2 \phi $ & $0$ & $-$ & $\<\tiny \myng{(1,1)}, \bullet\>$ & $8+\dots$ & ${\sigma_{\tiny \myng{(1,1)}}}$ \\
 $j^\mu_{\<\tiny \myng{(1,1)}, \bullet \>} \sim  \psi\,^{L,R}_{[i} \gamma^{\mu} \psi^{L,R}_{j]} $  & $1$ & $+$ & $\<\tiny \myng{(1,1)}, \bullet\>$ & $2$ & $j^{\mu}_{\myng{(1,1)}}$ \\
 $j^\mu_{\bifundamental_-} \sim  \psi\,^L_{i} \gamma^{\mu} \psi^R_{j} - \psi\,^R_{i} \gamma^{\mu} \psi_{j}^L $  & $1$ & $+$ & $\bifundamental_-$ & $2+\frac{16}{3\pi^2 N} + \dots$ & $-$ \\
\hline \hline
\end{tabular}
}
\end{center}
\caption{\label{tab:O(N/2)-large-N} Estimates for the large-$N$ scaling dimensions at the  $O(N/2)^2 \rtimes \Z_2$ GNY critical point. The last column shows which scaling dimensions in the $O(N)$ GNY critical point match the dimensions of some operator in the $O(N/2)^2 \rtimes \Z_2$ GNY critical point at low orders in the large-$N$ expansion.}
\end{table}

Next, we discuss some of the operators in the $\ONc$ model whose scaling dimensions are not found among operators in the $O(N)$ model even at leading order in the large-$N$ expansion. One example of such an operator is $\sigma_{\bifundamental_+} \sim  \psi\,^L_i \psi^R_j + \psi\,^R_i\psi_j^L $. The two-point function of such an operator up to order $O(1/N^2)$ is given by 
\be 
\label{eq:sigma-bifundamental-correction}
 D_{\sigma_{\bifundamental_+}} (0, x) &= 
  \begin{tikzpicture}[baseline={([yshift=-.5ex]current bounding box.center)}, scale=0.22]
 \pgftext{\includegraphics[scale=1.0]{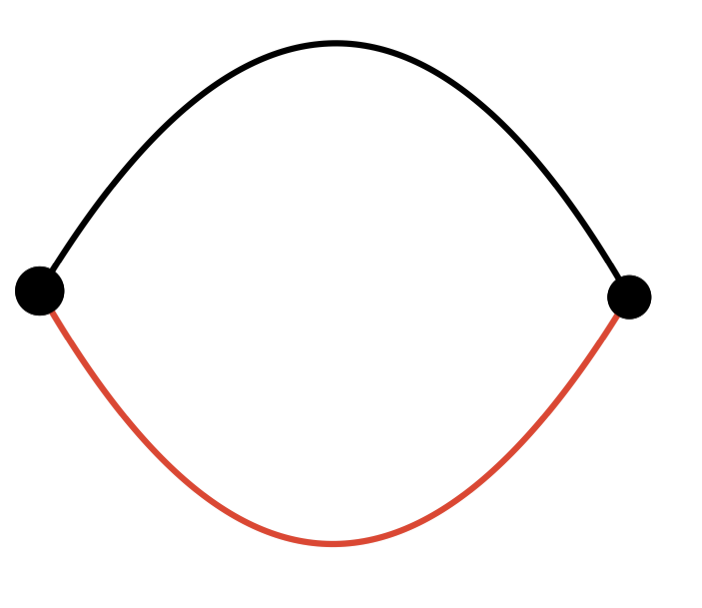}} at (0,0);
  \end{tikzpicture}  +   \begin{tikzpicture}[baseline={([yshift=-.5ex]current bounding box.center)}, scale=0.22]
 \pgftext{\includegraphics[scale=1.0]{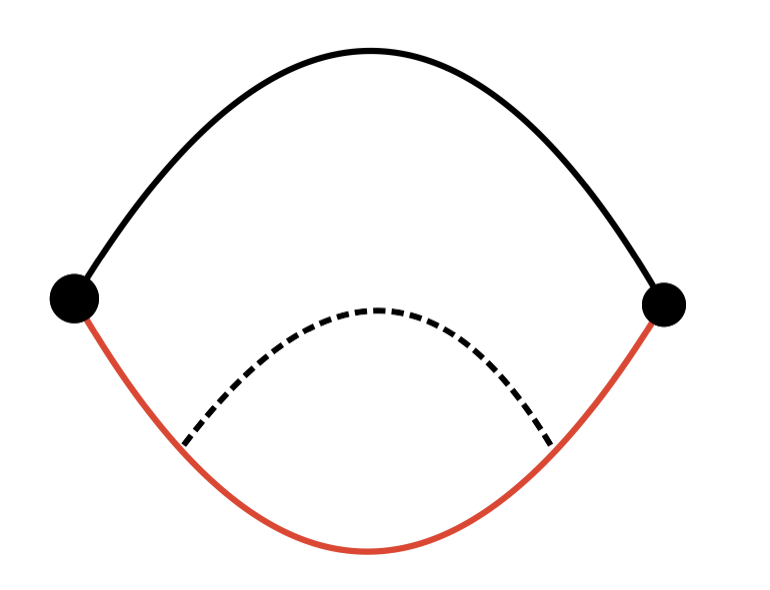}} at (0,0);
  \end{tikzpicture}  +  \begin{tikzpicture}[baseline={([yshift=-.5ex]current bounding box.center)}, scale=0.22]
 \pgftext{\includegraphics[scale=1.0]{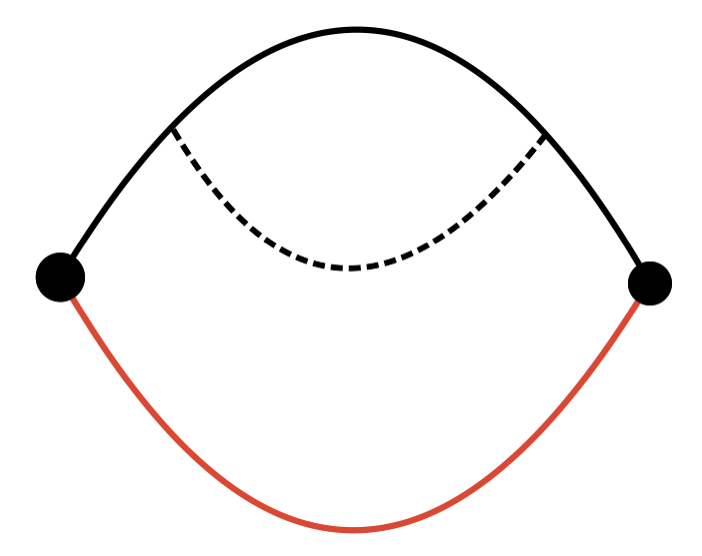}} at (0,0);
  \end{tikzpicture}  +  \begin{tikzpicture}[baseline={([yshift=-.5ex]current bounding box.center)}, scale=0.22]
 \pgftext{\includegraphics[scale=1.0]{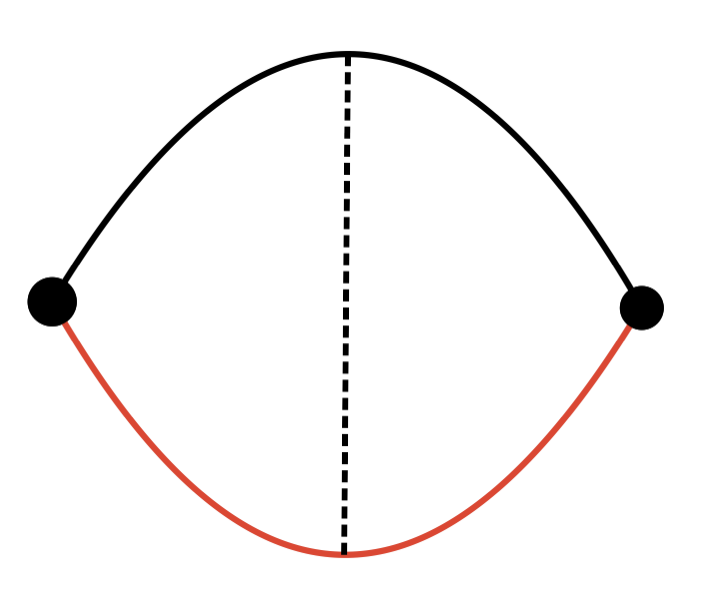}} at (0,0);
  \end{tikzpicture} \nn \\ & + O\left(\frac{1}{N^2}\right)\,.
\ee 
To extract the anomalous dimension of the operator, we need to read-off the coefficient of the logarithmic divergence in \eqref{eq:sigma-bifundamental-correction} for the last 
three diagrams $\delta_{\sigma_{\bifundamental_+}} = 2 \delta_\psi -\eta_\text{vertex}$ which are responsible for the $1/N$ correction. The first two diagrams contribute the same amount since $\delta_\psi = \delta_{\psi^L} = \delta_{\psi^R}$, while the last diagram was computed in appendix B of \cite{Iliesiu:2015qra} in order to compute the anomalous dimension $\delta_{\epsilon_{\tiny \myng{(2)}}}$ in the $O(N)$ model. The calculation is almost identical to there, the only difference is that the last diagram in    \eqref{eq:sigma-bifundamental-correction} has the opposite sign due to the difference in sign between the Yukawa coupling of $\psi^L$ and $\psi^R$. Therefore, 
\be 
\Delta_{\sigma_{\bifundamental_+}} = 2 + \delta_{\sigma_{\bifundamental_+}} =  2-\frac{16}{3\pi^2} \frac{1}N + O\left(\frac{1}{N^2}\right)\,.
\ee
Similarly, we see that the diagrams needed to compute the scaling dimension of $\epsilon_{\bifundamental_-} \sim \phi( \psi\,^L_i \psi^R_j + \psi\,^R_i\psi_j^L)$ and $\epsilon_{\bifundamental_+} \sim \sigma_+ ( \psi\,^L_i \psi^R_j + \psi\,^R_i\psi_j^L)$ from the $\ONc$ model are not among those found in the $O(N)$ model. 

Both the $O(N)$ GNY and $O(N/2)^2 \rtimes \Z_2$ GNY models have conserved spin-one currents, with the $O(N)$ model having $j^\mu_{\myng{(1,1)}} \sim  \psi_{[i}\gamma^\mu\psi_{j]}$ in the antisymmetric irrep of $O(N)$ and $j^\mu_{\antisymtensor} \sim   \psi\,^{L, R}_{[i}\gamma^\mu\psi_{j]}^{L,R}$ in the antisymmetric representation of each $O(N/2)$ subgroup.  However, to further distinguish the two models and to ensure that the $\ONc$ does not have a symmetry enhancement in the IR to a Lie group with a greater number of generators, we can compute the anomalous dimension of the bifundamental current $j^\mu_{\bifundamental_-} \sim  \psi\,^L_{i} \gamma^\mu \psi_{j}^R -  \psi\,^R_{i} \gamma^\mu \psi_{j}^L $. 
The computation for the anomalous dimension of this spin-1 operator follows from the computation which shows that the conserved current $j^\mu_{\antisymtensor}$ has no anomalous dimension. Specifically, 
\be 
\delta_{j^\mu_{\antisymtensor}} = 0 = 2\delta_\psi+ \tilde \eta_\text{vertex}\,, \qquad \delta_{j^\mu_{\bifundamental_-}}=  2\delta_\psi - \tilde \eta_\text{vertex}\,,
\ee
from which it follows that 
\be 
\Delta_{j^{\mu}_{\bifundamental_-}} = 2 + \frac{16}{3\pi^2} \frac{1}N + O\left(\frac{1}{N^2}\right)\,.
\ee
To summarize, while operators in the $O(N)$ GNY model have associated operators in the $O(N/2)^2 \rtimes \Z_2$ GNY model that have the same scaling dimension at low orders in the large-$N$ expansion, the reverse is not true. We review all discussed large-$N$ estimates in the  $O(N/2)^2 \rtimes \Z_2$ model in table \ref{tab:O(N/2)-large-N}.

\newpage
\bibliographystyle{JHEP}
\bibliography{refs}

\end{document}